\documentclass[12pt]{article}
\usepackage{amsmath}
\usepackage{jheppub}
\usepackage{amssymb,amsfonts}
\usepackage{latexsym}
\usepackage{mathtools}
\usepackage{tikz-cd}
\usepackage{tikz}
\usepackage{tikz-3dplot}
\usetikzlibrary{knots}
\usetikzlibrary{calc}
\usetikzlibrary{topaths}
\usetikzlibrary{decorations}
\usetikzlibrary{decorations.pathmorphing}
\usetikzlibrary{arrows,decorations.markings,cd}
\usetikzlibrary{calc,arrows,cd,decorations.markings,snakes}
\usetikzlibrary{arrows.meta, shapes.geometric}
\usetikzlibrary{backgrounds} 
\usetikzlibrary{braids}
\usetikzlibrary{tqft}

\usepackage{enumerate}
\usepackage{empheq}
\usepackage{dsfont}
\usepackage{hyperref}
\relax
\renewcommand{\theequation}{\arabic{section}.\arabic{equation}}

\def\be{\begin{equation}}
	\def\ee{\end{equation}}
\def\bea{\begin{eqnarray}}
	\def\eea{\end{eqnarray}}

\renewcommand{\vec}[1]{\boldsymbol{#1}}

\newcommand{\abs}[1]{\left|#1\right|}

\usepackage{amsthm}

\newcommand{\id}{\mathds{1}}

\renewcommand{\P}{\mathbb P}

\newcommand{\w}{\mathfrak w}

\newcommand{\tr}{\text{tr}}

\newcommand{\Z}{\mathcal Z}
\newcommand{\Hom}{\operatorname{Hom}}

\newcommand{\T}{\mathcal T}
\newcommand{\V}{\mathfrak V}
\newcommand{\SL}{\operatorname{SL}}

\newcommand{\Bord}{\boldsymbol{\operatorname{Bord}}}
\newcommand{\Vect}{\boldsymbol{\operatorname{Vect}}}

\tikzset{
	pics/torus/.style n args={3}{
		code = {
			\providecolor{pgffillcolor}{rgb}{1,1,1}
			\begin{scope}[
				yscale=cos(#3),
				outer torus/.style = {draw,line width/.expanded={\the\dimexpr2\pgflinewidth+#2*2},line join=round},
				inner torus/.style = {draw=pgffillcolor,line width={#2*2}}
				]
				\draw[outer torus] circle(#1);\draw[inner torus] circle(#1);
				\draw[outer torus] (180:#1) arc (180:360:#1);\draw[inner torus,line cap=round] (180:#1) arc (180:360:#1);
			\end{scope}
		}
	}
}

\newcommand{\hb}[1]{\vcenter{\hbox{\begin{tikzpicture}[fill=none,draw=black]
				\pic{torus={1cm}{2.8mm}{70}};
				\ifthenelse{\equal{#1}{0}}{}{\draw [yscale=cos(70),black,thick](1.05,0.08) arc (0:285:1.05cm);
				\draw [yscale=cos(70),black,thick](1.05,0.08) arc (0:-45:1.05cm);
				\node [black] at (0.5,-0.3) {$#1$};}
\end{tikzpicture}}}}

\newcommand{\defect}[1]{\vcenter{\hbox{\begin{tikzpicture}[fill=none,draw=black]
				\pic{torus={1cm}{2.8mm}{70}};
				\ifthenelse{\equal{#1}{0}}{}{\draw [dashed,yscale=cos(70),black,thick](1.05,0.08) arc (0:285:1.05cm);
					\draw [dashed,yscale=cos(70),black,thick](1.05,0.08) arc (0:-45:1.05cm);
					\node [black] at (0.5,-0.3) {$#1$};}
\end{tikzpicture}}}}

\title{A TQFT perspective on satellite constructions and toral decompositions}

\author{Nikolaos Angelinos}
\affiliation{Yau Mathematical Sciences Center, Tsinghua University, Beijing 100084, China} 
\emailAdd{angelinosn@mail.tsinghua.edu.cn}

\abstract{We consider the classical satellite and splice constructions of knot theory from the perspective of link-complement states in three-dimensional topological quantum field theory (TQFT). These states are prepared by a TQFT on a link-complement manifold, namely a manifold with multiple disjoint torus boundaries, obtained by removing a thickened link from a closed ambient three-manifold. In this setting, satellite and splice constructions are realized by gluing manifolds along torus boundaries, and the TQFT assigns to these building blocks multilinear operators acting on the corresponding torus Hilbert spaces. We give explicit prescriptions for these operators and show how they can be used to build link-complement states.
	This construction is closely related to the JSJ decomposition of link-complement manifolds. While splicing builds manifolds by gluing along tori, the JSJ decomposition cuts a manifold along essential tori into canonical pieces.
	 From the TQFT point of view, this expresses a link-complement state as a network of elementary operators associated with the JSJ pieces. 
	 We illustrate this framework by studying the entanglement entropy of Seifert-fibered link complements, Hopf-chain complements, and complements of Whitehead-doubled Hopf chains, showing how their different JSJ structures lead to distinct entanglement patterns.
	 }

\makeatletter
\gdef\@fpheader{}
\makeatother

\begin{document}
	
\maketitle

\section{Introduction}

Topological quantum field theory (TQFT) provides a bridge between topology and quantum mechanics. 
Knots and links embedded in three-manifolds provide a natural setting for exploring this connection. A basic insight, going back to Witten \cite{Witten:1988hf}, is that knot and link invariants can be evaluated by TQFT amplitudes. The same framework also assigns multipartite quantum states to links embedded in three-manifolds. These are called link-complement states, since they are prepared by the TQFT on the link complement, namely the manifold obtained by removing a tubular neighbourhood of the link from the ambient closed manifold. For example, the complements of a trefoil knot and a Hopf link in $S^3$ are manifolds with one and two torus boundaries, respectively:
$$\centering{\vcenter{\hbox{\includegraphics[width=0.22\textwidth]{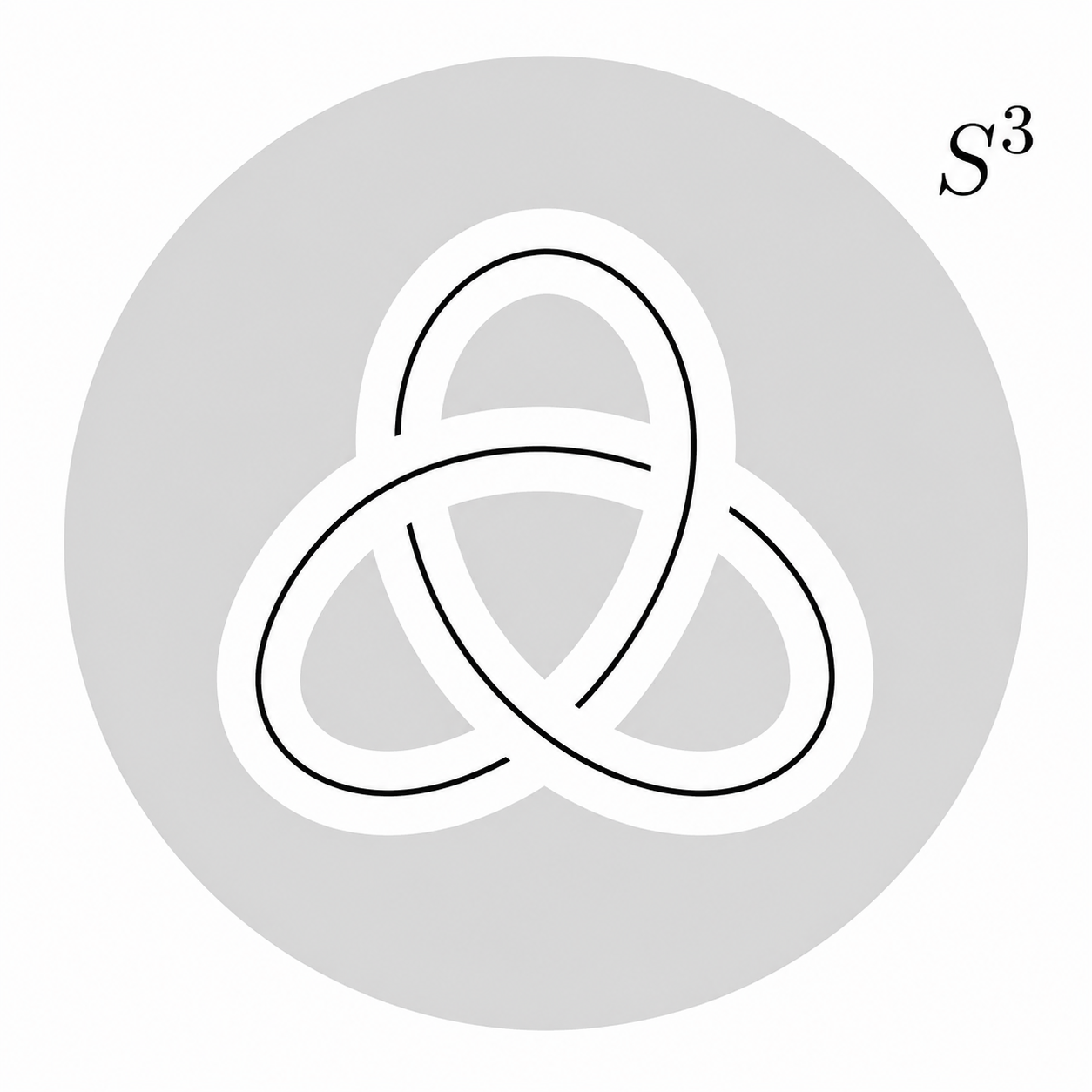}}}\hspace{50pt}\vcenter{\hbox{\includegraphics[width=0.22\textwidth]{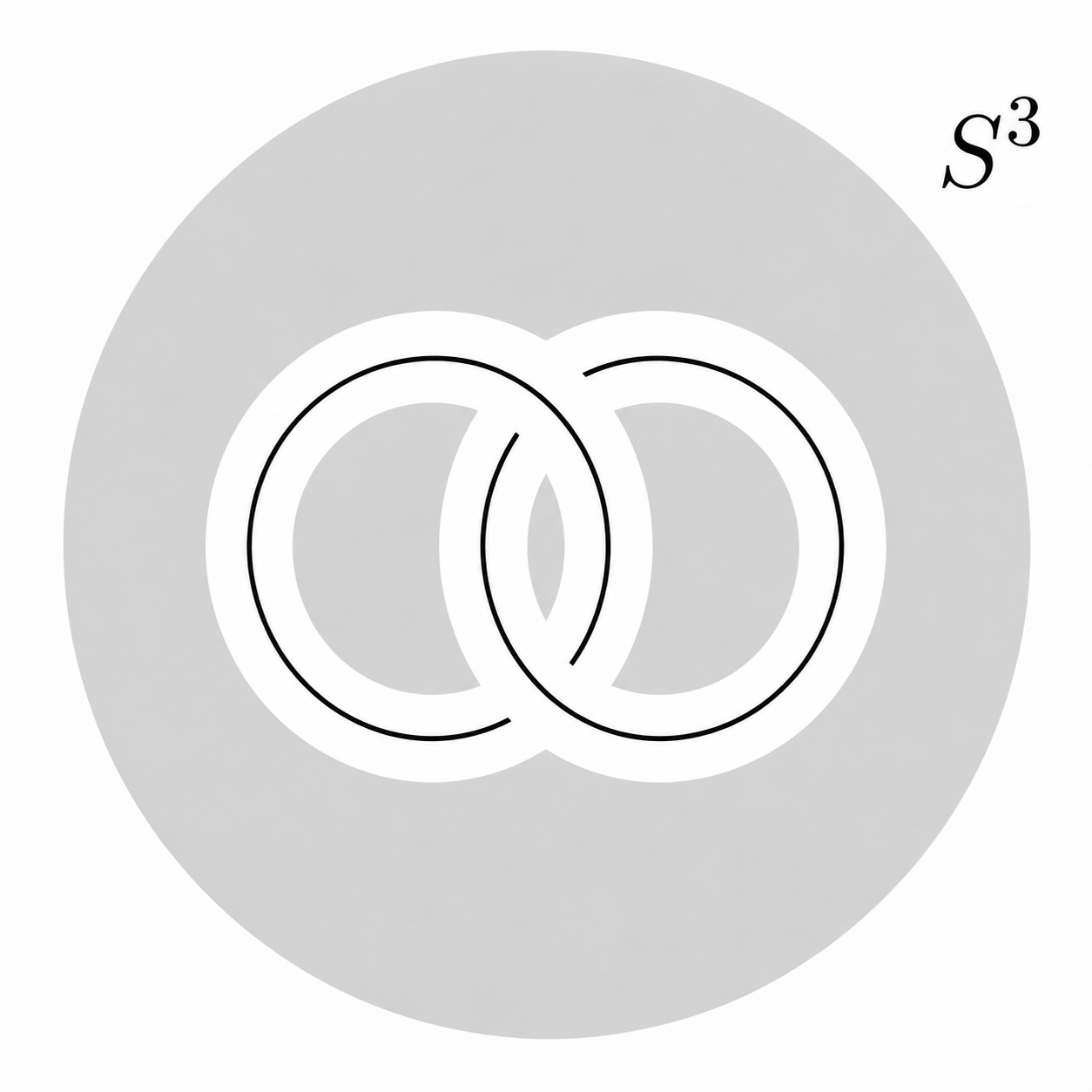}}}}$$
Above, the shaded region represents the manifold $S^3$, the lines represent knots, and the white regions represent the tubular neighborhoods of the knots that have been drilled out of $S^3$.

More generally, if a link has $n$ components, then its complement has $n$ disjoint torus boundaries. The TQFT assigns to such a manifold a multipartite state living in the $n$-fold tensor product of the torus Hilbert space. In this way, links give a topological realization of multipartite quantum states, in which properties of the state, such as its entanglement structure, are reflected in the topology of the link \cite{Salton_2017,Balasubramanian:2016sro}. This perspective has led to a growing body of work on link-complement states \cite{Balasubramanian_2018,Balasubramanian:2025kaf, yuan2025multientropylinkingchernsimonstheory,Large-party,Saini_2025, ramirezvaldez2024remarksentanglemententropyhopf,Hung_2018,Buican_2020, Dwivedi_2018,Dwivedi_2020,Dwivedi_2021,Cummings_2025,Camilo_2019}.

The central idea of this paper is to treat the classical satellite and splice constructions of knot theory as operators acting on link-complement states. This operator viewpoint is implicit in both the Chern-Simons path-integral and the axiomatic TQFT frameworks. Our goal is to make this observation explicit for satellite and splice constructions, to give practical formulas for the resulting operators, and to use them as building blocks for more complicated link-complement states. This provides a systematic way to go beyond the simple examples of prime links and connected sums that have mostly appeared in the literature. Conversely, since splicing is closely related to toral decompositions of three-manifolds, the same formalism gives a way to decompose link-complement states into networks of simpler multilinear maps.

We begin with the satellite construction, which is a standard way of building complicated knots from simpler ones. To construct a satellite, we first embed a knot, called the pattern, inside a solid torus. Then, we shape the entire solid torus into another knot, called the companion. Under this process, the pattern knot is carried to a new knot, called the satellite. For example:
\be \vcenter{\hbox{$\underset{\text{Companion knot}}{\includegraphics[width=0.23\textwidth]{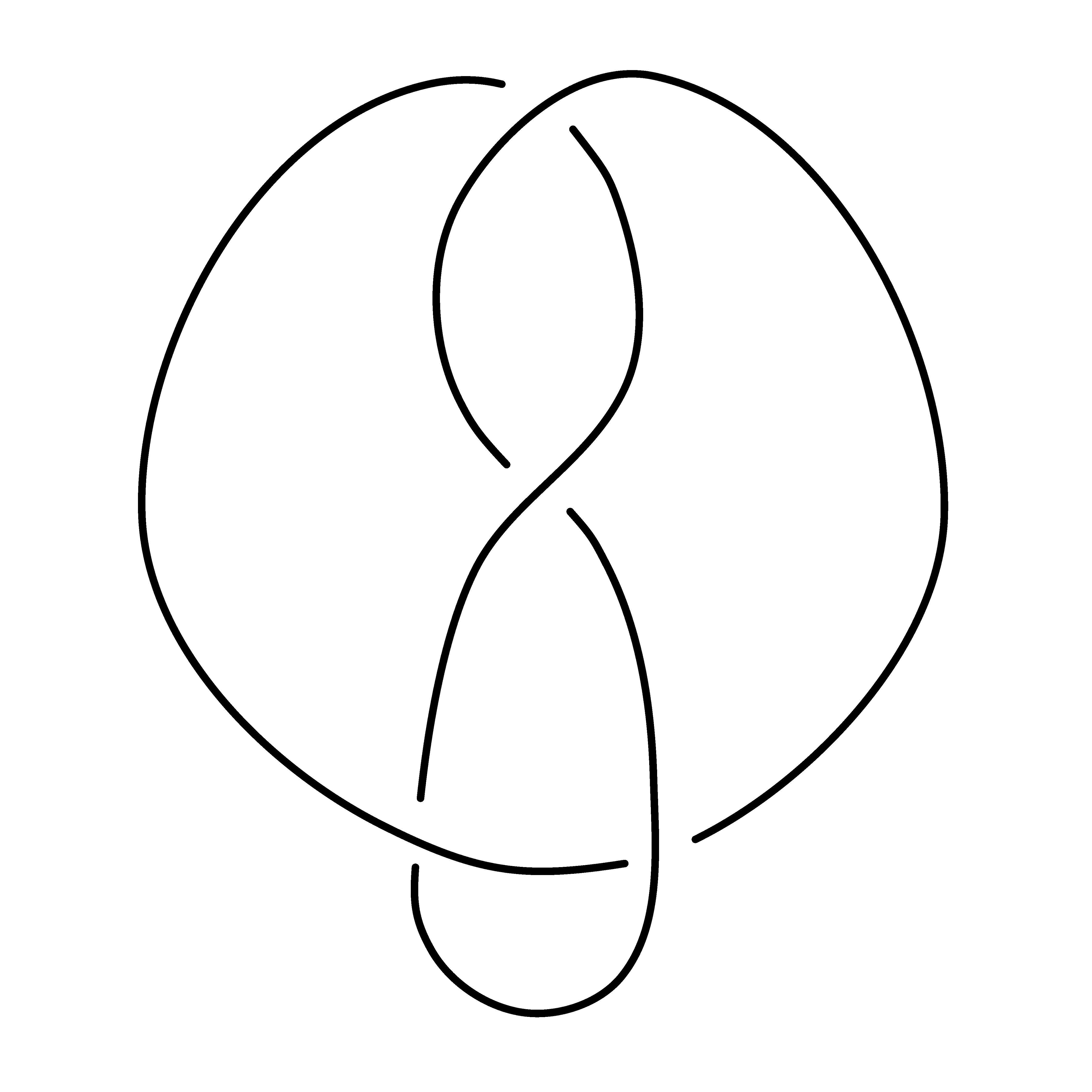}}$}}\circ
\vcenter{\hbox{$\underset{\text{Pattern knot}}{\includegraphics[width=0.25\textwidth]{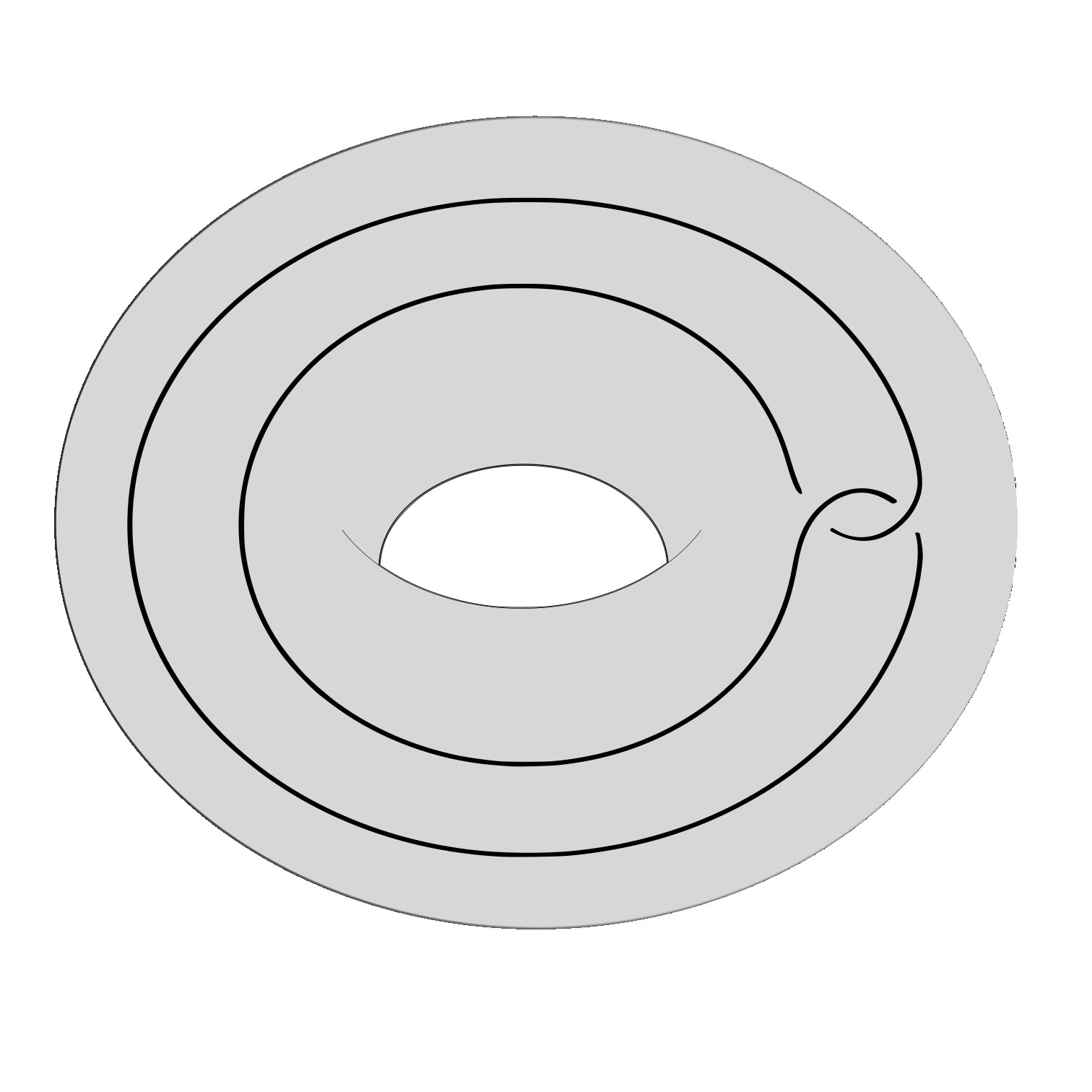}}$}}=
\vcenter{\hbox{$\underset{\text{Satellite knot}}{\includegraphics[width=0.25\textwidth]{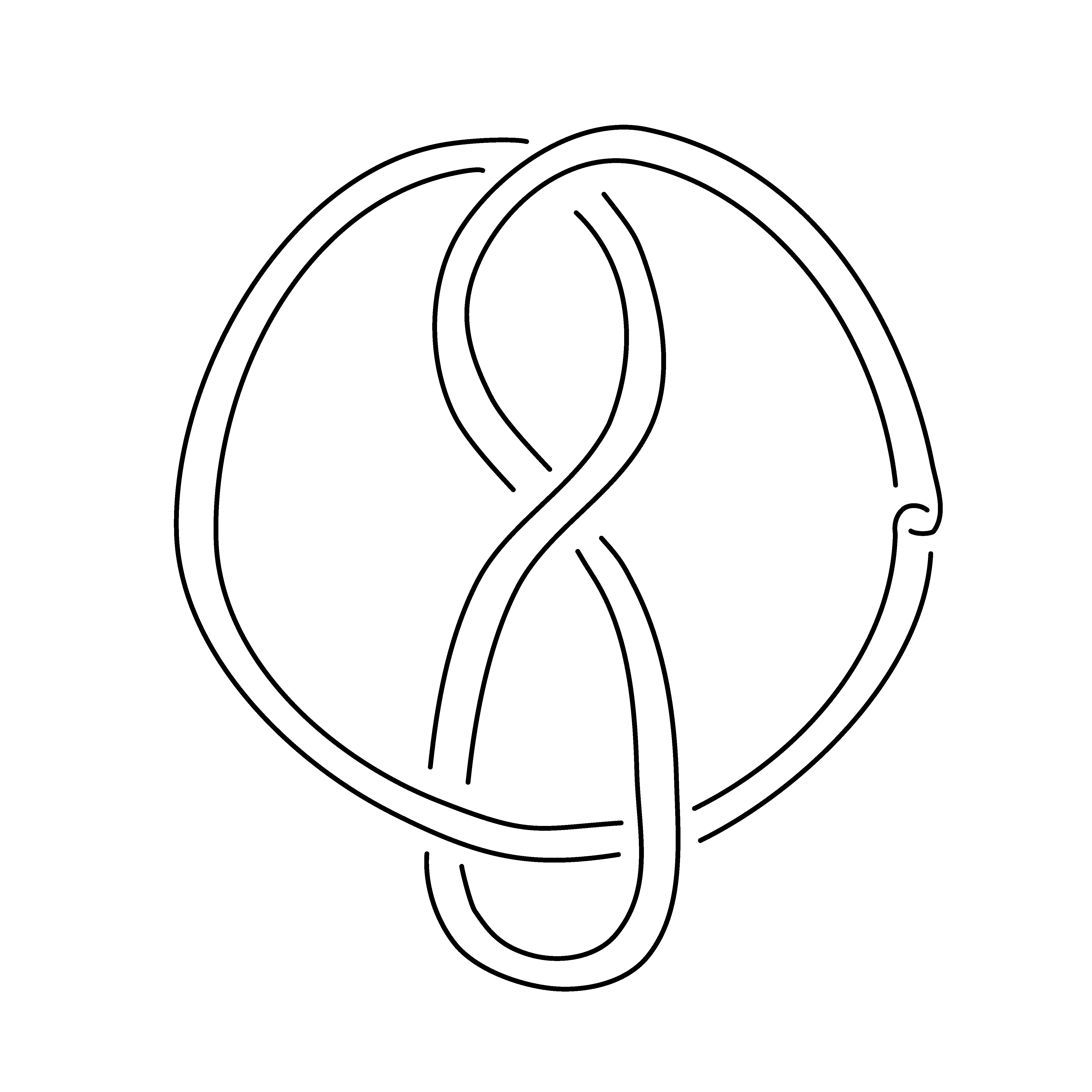}}$}}\label{trefdouble}\ee

The key property is that, at the level of link complements, the satellite operation is implemented by gluing manifolds along torus boundaries. The complement of the pattern inside the solid torus has two torus boundaries: an outer boundary coming from the solid torus itself, and an inner boundary coming from drilling out the pattern. To form the satellite complement, we glue the outer boundary to the complement of the companion knot. The remaining inner boundary then becomes the boundary of the satellite complement. Thus the satellite construction can be rephrased as a composition of bordisms:
\be \vcenter{\hbox{$\underset{\text{Complement of companion knot}}{\includegraphics[width=0.23\textwidth]{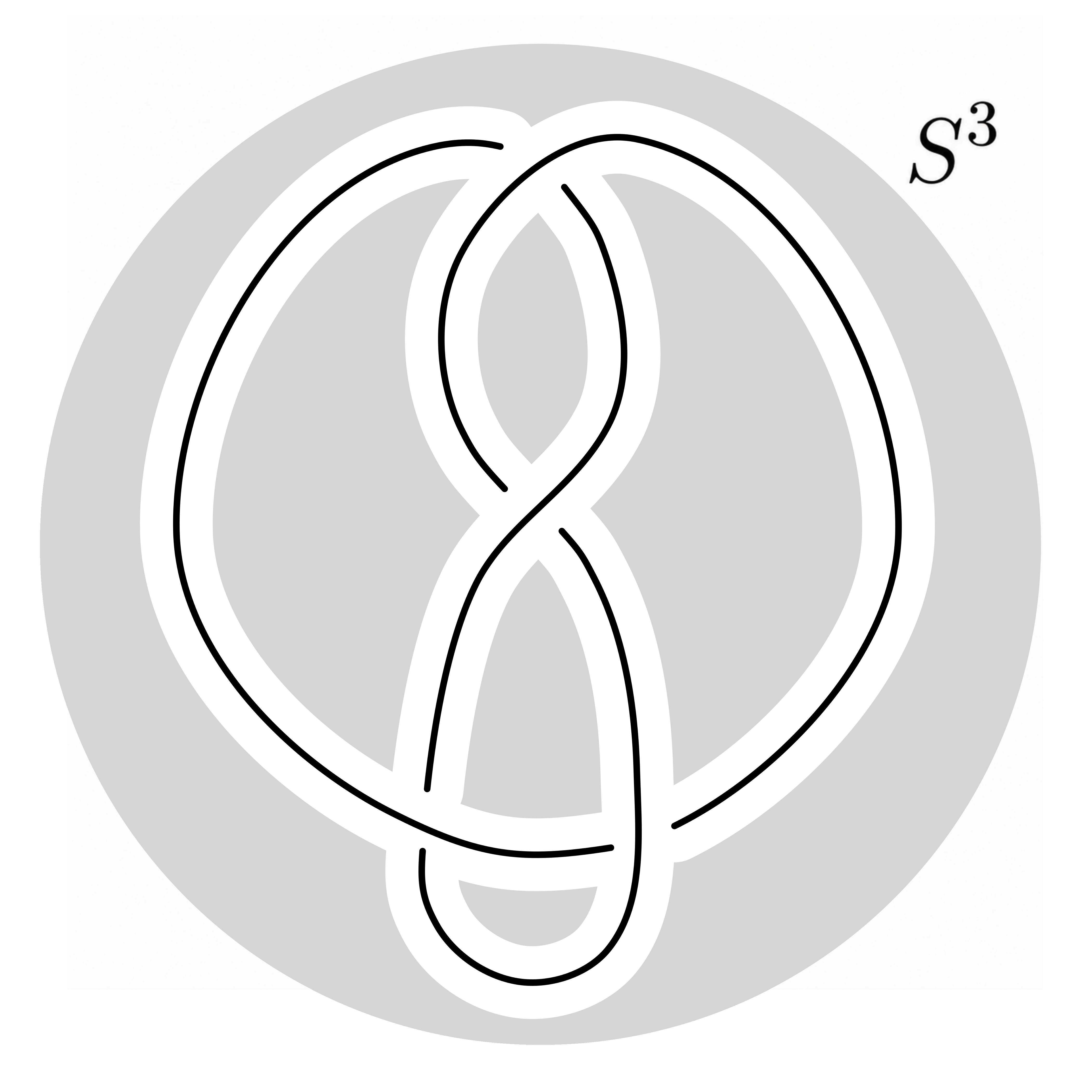}}$}}\circ
\vcenter{\hbox{$\underset{\text{Pattern bordism}}{\includegraphics[width=0.25\textwidth]{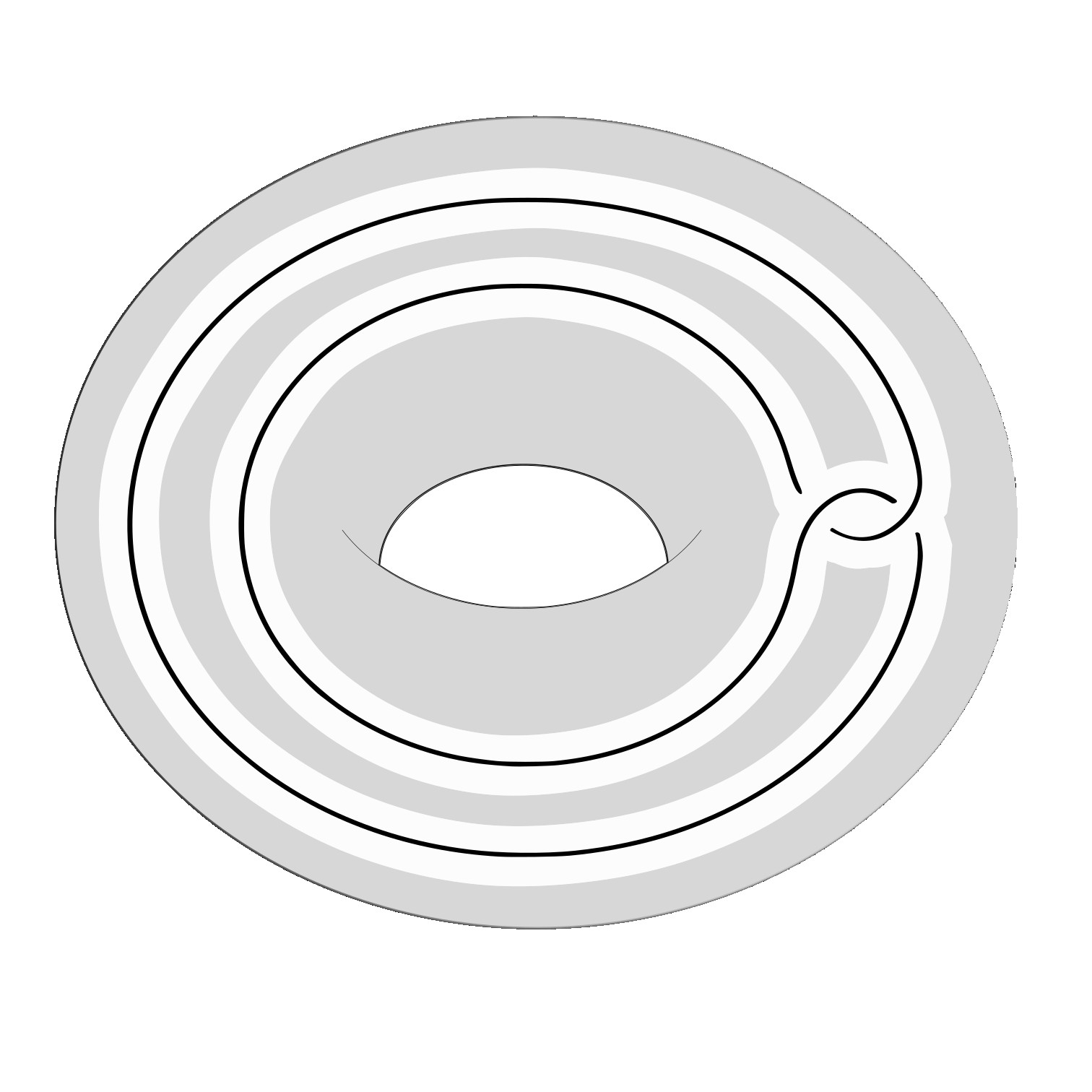}}$}}=
\vcenter{\hbox{$\underset{\text{Complement of satellite knot}}{\includegraphics[width=0.25\textwidth]{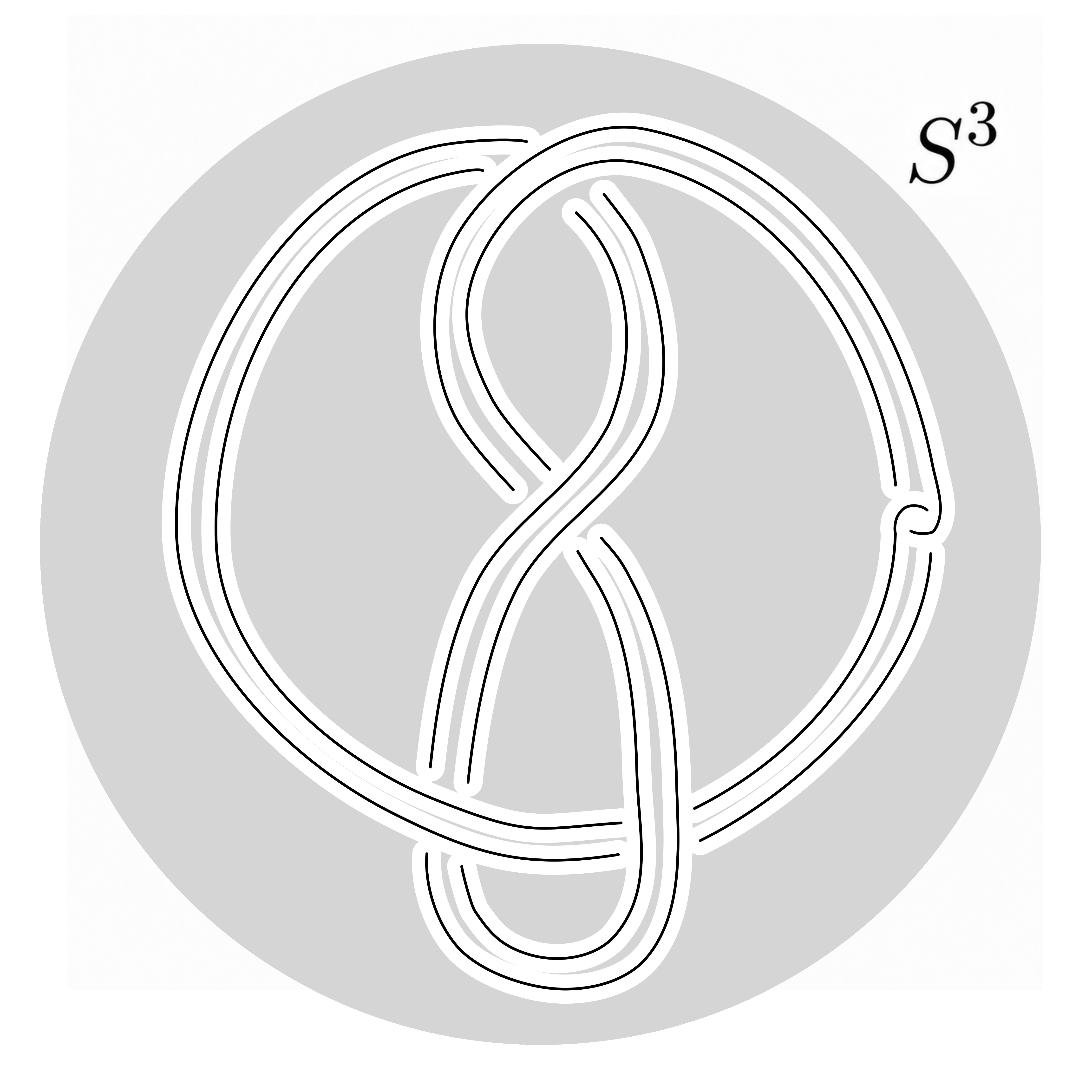}}$}}\label{trefdoublecompl}\ee
The TQFT turns the pattern bordism into an operator. This operator acts on the state of a knot complement and produces the state of the satellite complement. We describe this construction in detail in Section \ref{sec:sat}, where we introduce the satellite operator and give an explicit prescription for its matrix elements in \eqref{pattern_operator}, followed by several examples.

In Section \ref{sec:splice}, we pass from satellites to the more general splice construction. A basic example of a splice is the connected sum of two knots. As shown below, the connected sum can be constructed by splicing these knots along the two ``keys" of a Hopf keychain.
From the point of view of complement manifolds, splicing becomes a gluing operation along torus boundaries that may involve several incoming and outgoing boundary components.
In the example of a connected sum, the manifold associated with the Hopf keychain should be interpreted as a bordism with two ingoing boundaries, corresponding to the two ``keys", and one outgoing boundary corresponding to the ``keyholder". Gluing the complements of two knots to the ingoing boundaries produces the complement of their connected sum. The general TQFT prescription for splice operators is given in \eqref{splice_operator}. 
$$\vcenter{\hbox{\includegraphics[width=0.25\textwidth]{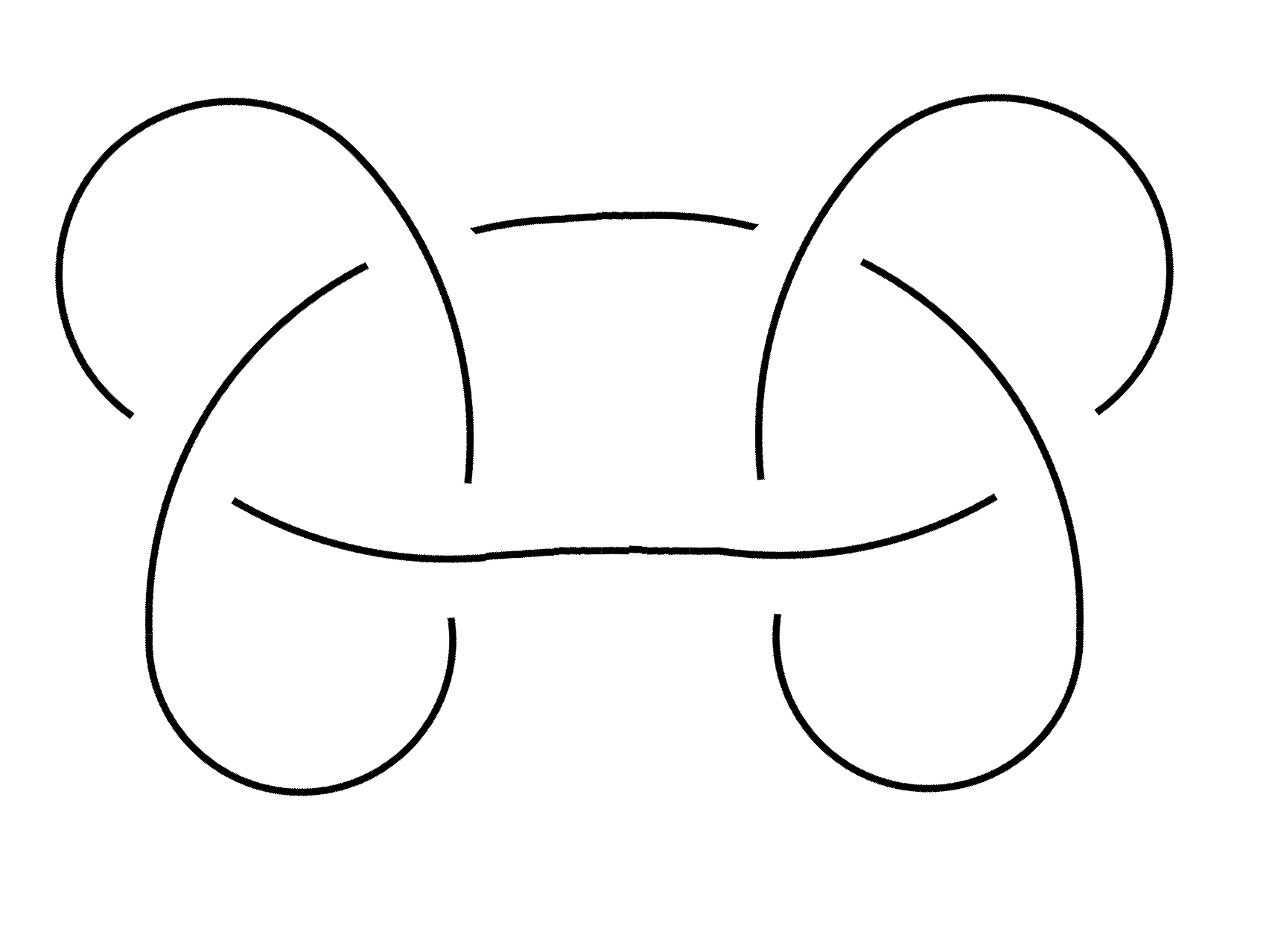}}}\begin{array}{c}
	\xrightarrow{\text{JSJ decomposition}} \\
	\xleftarrow[\text{splice construction}]{}
\end{array}\vcenter{\hbox{\includegraphics[width=0.4\textwidth]{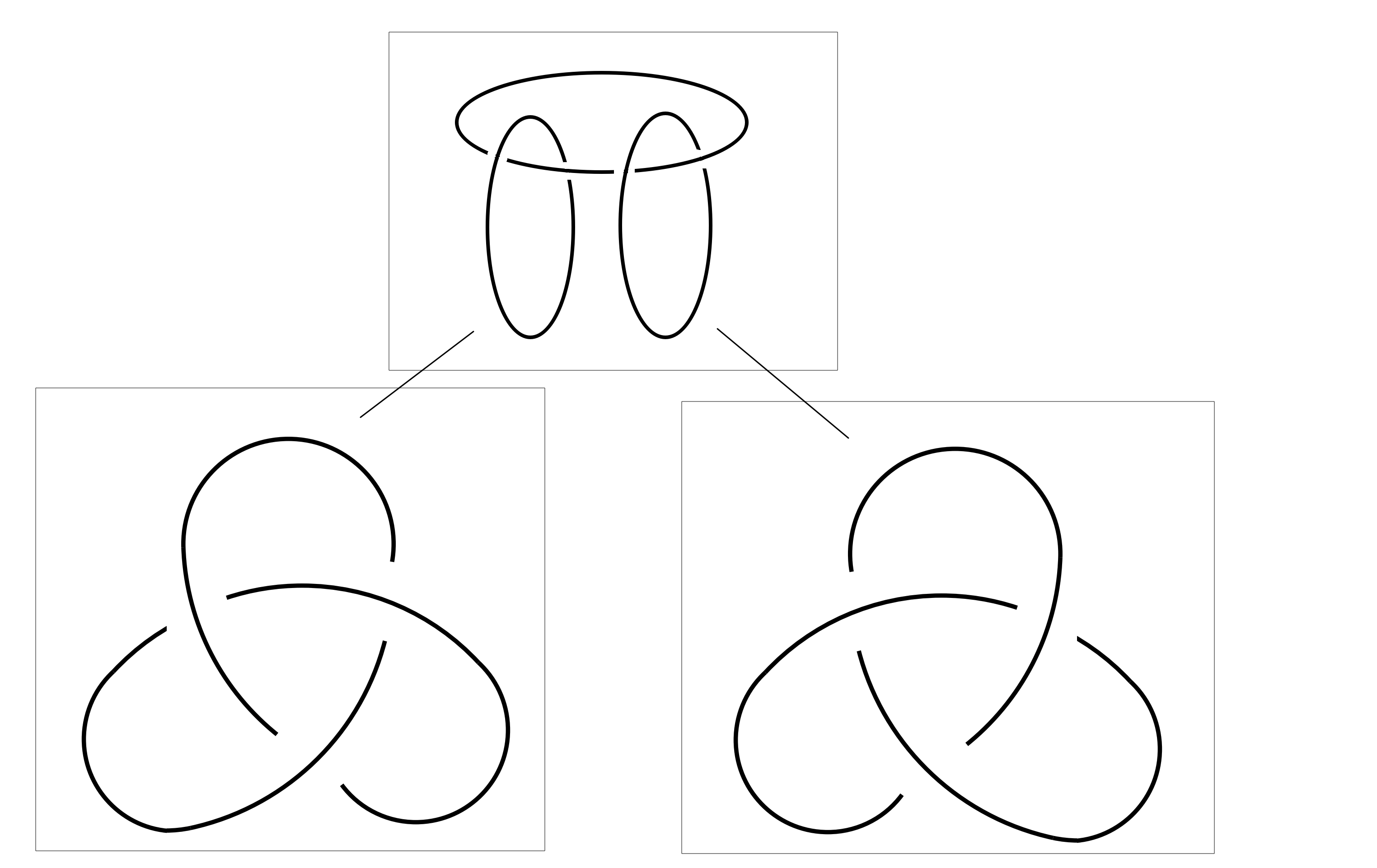}}}$$

This connects our discussion to the JSJ decomposition of three-manifolds, which is, in a sense, the inverse operation to splicing. The JSJ decomposition gives a canonical way to deconstruct a link-complement manifold into elementary manifolds with torus boundaries. Since a TQFT assigns operators to these pieces, the JSJ decomposition expresses the link-complement state as a network of multilinear operators:
$$\vcenter{\hbox{\includegraphics[width=0.4\textwidth]{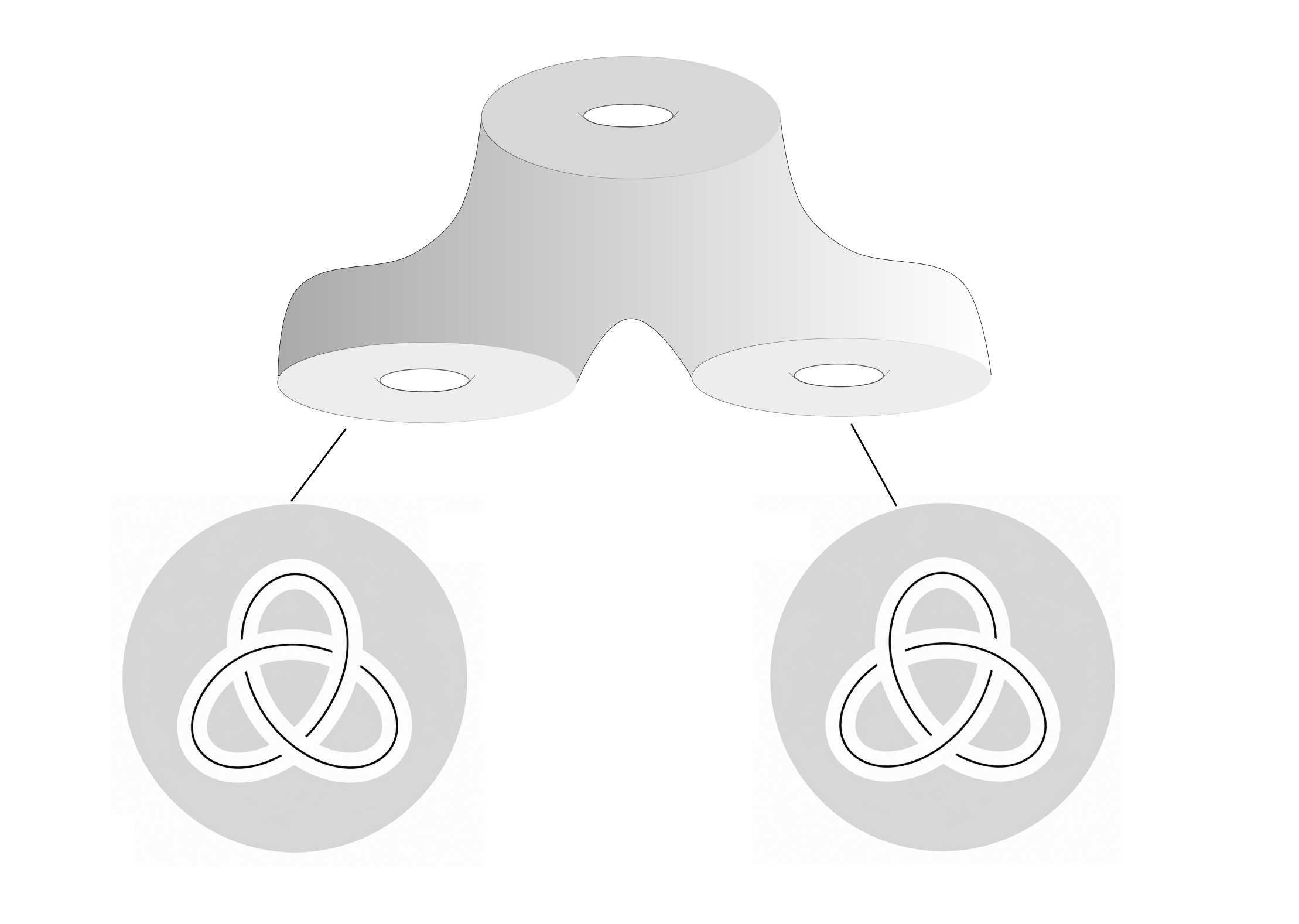}}} \xlongrightarrow{TQFT}\vcenter{\hbox{\scalebox{0.75}{\begin{tikzpicture}[
				tensor/.style={
					draw,
					rectangle,
					minimum width=4.2cm,
					minimum height=1.2cm,
					align=center
				},
				smalltensor/.style={
					draw,
					rectangle,
					minimum width=2.0cm,
					minimum height=1.0cm,
					align=center
				},
				line/.style={thick}
				]
				
				% Bottom boxes
				\node[tensor] (T1) at (-3,0) {Complement state of knot 1};
				\node[tensor] (T2) at ( 3,0) {Complement state of knot 2};
				
				% Top box
				\node[smalltensor] (P) at (0,2.5) {Connected sum operator};
				
				% Lines from bottom boxes to top box
				\draw[line] (T1.north) -- (P.south west);
				\draw[line] (T2.north) -- (P.south east);
				
				% Outgoing line above P_#
				\draw[line] (P.north) -- ++(0,1.2);
				
\end{tikzpicture}}}}$$

The pieces appearing in the JSJ decomposition belong to two broad families: Seifert-fibered pieces and hyperbolic pieces. Seifert-fibered manifolds have a preferred circle direction, and therefore admit a useful dimensional reduction. Reducing along these circles gives an effective 2D TQFT, described by a Frobenius algebra with additional point defects. Graph manifolds are obtained by gluing Seifert-fibered pieces together. They still admit a related 2D description, but with additional loop defects. This effective 2D description is introduced in section \ref{2Dgraph}.

Hyperbolic manifolds behave differently, as they do not admit the same kind of 2D reduction. Moreover,
their TQFT states generally depend on categorical data beyond the modular
$S$ and $T$ matrices. The Whitehead link provides a basic example of this
phenomenon \cite{beyond}. Including hyperbolic pieces therefore leads to richer
and qualitatively different behaviour of link-complement states.

We illustrate this perspective by studying the entanglement entropy of link complements consisting of different types of basic pieces. For Seifert-fibered link complements, the Frobenius algebra structure immediately implies GHZ-like states \cite{Balasubramanian:2025kaf} and explains the suppression of non-abelian sectors in the large-party limit, generalizing the results of \cite{Large-party}. As a graph-manifold example, we study the Hopf chain. Although it is built entirely from Seifert-fibered pieces, it is no longer a GHZ-like state. We show analytically that at the large-chain limit, in any unitary theory, the entanglement entropy converges to a finite value, confirming the predictions of \cite{ramirezvaldez2024remarksentanglemententropyhopf}. Moreover, we show that all sectors of the theory, both abelian and non-abelian, contribute to its entanglement entropy. Finally, we study Whitehead doubles of the Hopf chain as examples involving hyperbolic pieces. The Whitehead satellite operator suppresses abelian sectors and produces behaviour distinct from both the Seifert-fibered and graph-manifold examples. 

The paper is organized as follows. In section \ref{sec:Linkstates} we introduce link-complement states in TQFT and fix our conventions and notation. In section \ref{sec:sat} we describe the classical satellite construction of knot theory in terms of bordisms and present a prescription for calculating the matrix elements of the corresponding operators. In section \ref{sec:splice} we describe the classical splice construction in the language of TQFT, and discuss its relation to the JSJ decomposition of three-manifolds. In section \ref{2Dgraph} we explain the 2D TQFT description of graph manifolds obtained by dimensional reduction on a circle. Finally, in section \ref{sec:EE} we study the entanglement entropy of states built from different combinations of elementary pieces.

\section*{List of Symbols}
\begin{tabbing}
	\hspace*{3cm} \= \kill % Sets the column width spacing
	$\nu(\mathcal L)$ \> The open tubular neighborhood of a link $\mathcal L$\\
	$|\mathcal L|$ \> Number of components of the link $\mathcal L$\\
	$\mathcal M\setminus\nu(\mathcal L)$\> The complement of link $\mathcal L$ in the manifold $\mathcal M$\\
	$\mathcal L[a_1,\dots,a_n]$ \> A colored link\\
	$\mathcal M;\mathcal L[a_1,\dots,a_n]$ \> A manifold $\mathcal M$ with the insertion of a colored link\\
	$\mathcal O$ \> The unknot\\
	$\mathcal T_{(p,q)}$ \> The $(p,q)$ torus link\\
	$\mathcal S_{(p,q)}$ \> The $(p,q)$ Seifert link\\
	$\mathcal W$ \> The Whitehead link\\
	$\mathcal H^{n}$ \> The $n$-component Hopf keychain\\
	$\mathcal C^n$ \> The $n$-component Hopf chain\\
\end{tabbing}

\section{TQFT and link-complement states\label{sec:Linkstates}}

Let $\mathcal C$ be a unitary modular tensor category, and let
\be
\Z:\Bord^{\mathcal C}_3\longrightarrow \Vect_{\mathbb C}
\ee
denote the associated Reshetikhin-Turaev TQFT functor\footnote{$\Z$ may also be thought of as a Chern-Simons path-integral, normalized as in (\ref{Zsphere}).}. The objects of the monoidal bordism category
$\Bord^{\mathcal C}_3$ are compact oriented surfaces, possibly equipped with marked points labelled
by simple objects of $\mathcal C$. A morphism $\mathcal M:\Sigma_1\to \Sigma_2$ is a compact,
oriented three-manifold with
\be
\partial \mathcal M=\Sigma_2\sqcup \overline{\Sigma}_1,
\ee
possibly decorated by labelled ribbon graphs whose endpoints match labelled marked points on the boundary. Throughout the paper, we assume for notational simplicity that $\mathcal C$ is multiplicity-free ($N_{ab}^c\in\{0,1\}$).

Since, the TQFT is sensitive to framings, all drawings of links and ribbon graphs are equipped with blackboard
framing unless stated otherwise. We will suppress the framing anomaly of the three-manifold, as
it contributes only an overall phase to the states.

The main Hilbert spaces in this paper are
$
\V:=\Z(T^2)
$, its dual $\V_*$
and tensor products thereof.
A convenient basis of $\V$ is the anyon basis. For every simple
object $a\in\mathcal C$, let
\be |a\rangle:=\Z\left(\vcenter{\hbox{\begin{tikzpicture}[fill=lightgray]
			\begin{scope}[decoration={markings, mark=at position 0.8 with {\arrow{<}}}]
				\pic{torus={1cm}{2.8mm}{70}};
				\draw [postaction={decorate},yscale=cos(70),black,thick](1.05,0.08) arc (0:360:1.05cm);
				\node [black] at (0.5,-0.45) {$a$};
			\end{scope}
\end{tikzpicture}}}\right),\label{anyon_basis}\ee
where the diagram denotes a simple line $a$ inserted along the core of a solid torus.
A dual basis vector $\langle b|\in\V^*$ is represented by
the oppositely oriented solid torus with a $\bar b$-labelled line. Indeed, gluing these two solid tori by the identity map
\be
\langle b|a\rangle
=
\Z(S^2\times S^1;\bar b,a)
=
\dim_{\mathbb C} \Z(S^2;\bar b,a)
=
\delta_{ab},
\ee
where by $\Z(S^2;\overline b,a)$ we denote the vector space of the two-punctured 2-sphere.

The mapping class group of the torus $\SL(2,\mathbb Z)$ has a representation on $\V$ determined by the modular data of $\mathcal C$. We use the same symbols $S,T$ for the generators of $\SL(2,\mathbb Z)$ and for their
representations on $\V$, the distinction should be clear from context. We use the following convention for $S$-matrix
\be S_{ab}=\Z\left(S^3;\vcenter{\hbox{\begin{tikzpicture}\begin{scope}[decoration={markings, mark=at position 0.35 with {\arrow{<}}},scale=0.5]
				%\draw (2,5) circle (1.25cm);
				%\draw (4,5) circle (1.25cm);
				\draw[postaction={decorate}] (2, 5+1.25) arc (90:360:1.25cm);
				\draw (2, 5+1.25) arc (90:45:1.25cm);
				\draw (2+1.25, 5) arc (0:30:1.25cm);
				\draw (2-1.25,5) node[left] {\small $a$};
			\end{scope}
			
			\begin{scope}[decoration={markings, mark=at position 0.005 with {\arrow{>}}},scale=0.5]
				\draw[postaction={decorate}] (4-1.25, 5) arc (180:-90:1.25cm);
				\draw (4, 5-1.25) arc (270:225:1.25cm);
				\draw (4-1.25, 5) arc (180:210:1.25cm);
				\draw (4-1.25,5) node[left] {\small $b$};
\end{scope}\end{tikzpicture}}}\right).\label{Sdef}\ee
With this normalization
\be \Z(S^3)=S_{\id\id}={1\over \mathcal D},\label{Zsphere}\ee
where $\id$ is the identity object and $\mathcal D$ the total quantum dimension of $\mathcal C$.  The $T$-matrix is diagonal in the anyon basis
\be T_{ab}=e^{-{2\pi i\over 24}c}\delta_{ab}\theta_a,\ee
where $\theta_a$ is the topological twist of $a$ and $c$ is the chiral central charge. We follow the standard diagrammatic conventions, which can be found for example in Chapter 16 of \cite{Simon}.

It is often useful to evaluate configurations of lines by lifting them to $S^3$. By surgery, a closed three-manifold can be represented by a framed link in $S^3$, with the surgery components coloured by the Kirby colour. Thus a coloured link or ribbon graph in an arbitrary closed three-manifold can, in principle, be represented and evaluated by diagrammatic rules in $S^3$.

The manifold $S^3\setminus\mathcal M$, obtained by removing a solid torus $\mathcal M$ from $S^3$, is called the complement of $\mathcal M$ in $S^3$ and can be viewed as an element of $\Hom(T^2,\varnothing)$. The state $\Z(S^3\setminus\mathcal M)\in \V_*$ is called the complement state of $\mathcal M$. Note that the manifold $S^3\setminus\mathcal M$ depends on the way that $\mathcal M$ is embedded inside $S^3$. In particular, the torus may be knotted, in the sense that the line running along the central core of $\mathcal M$ may form a non-trivial knot. Only when the core is an unknot, the complement $S^3\setminus\mathcal M$ is another solid torus\footnote{This is the standard genus-$1$ Heegaard splitting of $S^3$ into two solid tori, identified along their boundaries by an $S$ transformation.}, and we define the basis 
\be \langle \tilde a|:=\Z(S^3\setminus\mathcal M;\mathcal O[a])=\langle a|S,\ee where the line $a$ is inserted along the non-contractible cycle of the complement solid torus with orientation chosen consistenly with (\ref{Sdef}). The corresponding ket basis is $|\tilde a\rangle= S^{-1}|a\rangle$.

A convention used throughout this paper is that link-complement states are written as bras. When a tubular neighbourhood of a knot is removed, the induced boundary orientation on the complement is opposite to the boundary orientation of the removed solid torus. We therefore regard the complement as a bordism $T^2\to\varnothing$, so that the corresponding state lies in $\V_*$. We will call the boundary of such a bordism outgoing. The natural state space for an $n$-component link complement will be $\V_*^{\otimes n}$, rather than $\V^{\otimes n}$ and we will write link-complement states as bras. This is opposite to the common convention in the literature, where link-complement states are often written as kets.

Diagrammatically, we represent a one-boundary complement state as
\be \Z(S^3\setminus \mathcal M)=\vcenter{\hbox{\scalebox{1}{\begin{tikzpicture}[
			tensor/.style={
				draw,
				rectangle,
				minimum width=3cm,
				minimum height=1cm,
				align=center
			},
			line/.style={thick}
			]
			
			\node[tensor] (M) at (0,0) {$\mathcal Z(S^3\setminus \mathcal M)$};
			
			\draw[line] (M.north) -- ++(0,1.2);
			
\end{tikzpicture}}}} .\ee
More generally, if $\mathcal W$ is a bordism with $n$ in-going and $m$ out-going boundaries, then $\Z(\mathcal W)$ is an operator from $\V_*^{\otimes n}$ to $\V_*^{\otimes m}$ and we draw it as
\be \Z(\mathcal W)=\vcenter{\hbox{\scalebox{1}{\begin{tikzpicture}[
				tensor/.style={
					draw,
					rectangle,
					minimum width=4.8cm,
					minimum height=1.2cm,
					align=center
				},
				line/.style={thick}
				]
				
				% Main tensor
				\node[tensor] (W) at (0,0) {$\mathcal Z(\mathcal W)$};
				
				% Top outgoing lines
				\draw[line] (-1.8,0.6) -- (-1.8,1.6);
				\draw[line] (-0.8,0.6) -- (-0.8,1.6);
				\draw[line] ( 1.8,0.6) -- ( 1.8,1.6);
				
				% Top ellipsis and label
				\node at (0.5,1.1) {$\cdots$};
				\node at (0.5,1.45) {$m$};
				
				% Bottom ingoing lines
				\draw[line] (-1.8,-1.6) -- (-1.8,-0.6);
				\draw[line] (-0.8,-1.6) -- (-0.8,-0.6);
				\draw[line] ( 1.8,-1.6) -- ( 1.8,-0.6);
				
				% Bottom ellipsis and label
				\node at (0.5,-1.1) {$\cdots$};
				\node at (0.5,-1.4) {$n$};
				
\end{tikzpicture}}}} .\ee

\subsection{Link complement states}

\begin{figure}
	\centering{\includegraphics[width=0.4\textwidth]{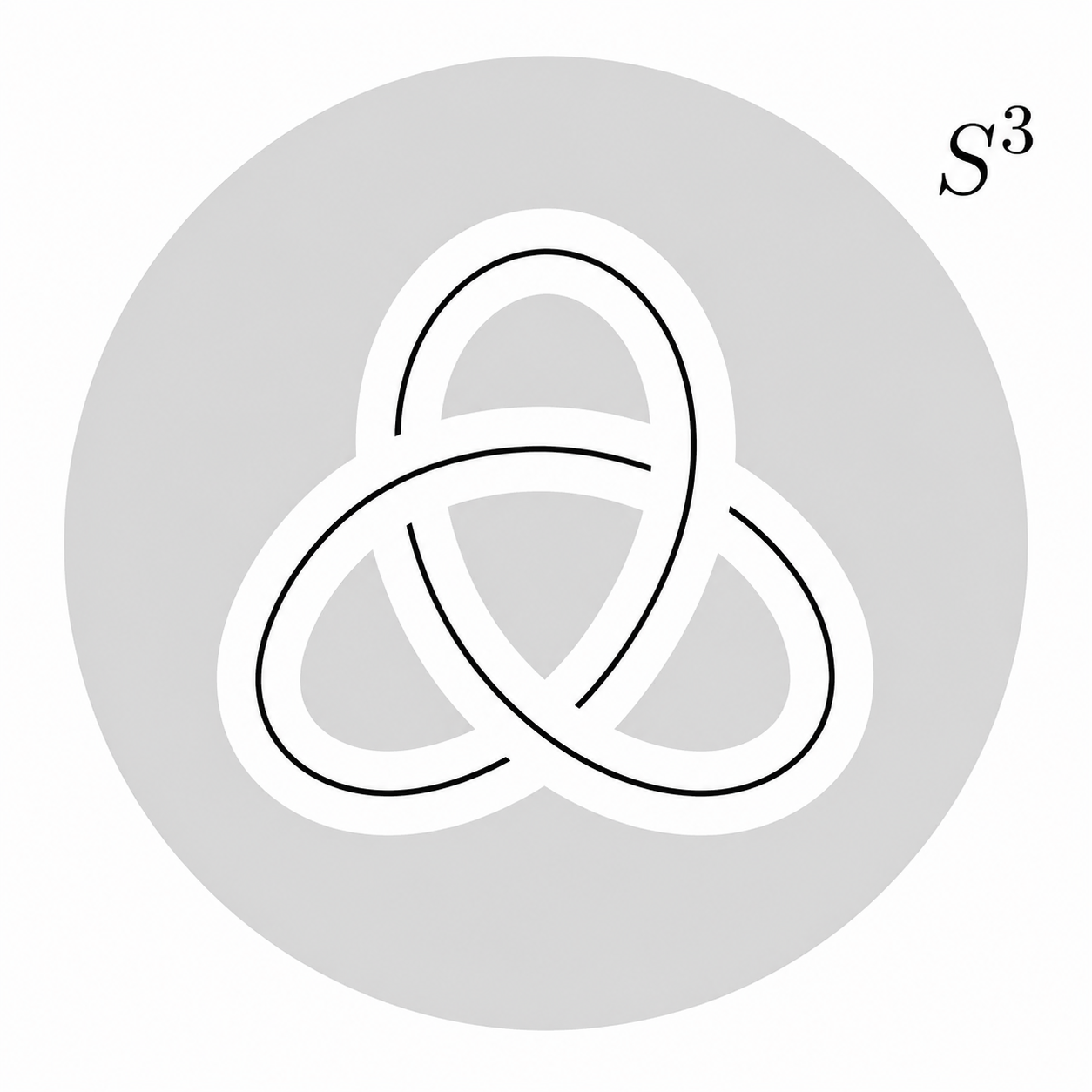}}
	\caption{A visualization of the complement of a trefoil knot in $S^3$. The tubular neighborhood of the knot is a torus, with the trefoil running along its core. The knot complement is obtained by removing this solid torus (along with the knot itself) from $S^3$.\label{knotcompl}}
\end{figure}

We now introduce link-complement states, which will be central in this paper. A link complement is constructed by removing a thickened link from some compact, closed, orientable 3-manifold.  This creates a manifold with a toroidal boundary for each link component, to which the TQFT assigns a multipartite quantum state. See figure \ref{knotcompl} for a visualization of a knot complement in $S^3$.

To be more precise, let $\mathcal L\subset S^3$ be an $n$-component link and $\nu(\mathcal L)$ its open tubular neighborhood. Thus $\nu(\mathcal L)$ consists of $n$ disjoint solid tori embedded in $S^3$, which are possibly linked and knotted. Removing $\nu(\mathcal L)$ from $S^3$ results in the link complement $S^3\setminus\nu(\mathcal L)$, which is a compact manifold with $n$ disjoint torus boundaries. With the conventions we introduced earlier, $S^3\setminus\nu(\mathcal L)$ is a bordism $\sqcup_{n}T^2\to\varnothing$ and hence produces a state $\Z(S^3\setminus\nu(\mathcal L))\in \V_*^{\otimes n}$. The state can be expanded in the dual anyon basis as follows
\be \Z(S^3\setminus\nu(\mathcal L))=\sum L_{a_1\dots a_n}\langle  a_1, a_2,\dots, a_n|.\ee
The coefficients are computed by gluing to each boundary component a solid torus with a line along its core carrying the corresponding anyon label. This reconstructs $S^3$ with the original link $\mathcal L$ colored by simple objects of $\mathcal C$
\be L_{a_1,\dots,a_n}=\Z(S^3\setminus\nu(\mathcal L))|a_1\dots a_n\rangle=\Z(S^3;\mathcal L[a_1,\dots,a_n]).\ee
Thus the components of the link complement state are precisely the colored link
invariants of $\mathcal L$.

As a first example, let $\mathcal L=\mathcal O$ be the $0$-framed unknot. The colored unknot invariant is 
\be\Z(S^3;\mathcal O[a])={d_a}\Z(S^3)={d_a\over \mathcal D}=S_{\id a}.\ee
Above, we first evaluated the contractible loop $\mathcal O[a]$ to its quantum dimension $d_a$ and used the normalization (\ref{Zsphere}). Equivalently, one could use (\ref{Sdef}) by setting $b=\id$. Hence the unknot complement state is
\be \Z(S^3\setminus\mathcal O)=\sum_a S_{\id a}\langle a|=\langle \tilde \id|.\label{unknot_state}\ee
This is consistent with the fact that the complement of an unknot in $S^3$ is again a solid torus.

For a two-component link $\mathcal L=\mathcal K_1\cup\mathcal K_2$, suppose first that the two components are split, so that a $S^2$ separates $\mathcal K_1$ from $\mathcal K_2$. By surgery on that $S^2$ surface, it follows that the invariant factorizes as
\be \Z(S^3;\mathcal K_1[a_1]\cup_{\text{unlinked}}\mathcal K_2[a_2])={\Z(S^3;\mathcal K_1[a_1])\Z(S^3;\mathcal K_2[a_2])\over \Z(S^3)}.\ee
In particular, for the two-component unlink
\be \Z(S^3;\mathcal O[a_1]\cup_{\text{unlinked}}\mathcal O[a_2])=S_{\id a_1}S_{\id a_2}/S_{\id\id}.\ee 
The corresponding state is
\be \Z(S^3\setminus \mathcal O\cup_{\text{unlinked}}\mathcal O)=\sum_{a,b} {S_{a\id}S_{b\id}\over S_{\id\id}}\langle a,b|=\mathcal D\langle \tilde 1,\tilde 1|.\ee
This is a product state.

On the other hand, if the two components form a Hopf link, we have from (\ref{Sdef})
\be\Z(S^3;\mathcal O[a_1]\cup_{\text{Hopf}}\mathcal O[a_2])=S_{a_1a_2}.\ee
The Hopf-link complement state is therefore
\be \Z(S^3\setminus \mathcal O\cup_{\text{Hopf}}\mathcal O)=\sum_{a,b} S_{ab}\langle a,b|=\sum_a \langle a,\tilde a|.\ee
Thus the unlink gives a tensor-product state, whereas the Hopf link gives an entangled state. In this sense, links in $S^3$ provide a topological visualization of multipartite entanglement in the corresponding TQFT states.

This is the standard link-complement state construction studied previously in Chern-Simons theory \cite{Salton_2017,Balasubramanian:2016sro}. In the present paper we use the same construction for general RT TQFT. We will later explain how more complex link-complement states can be built by classical knot-theoretic constructions, which can be interpreted as compositions of bordisms with multiple incoming and outgoing torus boundaries.

\section{The satellite construction\label{sec:sat}}

In this section we review the satellite construction from classical knot theory and its interpretation as a composition of bordisms with torus boundaries. The main result of the section is the formula \eqref{pattern_operator}, which expresses the TQFT operator associated with a satellite construction in terms of coloured link invariants. We then discuss several examples and prescriptions for calculating the matrix elements of these operators.

\subsection{The classical satellite construction}

The satellite construction is one of the basic operations in classical knot theory. It gives a way of producing complex knots from simpler ones by inserting a pattern into the neighborhood of a companion knot.  Conversely, the complement of a non-trivial satellite knot in $S^3$ contains an essential torus\footnote{Roughly speaking, an essential torus is an embedded torus whose noncontractible loops remain noncontractible inside the 3-manifold, and which is not just a copy of a boundary torus pushed into the interior. See the notes by Hatcher \cite{Hatcher2001NotesOB} for precise definitions.} and cutting the manifold along this torus separates it into these two pieces. In this way, the satellite construction is not only a method for producing complicated knots, but also offers a decomposition principle for some 3-manifolds.

\begin{figure}
	$$\vcenter{\hbox{\includegraphics[width=0.25\textwidth]{fig8-knot}}}\circ
	\vcenter{\hbox{\includegraphics[width=0.25\textwidth]{Whitehead_pattern}}}=
	\vcenter{\hbox{\includegraphics[width=0.3\textwidth]{fig8-double}}}$$
	$$\vcenter{\hbox{\includegraphics[width=0.25\textwidth]{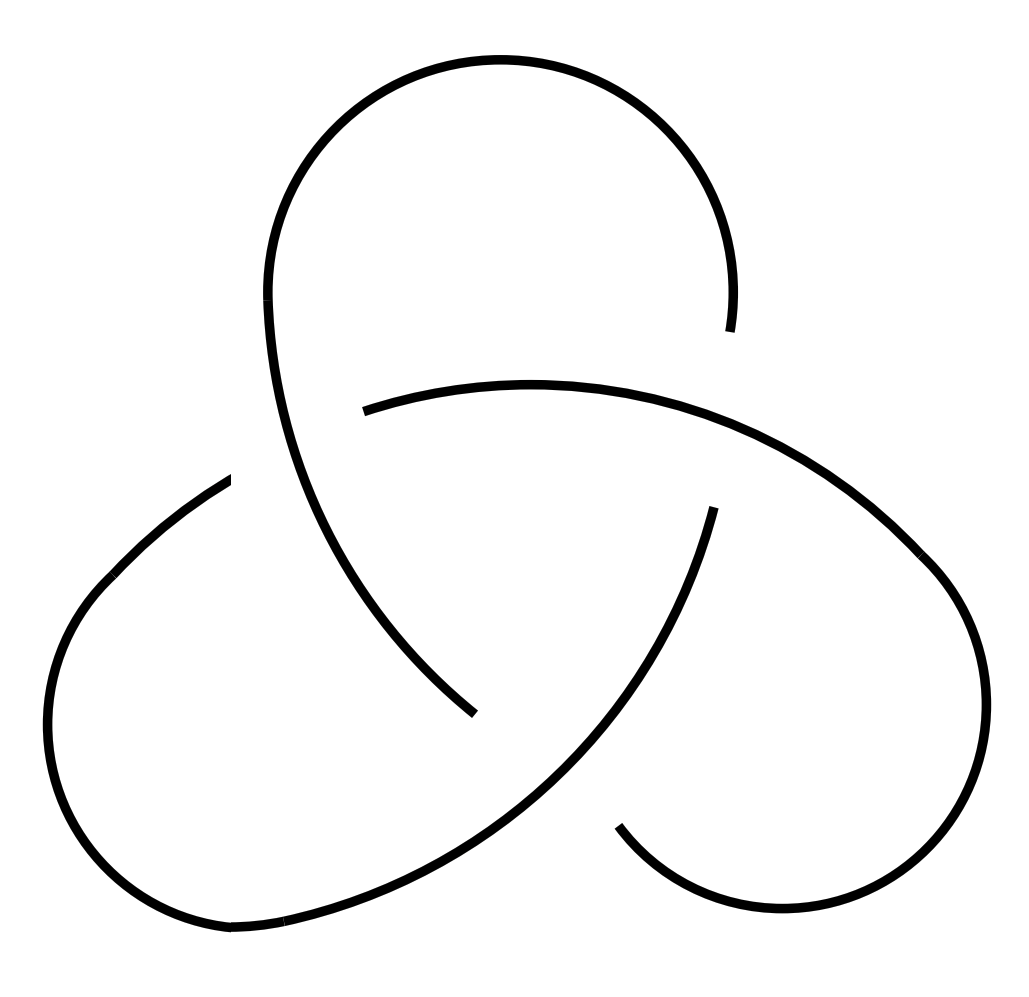}}}\circ
	\vcenter{\hbox{\includegraphics[width=0.25\textwidth]{Whitehead_pattern}}}=
	\vcenter{\hbox{\includegraphics[width=0.3\textwidth]{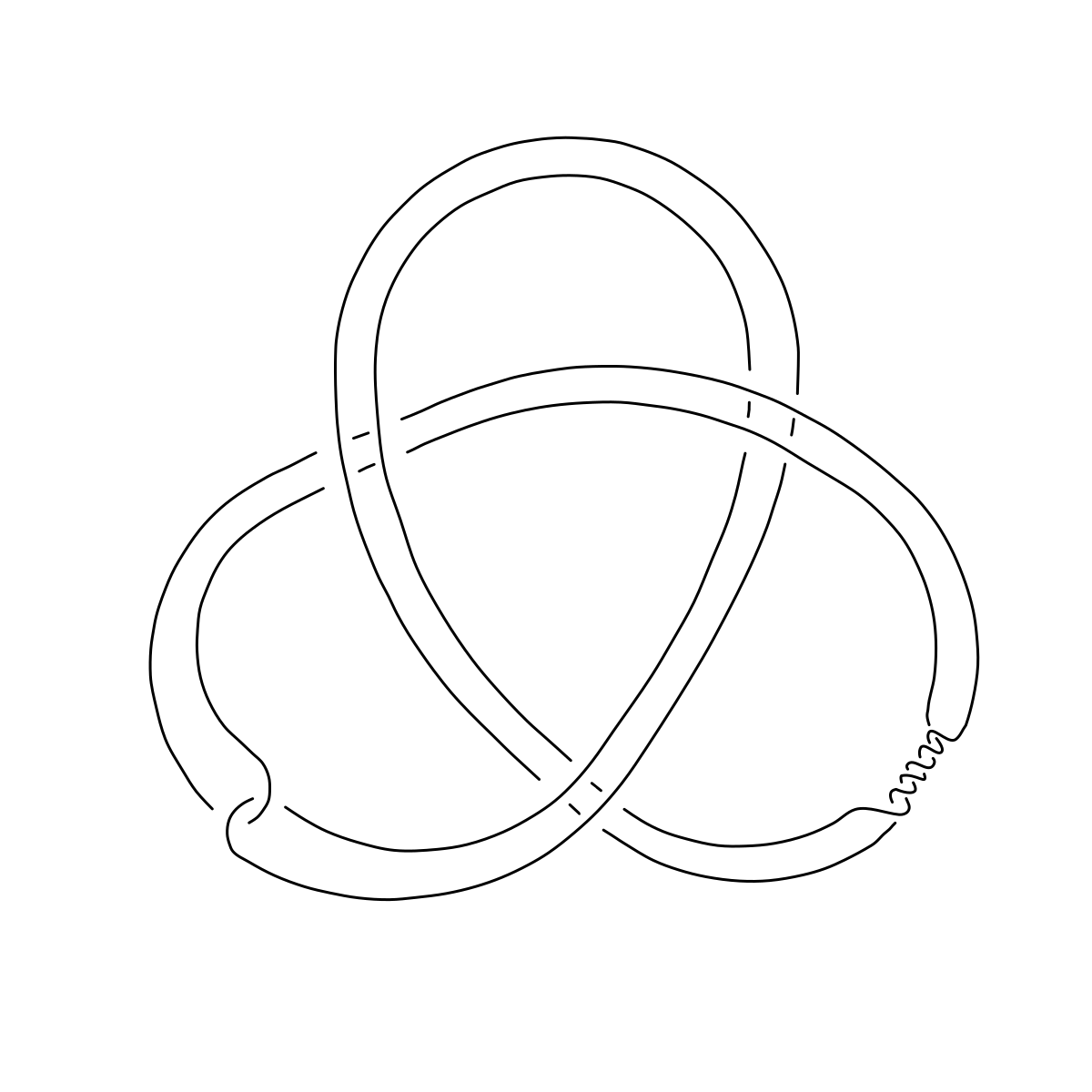}}}$$
	\caption{The Whitehead doubling satellite operation. The pattern (middle image) is a knot or link embedded in a solid torus. Knotting the solid torus into the shape of the companion (left image) we obtain the satellite knot or link (right image). This pattern is closely related to the Whitehead link.
	On the bottom the trefoil is equipped with blackboard framing (with framing $-3$) and the resulting Whitehead double contains three positive full twists.\label{whitedoubling}}
\end{figure}

Let $\mathcal M \cong S^{1}\times D^{2}$ be a standard solid torus, and let
$
\mathcal P \subset \mathcal M
$
be a knot embedded in its interior. The pair\footnote{As a shorthand notation, in most cases, we denote the pattern simply by $\mathcal P$.} $(\mathcal M,\mathcal P)$ is called a \emph{pattern}. Given a knot
$
\mathcal K \subset S^{3},
$
called the \emph{companion}, choose a tubular neighbourhood $\nu(\mathcal K)$. A satellite knot $\mathcal K\circ\mathcal P$ is obtained by identifying the solid torus $\mathcal M$ with $\nu(\mathcal K)$, and then taking the image of $\mathcal P$ under this identification.
 In this way the knot $\mathcal P$, originally living inside an unknotted solid torus, is inserted into the tubular neighbourhood of the possibly knotted companion $\mathcal K$. See figure \ref{whitedoubling} for examples.

There are two degenerate cases worth keeping in mind. If $\mathcal P$ is isotopic to the core of the torus $\mathcal M$, then the satellite operation is trivial $\mathcal K\circ\mathcal P=\mathcal K$. If $\mathcal P$ is contained in a $3$-ball inside $\mathcal M$, then the satellite is independent of the companion knot and is isotopic to pattern knot $\mathcal K\circ\mathcal P=\mathcal P$.

\begin{figure}
	$$\vcenter{\hbox{\includegraphics[width=0.25\textwidth]{trefoil}}}\circ
	\vcenter{\hbox{\includegraphics[width=0.25\textwidth]{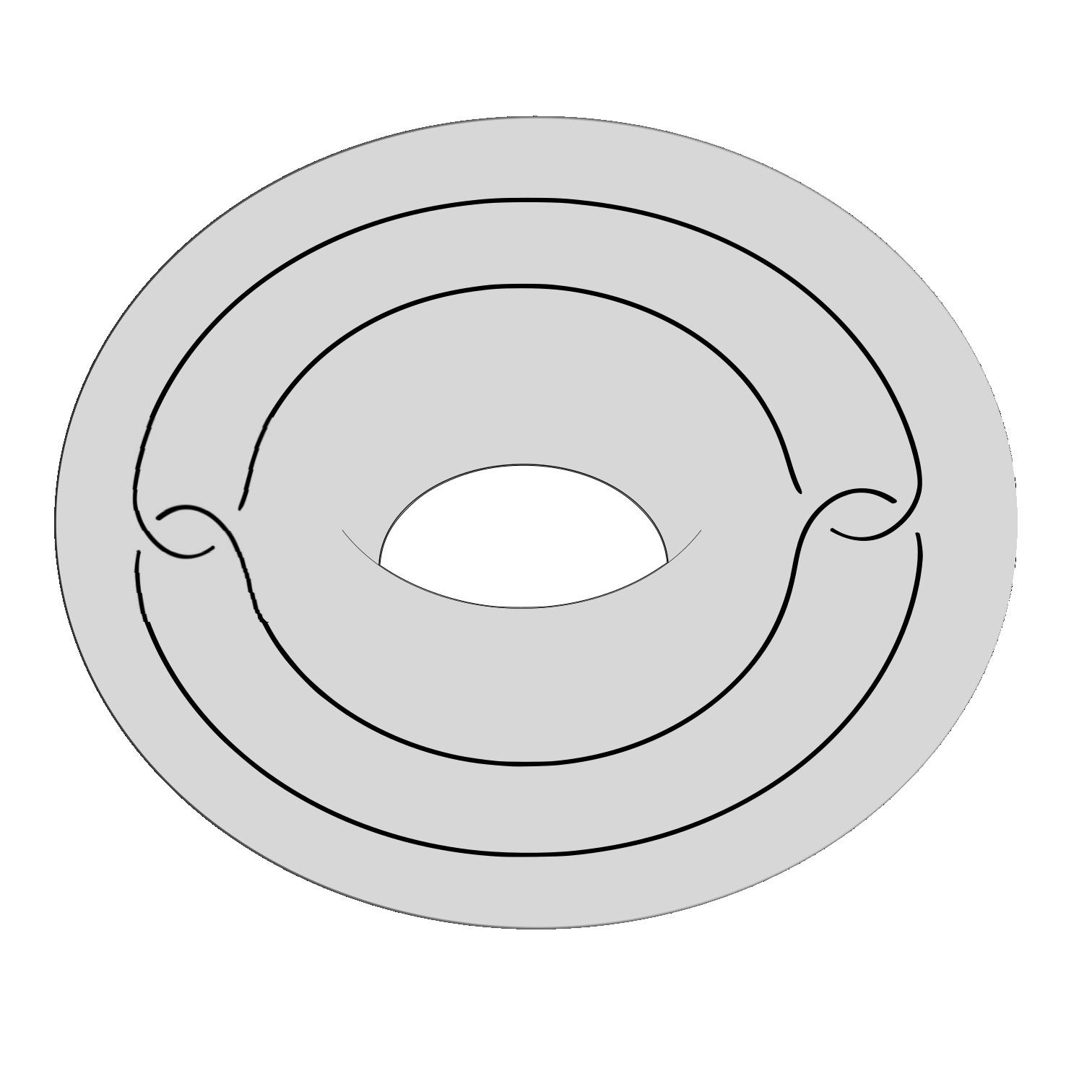}}}=
	\vcenter{\hbox{\includegraphics[width=0.3\textwidth]{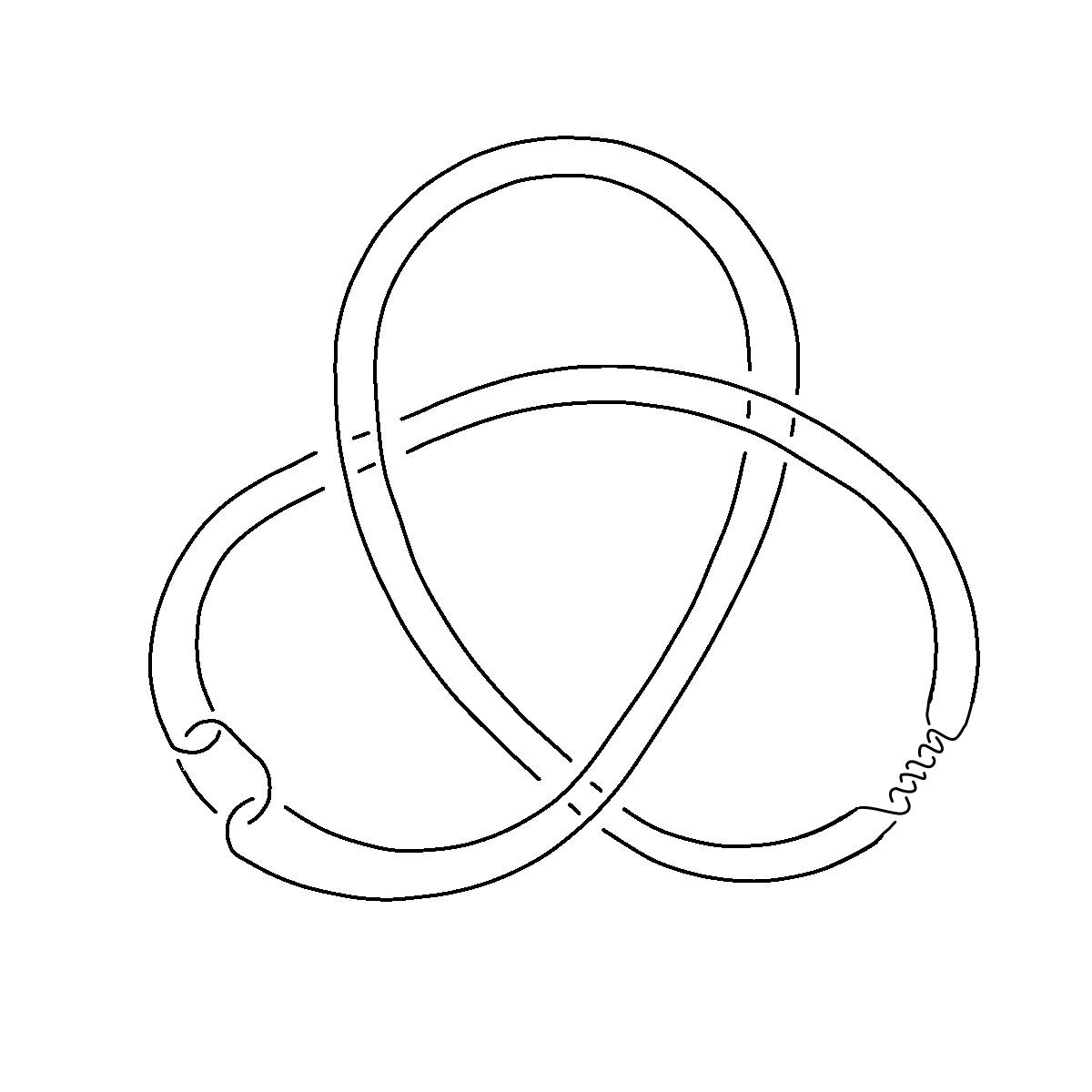}}}$$
	\caption{The satellite construction for a link pattern. This pattern is closely related to the Borromean rings.\label{bing-doubling}}
\end{figure}

The same construction applies to link patterns (see figure \ref{bing-doubling}). We may take $\mathcal P\subset\mathcal M$ to be a link rather than a single knot. Applying the above construction to a companion knot $\mathcal K$ produces a satellite link with $|\mathcal P|$ components.

Two numerical invariants of a pattern will be useful. The wrapping number is the geometric quantity
\be
\w(\mathcal P):=
\min_D |D\cap \mathcal P|,
\ee
where the minimum is taken over meridional disks $D\subset \mathcal M$, after isotoping
$\mathcal P$ in $\mathcal M$. Thus, $\w(\mathcal P)$ measures the minimal geometric
intersection with a meridian of the solid torus.
The other invariant is the winding number, which is the homology class
\be
\omega(\mathcal P)\in H_1(\mathcal M;\mathbb Z)\cong \mathbb Z.
\ee

Several familiar operations are special cases of satellite construction. Cabling is an operation where the pattern $\mathcal P$ is a torus knot (or torus link)  on $\partial\mathcal M$, pushed into its interior. If the pattern is is the $(p,q)$-torus link pattern, then $\mathcal K\circ\mathcal P$ is called the $(p,q)$-cable of $\mathcal K$. 

Another important example is Whitehead doubling. Here $\mathcal P$ is an unknot embedded in a non-trivial way inside the solid torus $\mathcal M$ (with $\w=2$ and $\omega=0$). The resulting satellite with companion $\mathcal K$ is called the Whitehead double of $\mathcal K$. Depending on the choice of clasp and the number of twists in the pattern, one obtains positive or negative, twisted or untwisted Whitehead doubles. Whitehead doubles are especially important because many of their classical invariants are insensitive to the companion knot. For example, the untwisted Whitehead double has trivial Alexander polynomial, independently of $\mathcal K$. We will see later that this property has an interesting manifestation in TQFT: Whitehead doubling suppresses the abelian sectors of the theory.

\begin{figure}
	$$\vcenter{\hbox{\includegraphics[width=0.25\textwidth]{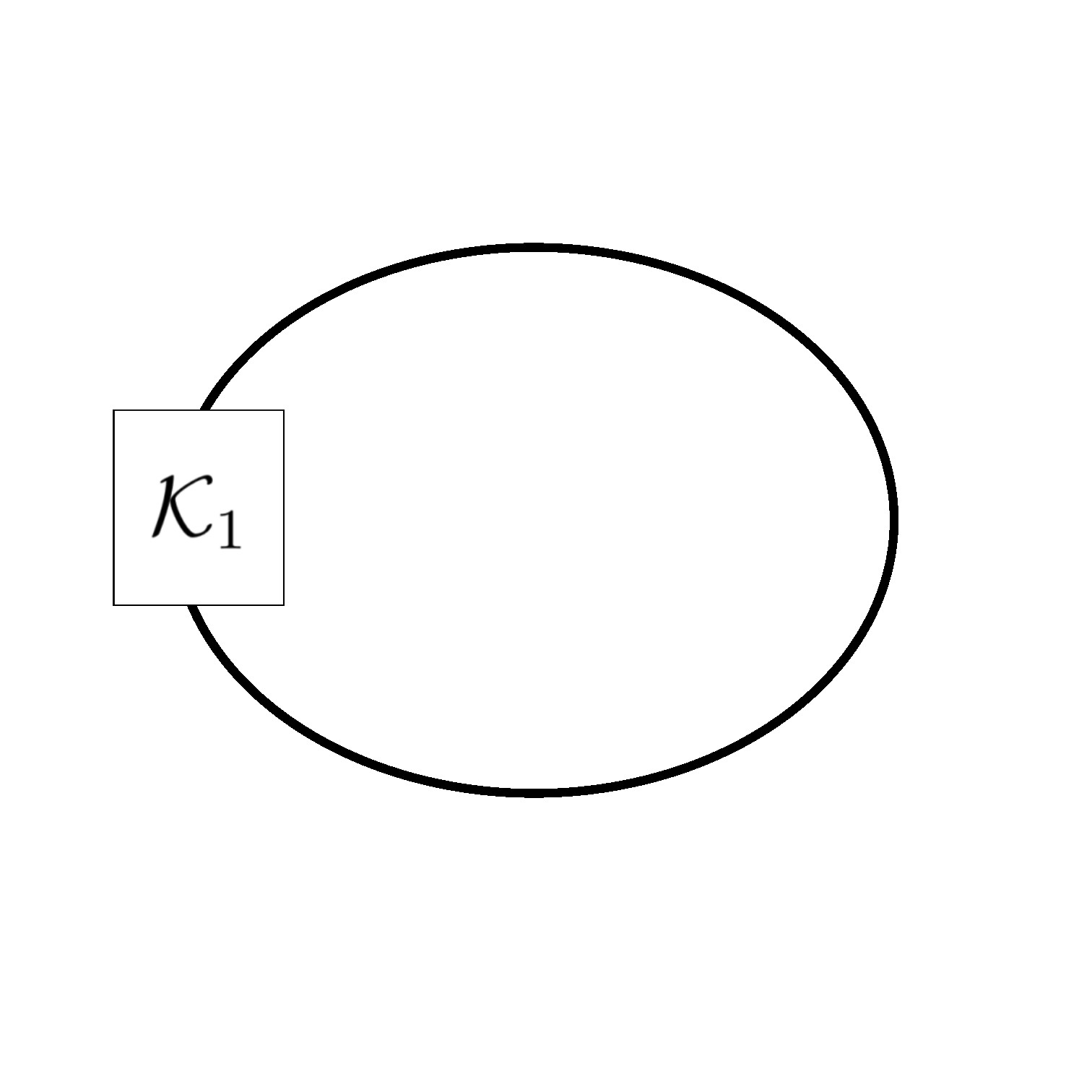}}}\circ
	\vcenter{\hbox{\includegraphics[width=0.25\textwidth]{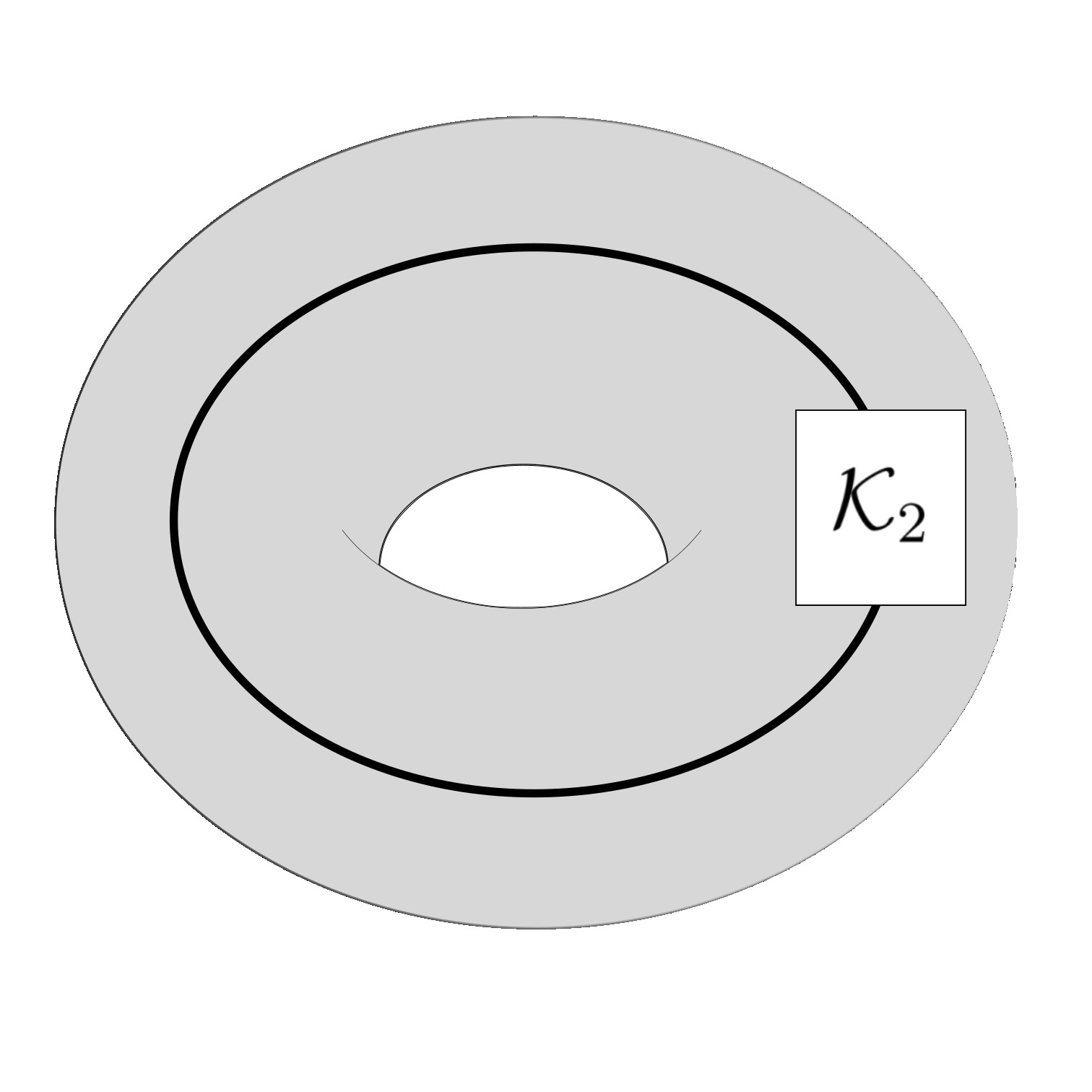}}}=
	\vcenter{\hbox{\includegraphics[width=0.3\textwidth]{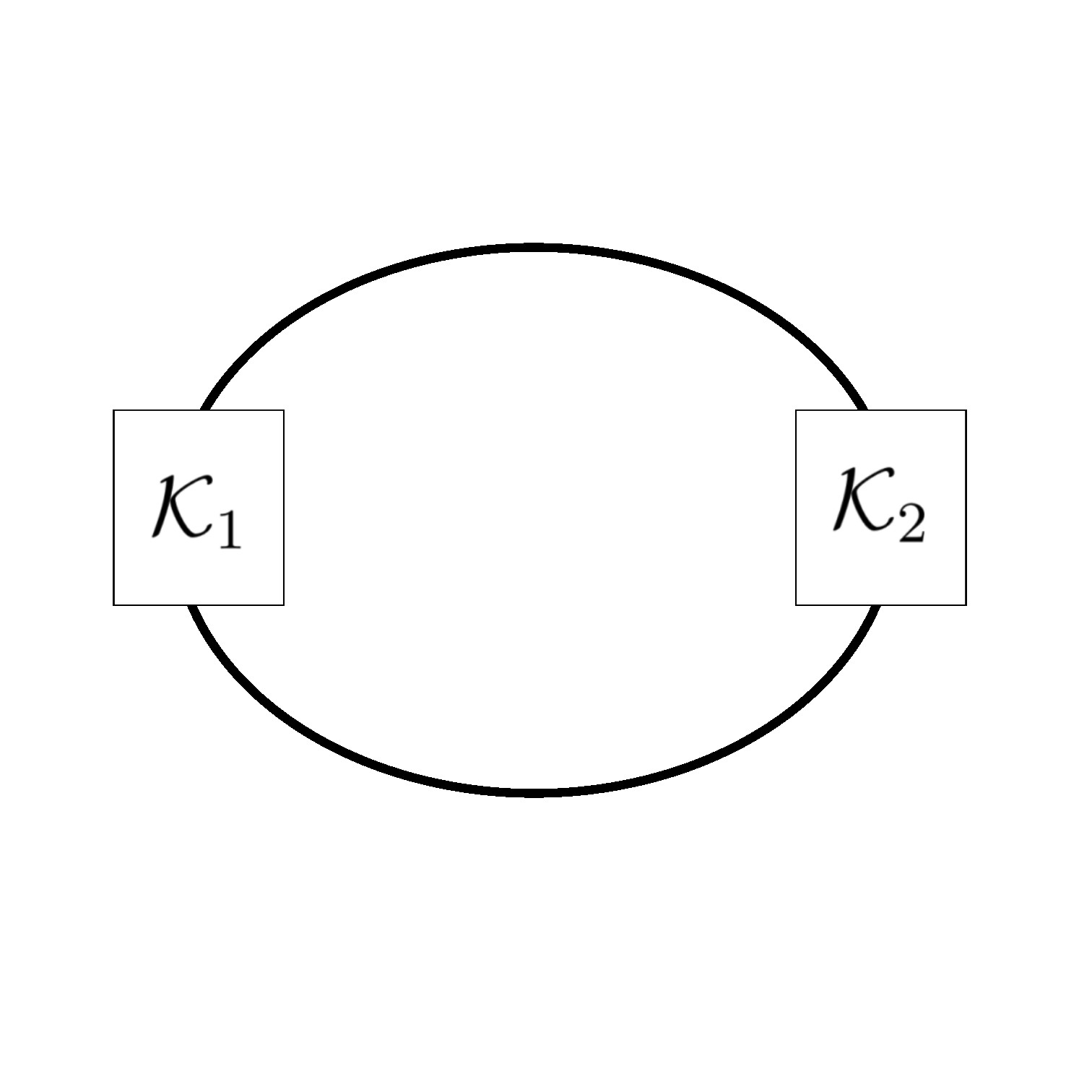}}}$$
	\caption{The satellite operation with a companion of wrapping number $1$ produces a connected sum. The boxes above represent two tangles $\mathcal K_1,\mathcal K_2$.\label{connectedsumsat}}
\end{figure}

Connected sum can also be realized as a satellite construction. If the pattern $\mathcal P$ has wrapping number $\w=1$ in $\mathcal M$, then the satellite construction produces the connected sum $\mathcal K\circ\mathcal P = \mathcal K\# \mathcal P$. See figure \ref{connectedsumsat}.

\subsection{Satellites as composition of bordisms}

The satellite construction has a particularly clear interpretation in terms of bordisms with torus boundaries.
Let $\mathcal M=S^1\times D^2$ be a standard solid torus. Let $\mathcal L\subset \mathcal M$ be a framed $n$-component link pattern. Removing an open tubular neighbourhood of
$\mathcal L$ we obtain a bordism
\be
\mathcal M\setminus \nu(\mathcal L):
\bigsqcup_{i=1}^n T^2 \longrightarrow T^2 
\ee
with $n+1$ torus boundaries. We regard the outer boundary $\partial\mathcal M$ as the ingoing boundary. The rest of the boundaries, coming from the components of $\mathcal L$, are regarded as outgoing boundaries. The TQFT assigns a linear map
\be
\P_{\mathcal L}:=\Z( \mathcal M\setminus \nu(\mathcal L)):\V_*\to\V_*^{\otimes n},
\ee
which we call a satellite operator.

This operator implements the satellite construction at the level of link-complement states. Indeed, let $\mathcal K\subset S^3$ be a companion knot. Gluing the outer boundary of $\mathcal M\setminus \nu(\mathcal L)$ to the boundary of the companion complement $S^3\setminus \nu(\mathcal K)$ gives  $S^3\setminus(\mathcal K\circ\mathcal L)$, i.e., the complement of the satellite link. Hence, from functoriality of $\Z$:
\be \Z\left(S^3\setminus \nu(\mathcal K\circ\mathcal L)\right) =   \mathcal Z(S^3\setminus \nu(\mathcal K))\P_{\mathcal L}.\ee This way, the classical satellite operation becomes a linear map on TQFT states.

The pattern operator depends not only on the isotopy class of
$\mathcal L\subset \mathcal M$, but also on the parametrization of the outer boundary $\partial \mathcal M$. We fix the
standard meridian-longitude basis on $\partial \mathcal M$. Changing this parametrization by
elements of $\SL(2,\mathbb Z)$ pre-composes the pattern operator by the
corresponding mapping-class-group representation\footnote{For one-component patterns, satellites form a monoid under composition. This monoid contains the mapping class group $\SL(2,\mathbb Z)$, realized by the cylinder $T^2\times [0,1]$ with one boundary parametrized by $U\in \SL(2,\mathbb Z)$. More general satellite operators are obtained by allowing nontrivial patterns inside the solid torus. In this sense, these satellite operators generalize the mapping class group \cite{Levine_2001}.}.

\begin{figure}
	\centering\includegraphics[width=0.4\textwidth]{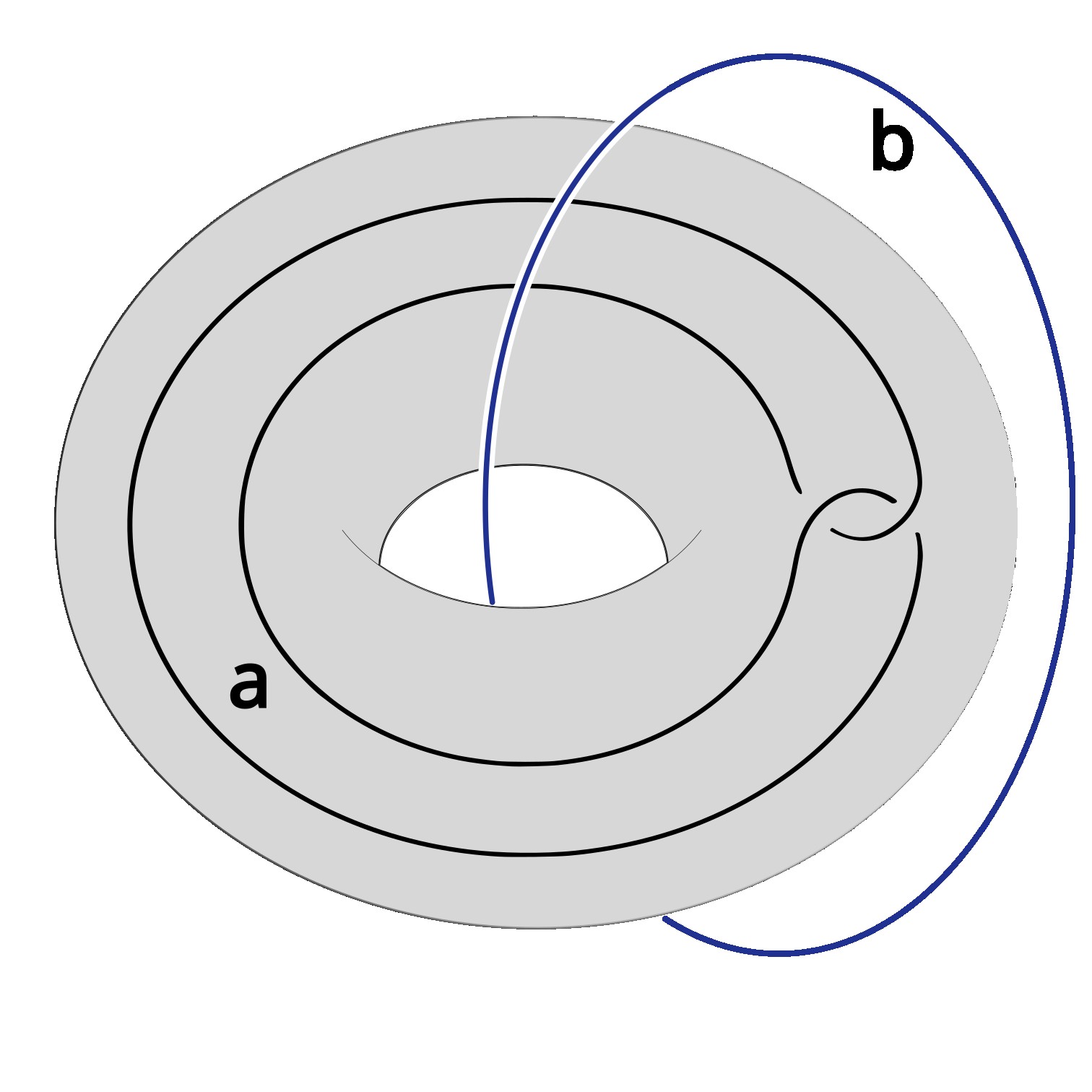}
	\caption{Closing a satellite operator in $S^3$. The extra unknot component $\mathcal O$, labelled by $b$, is the core of the complementary solid torus in the standard Heegaard splitting $S^3=\mathcal M\cup_S\mathcal M'$. The linking of $\mathcal O$ with $\mathcal L$ is determined by how the pattern sits inside the solid torus.\label{closingpattern}}
\end{figure}

We now compute the matrix elements of $\P_{\mathcal L}$. Let $S^3=\mathcal M\cup_S \mathcal M'$ be the standard genus-one Heegaard splitting, where $\mathcal M'$ is the complementary solid torus. Closing the outer (ingoing) boundary of $\mathcal M\setminus \nu(\mathcal L)$ by $\mathcal M'$ with a $b$-labeled line along its core, and closing the $n$ inner (outgoing) boundaries by solid tori with lines $(a_1,\dots,a_n)$ along their cores produces the colored link $\mathcal O[b]\cup \mathcal L[a_1,\dots,a_n]\subset S^3$.  We obtain
\be
\langle b|S\,\P_{\mathcal L}|a_1,\ldots,a_n\rangle
=
\Z(S^3;\mathcal O[b]\cup \mathcal L[a_1,\ldots,a_n]).\label{pattern_matrix_element}
\ee
 See figure \ref{closingpattern} for an example.

Using expression (\ref{pattern_matrix_element}), we can write the pattern operator as
\be
\boxed{
	\P_{\mathcal L}
	=
	\sum_{b,a_1,\ldots,a_n}
	|\widetilde b\rangle
	\langle a_1,\ldots,a_n|\,
	\Z(S^3;\mathcal O[b]\cup \mathcal L[a_1,\ldots,a_n])
}.
\label{pattern_operator}
\ee
Formula (\ref{pattern_operator}) expresses the pattern operator in terms of link invariants in $S^3$.

%To illustrate the next point, let us consider the operator with a single knot pattern $\mathcal K$. In this case (\ref{pattern_operator}) takes the form
%\be \P_{\mathcal K}=\sum_{a,b}	|\widetilde b\rangle
%\langle a|\,
%\Z(S^3;\mathcal O[b]\cup \mathcal K[a]).\ee
%From this expression, and using the fact that the invariant of a link is related to its mirror image by complex conjugation, we can express the Hermitian conjugate operator as follows
%\be \mathcal S\circ\P_{\mathcal K}^\dagger\circ \mathcal S^{-1}= S\;\P_{\overline{\mathcal K}}\;S^{-1}.\ee
%Thus Hermitian conjugation of a satellite operator amounts to taking the mirror image of the pattern, conjugating by $S$ and then reversing the role of the in-going and out-going boundaries. This extends straightforwardly to the general operator (\ref{pattern_operator}).

Let us consider two degenerate examples.
First, suppose that $\mathcal L\subset \mathcal M$ is the core of the solid torus, with zero framing. Then
$\mathcal M\setminus \nu(\mathcal L)\cong T^2\times[0,1]$ and $\P_{\mathcal O}=\id_{\V}$ is the identity operator. As expected, this is a trivial satellite construction, leaving the companion unchanged.

Now suppose that $\mathcal L\subset \mathcal M$ is contained in a three-ball inside $\mathcal M$. In this case, the auxiliary unknot $\mathcal O$ appearing in (\ref{pattern_operator}) is split from $\mathcal L$ and hence 
\be \Z(S^3;\nu(\mathcal O[b]\cup\mathcal L[a_1,\dots,a_n]))={\Z(S^3;\mathcal O[b])\Z(S^3;\mathcal L[a_1,\dots,a_n])\over \Z(S^3)}.\ee
Therefore, as a linear map $\V_*\to \V_*^{\otimes n}$, we find
\be \P_{\mathcal L}=|\id\rangle{1\over S_{\id\id}}\Z(S^3\setminus\nu(\mathcal L)) .\ee
Thus a pattern contained in a three-ball forgets the companion and the operator projects onto the link-complement state of $\mathcal L$.
In particular, if $\mathcal L=\mathcal O$ this projects onto the unknot complement state
\be
\P_{\mathcal O}
={1\over S_{\id\id}}
|{\id}\rangle\langle \widetilde{\id}|
.\ee

In summary, $\P_{\mathcal L}$ rephrases the satellite construction as
a linear map acting on a knot complement. Conversely, whenever a link complement contains an essential torus arising from a satellite construction, the corresponding TQFT state decomposes into simpler linear maps. This is an example of the general decomposition principle we will discuss later.

\subsection{Examples}

\subsubsection{Example I: Connected sums}

Formula (\ref{pattern_operator}) simplifies considerably for patterns with $\w=1$. This reflects the fact that the satellite of a knot $\mathcal K$ with a pattern $\mathcal L$ of $\w=1$ produces the connected sum $\mathcal K\#\mathcal L$. Thus $\P_{\mathcal L}$ acting on a knot complement performs the ``connected sum with $\mathcal L$" operation. As we will see later, from the point of view of splicing, this operator is itself composite and can be further decomposed.

Let $\mathcal L$ be an $n$-component link pattern in the solid torus $\mathcal M$. Suppose that the component $\mathcal L_1$ has wrapping number $1$ in $\mathcal M$, while
	$\mathcal L_2,\ldots,\mathcal L_n$ have wrapping number $0$. Let
	$\mathcal O\cup\mathcal L\subset S^3$ be the link obtained by closing the pattern in $S^3$ in the same way as in the derivation of (\ref{pattern_operator}), and let the label $a_1$ be assigned to the
	component $\mathcal L_1$. Then
	\be
	\P_{\mathcal L}
	=
	\sum_{a_1,\ldots,a_n}
	|a_1\rangle
	\langle a_1,\ldots,a_n|\,
	\frac{\mathcal Z(S^3;\mathcal L[a_1,\ldots,a_n])}{S_{\id a_1}}.
	\label{connected_sum_operator}
	\ee
This follows directly from the fact that $\mathcal L_1$ has wrapping number one in $\mathcal M$. Indeed, the link $\mathcal O[b]\cup \mathcal L[a_1,\dots,a_n]$ is a connected sum of $\mathcal O[b]\cup_{\text Hopf}\mathcal O[a_1]$ with $\mathcal L[a_1,\dots,a_n]$, connected along the components with label $a_1$.
	Therefore the invariant factorizes as
	\be
	\mathcal Z(S^3;\mathcal O[b]\cup\mathcal L[a_1,\ldots,a_n])
	=
	{
		\Z(S^3;\mathcal O[b]\cup_{\text{Hopf}}\mathcal O[a_1])
		\Z(S^3;\mathcal L[a_1,\ldots,a_n])
		\over
		\mathcal Z(S^3;\mathcal O[a_1])
	}.
	\ee
	Using (\ref{Sdef}), we immediately obtain (\ref{connected_sum_operator}).

\subsubsection{Example II: cabling operators}

Torus links provide a useful class of patterns whose TQFT operators can be expressed explicitly in terms of modular data. A torus link is a link which can be embedded without self-intersections on the surface of a torus. It is specified by a pair of integers $(p,q)$, which denote the homology class of the link on the torus. If $\gcd(p,q)=1$, the link has one component and is a torus knot. If $n=\gcd(p,q)>1$, then the torus link has $n$ components, each isotopic to $(p/n,q/n)$. As unoriented links in $S^3$, exchanging $p$ and $q$ gives the same torus link.

For satellite constructions, however, we must choose an embedding of the torus link as a pattern inside a solid torus $\mathcal M$. We use the canonical embedding obtained by starting with the $(p,q)$ link on the surface $\partial\mathcal M$ and pushing it slightly into the interior of $\mathcal M$.
The $(p,q)$ link has winding number $p$ in the solid torus, hence generally $(p,q)$ and $(q,p)$ define distinct patterns. Other embeddings are also possible. For example, if one wants to realize connected sum with a torus link, the same torus link must instead be embedded as a $\w=1$ pattern.

\begin{figure}
	\centering\includegraphics[width=0.4\textwidth]{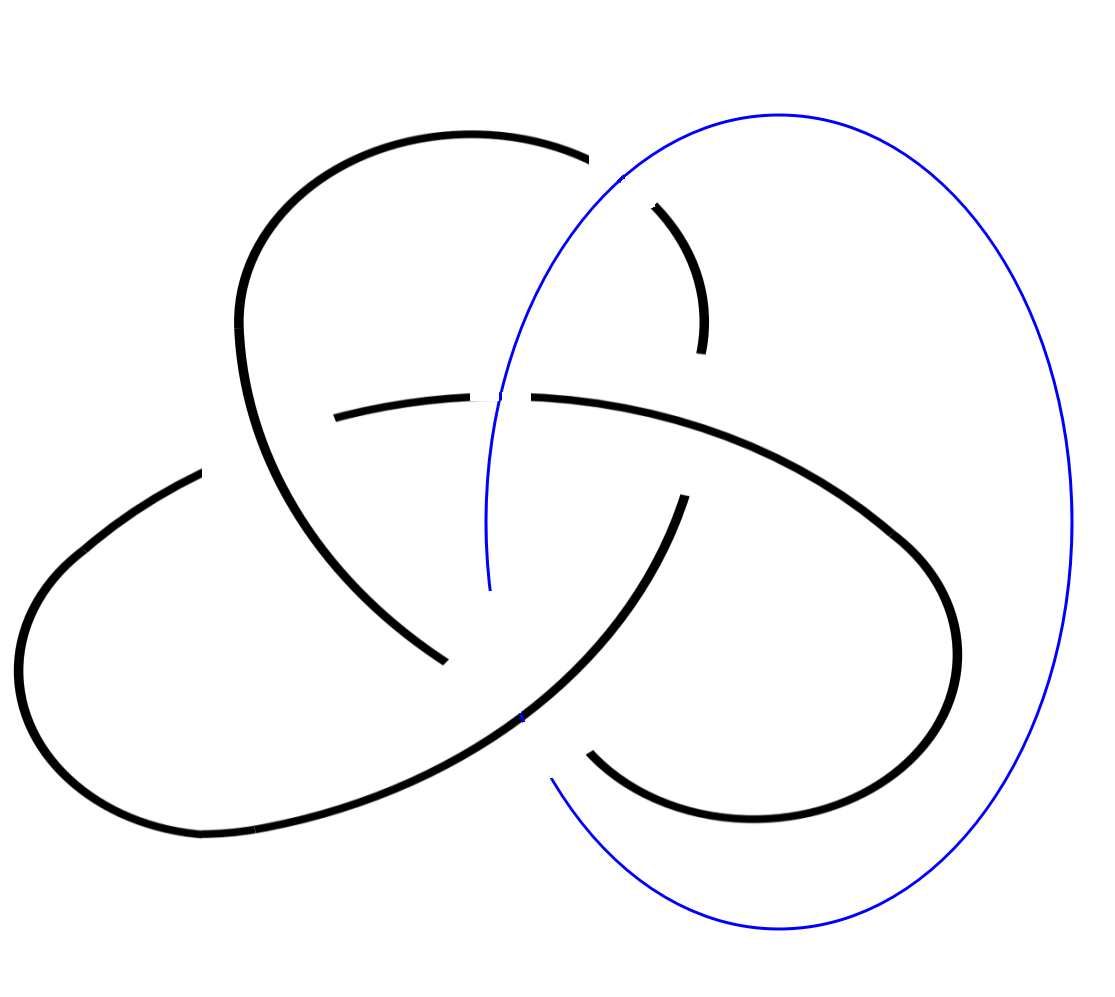}
	\caption{The Seifert link $\mathcal S_{(2,3)}$ consisting of an unknot linked with the trefoil.\label{seifertlink}}
\end{figure}

Let us now compute the satellite operator for the $(p,q)$ torus link pattern. According to (\ref{pattern_operator}), this amounts to an evaluation of the colored link $\mathcal O\cup \mathcal T_{(p,q)}$ in $S^3$, where $\mathcal O$ is the auxiliary unknot (along the complementary torus). This link is also known as the Seifert link $\mathcal S_{(p,q)}$. See figure \ref{seifertlink}. 

One direct way to calculate $\mathcal S_{(p,q)}$ is to represent the torus link $(p,q)$ as the closure of a braid and then use the graphical calculus of the modular tensor category. For example, the trefoil is the closure of $\sigma_1^3$ and one obtains
\be \Z(S^3;\T_{(2,3)}[a]\cup\mathcal O[b])=\sum_c S_{bc}(R_c^{aa})^3=\sum_c S_{bc}R_c^{aa}{\theta_c\over \theta_a^2}.\label{braidt23}\ee
Though not obvious, this link invariant can be fully expressed in terms of modular data \cite{beyond,Rmodular}
\be R_c^{aa}={\theta_c^{1/2}\over \theta_a}\Lambda_{a,c},~~~\Lambda_{a,c}=\theta_c^{-1/2}\sum_{x,y}\theta_y^2\theta_x^{-2}S_{\id y}N_{yx}^a \bar S_{cx} .\ee
Here we made a choice of square root of $\theta_c$, and the sign $\Lambda_{a,c}\in \{\pm 1\}$ is determined by modular data.

A more geometric derivation\footnote{Yet another way is to start with a solid torus with a line wrapping $p$ times along its core and perform a ``fractional twist" $T^{q/p}$ \cite{Stevan_2010}. This leads to an expression in terms of the ``Adams operator", which expresses a line looping $p$ times along the core of the torus as a linear combination of simple loops. In our terms, the Adams operator can be constructed by removing the neighborhood of this loop from the interior of the solid torus.}, which guarantees an expression in terms of modular data, begins with a cylinder $T^2\times I$. Insert $n$ parallel lines, labelled by $a_1,\dots,a_n$, each isotopic to the core of the solid torus obtained by filling the inner boundary. Using the Verlinde formula we can write this operator in a diagonal form
\be \Z(T^2\times I;\mathcal O[a_1],\dots\mathcal O[a_n])=\sum_l {S_{a_1l}\cdots S_{a_nl}\over S_{\id l}^{n}}|\tilde l\rangle\langle\tilde l| .\ee
 Now choose $U\in \SL(2,\mathbb Z)$, such that $(1,0)U=(p/n,q/n)$. Conjugating the above operator by $U$ sends the $n$ parallel lines to the $n$ components of the $(p,q)$ torus link pattern. Filling in the interior hollow torus and then drilling out the tubular neighborhood of the link, we obtain the $(p,q)$-cabling operator
\be \P_{(p,q)}=\sum_l {\langle \tilde l|U^\dagger|\id\rangle\over S_{\id l}^n} U|\tilde l\rangle\langle \tilde l|^{\otimes n}.\label{torus_pattern}\ee
When comparing \eqref{torus_pattern} with a braid calculation such as \eqref{braidt23}, we must account for the choice of basis for the torus cycles and for the framing of the torus-link components. In the trefoil example, the framing differs by three twists.

\subsubsection{Example III: Whitehead doubles\label{subsec:W}}

The Whitehead pattern is obtained by embedding an unknot in a solid torus in a nontrivial way. It has winding number $\omega=0$ and wrapping number $\w=2$. Its name comes from its close relation to the Whitehead link $\mathcal W$. Specifically, after closing the torus pattern in $S^3$, the auxiliary core $\mathcal O$ together with the pattern knot forms the Whitehead link (see figure \ref{closingpattern}). Unlike the torus-link patterns discussed above, the Whitehead pattern is not determined, in general, by modular data alone \cite{beyond}. We follow \cite{beyond} for the calculation of the coloured Whitehead link invariant.

With the labelling and framing convention of Figure \ref{closingpattern}, the Whitehead link invariant can be written as
\be \Z(S^3;\mathcal W[a,b])= \sum_c { S_{bc} S_{aa}^{{c}}\over S_{\id a}}.\label{Winvariant}\ee
Here $a$ labels the Whitehead pattern component, $b$ labels the auxiliary unknot $\mathcal O$, and $S^{c}_{aa}$ is the punctured $S$-matrix.
Therefore, the Whitehead pattern operator (\ref{pattern_operator}) is
\be \P_{\mathcal W}=\sum_{a,c}|c\rangle\langle a|{S_{aa}^{c}\over S_{\id a}}.\label{Pw}\ee

The punctured $S$-matrix is defined diagrammatically by
\be S_{ab}^{c}:={1\over \mathcal D\sqrt{d_c}}\begin{tikzpicture}[baseline=0, thick,scale=.5, shift={(0,-4.8)}]
	%	\draw[help lines] (0,0) grid (5,10);
	
	\begin{scope}[decoration={markings, mark=at position 0.35 with {\arrow{<}}}]
		%\draw (2,5) circle (1.25cm);
		%\draw (4,5) circle (1.25cm);
		\draw[postaction={decorate}] (2, 5+1.25) arc (90:360:1.25cm);
		\draw (2, 5+1.25) arc (90:45:1.25cm);
		\draw (2+1.25, 5) arc (0:30:1.25cm);
		\draw (2-1.25,5) node[left] {\small $a$};
	\end{scope}
	
	\begin{scope}[decoration={markings, mark=at position 0.005 with {\arrow{>}}}]
		\draw[postaction={decorate}] (4-1.25, 5) arc (180:-90:1.25cm);
		\draw (4, 5-1.25) arc (270:225:1.25cm);
		\draw (4-1.25, 5) arc (180:210:1.25cm);
		\draw (4-1.25,5) node[left] {\small $b$};
	\end{scope}
	
	\begin{scope}[decoration={markings, mark=at position 0.5 with {\arrow{<}}}]
		\draw[black, postaction=decorate] (2,5-1.25) to [out=-90, in=-90] (4, 5-1.25); 
		\draw (3,3-0.2) node {\small $c$};
	\end{scope}
\end{tikzpicture}.\ee
The punctured $S$-matrix can be expressed in terms of twists and $F$ symbols as (see appendix \ref{app:SRF})
\be S_{aa}^c={d_a\over \mathcal D}\sum_d \sqrt{d_d\over d_c} N^d_{a\bar a}{\theta_d\over \theta_a^2} F_{\bar aac}^{\bar aa\bar d}.\ee

The Whitehead-link invariant satisfies the nontrivial symmetry relations \cite{beyond}
\be \Z(S^3;\mathcal W[a,b])={\theta_a^2\over \theta_b^2}\Z(S^3;\mathcal W[b,a])=\Z(S^3;\mathcal W[a,\overline b]).\label{Wrel}\ee
These identities are useful consistency checks on the operator $\P_{\mathcal W}$. Additionally, they imply that the left and right nullspaces of $\P_{\mathcal W}$ are isomorphic.

A crucial property of the Whitehead operator is its action on abelian anyons. If $b$ is abelian, then
\be \P_{ \mathcal W}|b\rangle=|\id\rangle {S_{bb}\over S_{\id b}}.\ee
Since its left and right nullspaces are isomorphic, viewing $\P_{\mathcal W}$ as a map $\V_*\to\V_*$, its image has at most dimension $1+\mathcal N$, where $\mathcal N$ is the number of non-abelian anyons. This reflects the fact that abelian anyons detect only linking data. Since the Whitehead link has vanishing pairwise linking number, abelian sectors cannot distinguish its nontrivial structure from the unlink.

In Appendix \ref{appWH} we compute the Whitehead doubling operator explicitly for the Fibonacci and Ising modular tensor categories.

\section{The splice construction and JSJ decomposition\label{sec:splice}}

\subsection{Splicing as composition of bordisms}

The satellite construction is a special case of a broader family of operations on links known as \emph{splicing} \cite{budney2007jsjdecompositionsknotlinkcomplements,EisenbudNeumann1985}. The splice construction is closely related to the theory of toral decompositions of three-manifolds, culminating in the JSJ decomposition, named after Jaco-Shalen and Johannson \cite{JacoShalen1978,Johannson1979,NeumannSwarup1997}.

Informally, to construct a splice, consider two links $\mathcal L$ and $\mathcal L'$ embedded in two distinct $S^3$ manifolds, with distinguished components $\mathcal K\subset \mathcal L$ and $\mathcal K'\subset \mathcal L'$. The splice is formed by drilling out the neighborhoods of $\mathcal K$ and $\mathcal K'$, and then gluing the two link complements so that the meridians and longitudes of the distinguished components are exchanged. The remaining components become the components of the spliced link, now embedded in a new ambient closed manifold (which may not be $S^3$).

We will not introduce the full formal definition of link splicing. For the precise construction in $S^3$, and many illuminating examples, we refer the reader to Budney \cite{budney2007jsjdecompositionsknotlinkcomplements}. 
 Here we give a more general definition of this construction which is natural from the TQFT perspective: a splice piece can be viewed as a bordism with multiple incoming and outgoing torus boundaries. Thus, the TQFT turns a splice piece into a multilinear map between torus Hilbert spaces.

Let the splice piece $\mathcal X^{r,s}$ be a compact oriented manifold with $r+s$ torus boundaries, where $r$ of these boundaries are incoming and the remaining $s$ as outgoing. In other words, this is a bordism
\be \mathcal X^{r,s}: \bigsqcup_{i=1}^s T_i^2 \longrightarrow \bigsqcup_{j=1}^r T_j^2 . \ee
With our conventions, the TQFT produces a linear map
\be\Z(\mathcal X):
\V_*^{\otimes r}
\longrightarrow
\V_*^{\otimes s}.
\ee
The satellite construction is the special case in which $r=1$.

Every splice map is naturally related to an ordinary link-complement manifold. For simplicity, we focus on $S^3$. Start with an $n$-component link $\mathcal L\subset S^3$, with $n=r+s$, and consider its complement $S^3\setminus \nu(\mathcal L)$. If all boundary components are treated as outgoing, the TQFT assigns the link-complement state 
\be \Z(S^3\setminus\nu(\mathcal L)) \in \V_*^{\otimes n}. \ee
 To reinterpret some of these boundaries as ingoing, we use the pairing supplied by the modular $S$-matrix. Define the $\mathcal S$-map
 \be\mathcal S = \sum_{a,b} S_{ab}\,\langle a,b| : \V\longrightarrow \V_*, \label{Smap}\ee
  and its inverse 
  \be \mathcal S^{-1} = \sum_{a,b}\overline S_{ab}\,|a,b\rangle : \V^*\longrightarrow \V. \label{Smapinv} \ee
     In diagrammatic notation we write
  \be \mathcal S=\vcenter{\hbox{\begin{tikzpicture}[
  			tensor/.style={
  				draw,
  				circle,
  				minimum size=0.4cm,
  				align=center
  			},
  			line/.style={thick}
  			]
  			
  			% Circular tensor
  			\node[tensor] (S) at (0,0) {$\mathcal S$};
  			
  			% Two outgoing lines
  			\draw[line] (S.north west) - +(-0.8,1.2);
  			\draw[line] (S.north east) - +( 0.8,1.2);
  			
  \end{tikzpicture}}},~~~\mathcal S^{-1}=\vcenter{\hbox{\begin{tikzpicture}[
  			tensor/.style={
  				draw,
  				circle,
  				minimum size=0.4cm,
  				align=center
  			},
  			line/.style={thick}
  			]
  			
  			% Circular tensor
  			\node[tensor] (S) at (0,0) {$\mathcal S^{-1}$};
  			
  			% Two outgoing lines
  			\draw[line] (S.south west) - +(-0.8,-1.2);
  			\draw[line] (S.south east) - +( 0.8,-1.2);
  			
  \end{tikzpicture}}}\ee
   Here $\mathcal S$ is the Hopf-link complement state in $S^3$. Geometrically, it is the operation which converts the torus cycle convention appropriate for a drilled-out link component into the convention appropriate for gluing another splice piece along that boundary. This is the same general idea we used in the derivation of the satellite operator (\ref{pattern_operator}).

Now to construct the splice piece $\mathcal X_{\mathcal L}^{r,s}$, we treat the first $r$ boundary components of $S^3\setminus\nu(\mathcal L)$ as incoming, while the remaining $s$ boundary components are treated as outgoing. The corresponding splice operator is obtained by applying $\mathcal S^{-1}$ to the first $r$ tensor factors of the link-complement state
\be \Z(\mathcal X_{\mathcal L}^{r,s}) = \Z(S^3\setminus\nu(\mathcal L)) \left( (\mathcal S^{-1})^{\otimes r}\otimes \id^{\otimes s} \right) \in \V^{\otimes r}\otimes\V_*^{\otimes s}. \ee
Equivalently, this defines a linear map
\be\Z(\mathcal X_{\mathcal L}^{r,s}):\V_*^{\otimes r} \longrightarrow \V_*^{\otimes s}. \ee
   
In components, write the original link-complement state as 
\be\Z(S^3\setminus\nu(\mathcal L)) = \sum_{a_1,\ldots,a_r,b_1,\ldots,b_s} \Z\left( S^3; \mathcal L[a_1,\ldots,a_r,b_1,\ldots,b_s] \right) \langle a_1,\ldots,a_r,b_1,\ldots,b_s|. \ee 
Then the corresponding splice operator is 
\be \boxed{ \Z(\mathcal X_{\mathcal L}^{r,s}) = \sum_{a_1,\ldots,a_r,b_1,\ldots,b_s} \Z\left( S^3; \mathcal L[a_1,\ldots,a_r,b_1,\ldots,b_s] \right)  |\widetilde a_1,\ldots,\widetilde a_r\rangle  \langle b_1,\ldots,b_s| . } \label{splice_operator} \ee 
This is the basic algebraic operation that turns a link-complement state into a splice map.

The splice construction is especially important for the JSJ decomposition. Recall that the JSJ decomposition cuts a compact, irreducible\footnote{A compact manifold is irreducible if every embedded $S^2$ bounds a three-ball. The only prime compact connected orientable manifold that is not irreducible is $S^2\times S^1$.} manifold with torus boundary along a minimal family of disjoint incompressible tori, so that each component is either Seifert-fibered or atoroidal. The Seifert-fibered pieces are not decomposed further, even though they may contain incompressible tori, because this decomposition is not canonical. By Thurston hyperbolization, the non-Seifert-fibered atoroidal pieces are hyperbolic. The original link complement is recovered by splicing the JSJ pieces back together.

From this description it is clear that a splice piece obtained by cutting a link complement in $S^3$ along incompressible tori may not literally be a link complement in $S^3$. Nevertheless, one can canonically re-embed these pieces in $S^3$ as link complements by acting with the $\mathcal S$-map on the in-going boundaries. This implements the ``untwisted re-embedding'' map described by Budney \cite{budney2007jsjdecompositionsknotlinkcomplements}. 

More generally, for links embedded in other closed manifolds, one may represent the ambient manifold by Dehn surgery on a framed link in $S^3$. Leaving the surgery boundaries open, decomposing the complement manifold, and then refilling them allows the same TQFT description to be used in that setting as well. Strictly speaking, this is not the JSJ decomposition of the manifold itself, which is intrinsic and independent of any surgery presentation. Rather, after presenting the ambient manifold by surgery in $S^3$, we construct an auxiliary link-complement in $S^3$ by removing the tubular neighbourhoods of both the embedded link and the surgery link. We may decompose this auxiliary manifold into splice pieces and then recover a decomposition of the original manifold by Dehn filling the surgery boundary components with the prescribed surgery data. The surgery presentation is used here only as a computational device for the TQFT evaluation. Thus, for the purposes of the TQFT construction, splice pieces may be represented by link complements in $S^3$, together with boundary parametrizations and, when necessary, Dehn-filling data.

\begin{figure}
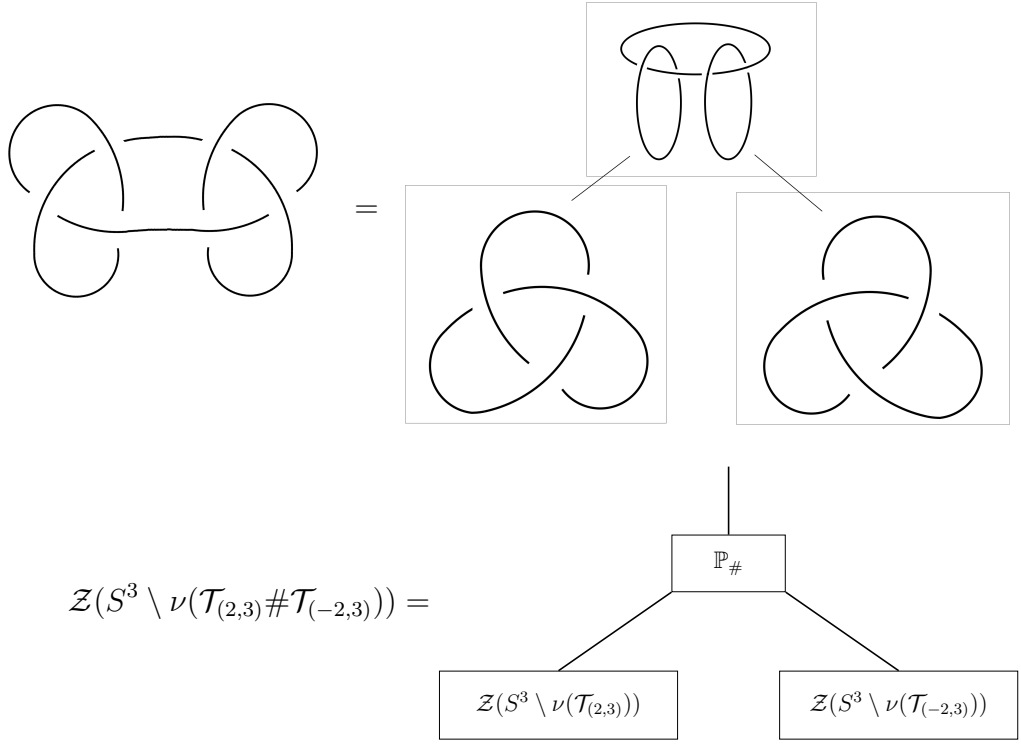

	$$\vcenter{\hbox{\includegraphics[width=0.3\textwidth]{connectedsum_trefoils}}}=\vcenter{\hbox{\includegraphics[width=0.6\textwidth]{connectedsum_splices}}}$$
	$$ \Z(S^3\setminus\nu(\mathcal T_{(2,3)}\#{\mathcal T}_{(-2,3)}))= \vcenter{\hbox{\scalebox{0.75}{\begin{tikzpicture}[
		tensor/.style={
			draw,
			rectangle,
			minimum width=4.2cm,
			minimum height=1.2cm,
			align=center
		},
		smalltensor/.style={
			draw,
			rectangle,
			minimum width=2.0cm,
			minimum height=1.0cm,
			align=center
		},
		line/.style={thick}
		]
		
		% Bottom boxes
		\node[tensor] (T1) at (-3,0) {$\mathcal Z(S^3\setminus \nu(\mathcal T_{(2,3)}))$};
		\node[tensor] (T2) at ( 3,0) {$\mathcal Z(S^3\setminus \nu(\mathcal T_{(-2,3)}))$};
		
		% Top box
		\node[smalltensor] (P) at (0,2.5) {$\P_{\#}$};
		
		% Lines from bottom boxes to top box
		\draw[line] (T1.north) -- (P.south west);
		\draw[line] (T2.north) -- (P.south east);
		
		% Outgoing line above P_#
		\draw[line] (P.north) -- ++(0,1.2);
		
	\end{tikzpicture}}}}$$
	\caption{Top: The JSJ decomposition of a connected sum into splice pieces. Bottom: The corresponding decomposition of the knot-complement state as a network of TQFT bordisms/linear maps.\label{connsumfig}}
\end{figure}

The TQFT turns the JSJ decomposition into a network decomposition of the link-complement state. The vertices of the network are the TQFT maps associated with the JSJ pieces, and the internal edges are the torus Hilbert spaces associated with the tori along which the pieces are glued. See figure \ref{connsumfig} for an example.

A simple splice operation which is not a satellite operation is the connected-sum map (see figure \ref{connsumfig})
\be\P_{\#}: \V_*\otimes\V_* \longrightarrow \V_*. \ee
Topologically, $\P_{\#}$ is related to the complement of the three-component Hopf keyring $\mathcal H^3$, with the two ``keys" treated as incoming boundaries and the central component treated as outgoing. When this splice map is applied to two link-complement states in $S^3$, it produces the complement state of the connected-sum link. In particular, for knots $\mathcal K_1$ and $\mathcal K_2$
\be\Z(S^3\setminus\nu(\mathcal K_1\#\mathcal K_2)) =  \left( \Z(S^3\setminus\nu(\mathcal K_1)) \otimes \Z(S^3\setminus\nu(\mathcal K_2)) \right)\P_{\#}. \ee

In terms of the general prescription above, the connected-sum map is obtained by starting with the three-boundary link-complement state of the Hopf keyring $\mathcal H^3$ and applying $\mathcal S^{-1}$ to the two boundary components corresponding to the keys 
\be \P_{\#} = \Z\left(S^3\setminus\nu(\mathcal H^3)\right) \left( \mathcal S^{-1}\otimes  \id\otimes \mathcal S^{-1} \right). \label{csum}\ee Diagrammatically,
\be \vcenter{\hbox{\scalebox{0.75}{\begin{tikzpicture}[
				tensor/.style={
					draw,
					rectangle,
					minimum width=4.2cm,
					minimum height=1.2cm,
					align=center
				},
				smalltensor/.style={
					draw,
					rectangle,
					minimum width=2.0cm,
					minimum height=1.0cm,
					align=center
				},
				line/.style={thick}
				]

				% Top box
				\node[smalltensor] (P) at (0,2.5) {$\P_{\#}$};
				
				% Lines from bottom boxes to top box
				\draw[line] (T1.north) -- (P.south west);
				\draw[line] (T2.north) -- (P.south east);
				
				% Outgoing line above P_#
				\draw[line] (P.north) -- ++(0,1.2);
				
\end{tikzpicture}}}}=\vcenter{\hbox{\begin{tikzpicture}[
			box/.style={
				draw,
				rectangle,
				minimum width=5.6cm,
				minimum height=1.2cm,
				align=center
			},
			circ/.style={
				draw,
				circle,
				minimum size=0.6cm,
				inner sep=1pt,
				align=center
			},
			line/.style={
				thick,
				line cap=round,
				line join=round
			}
			]
			
			% Main box
			\node[box] (H) at (0,0.8)
			{$\mathcal Z\bigl(S^3\setminus \nu(\mathcal H^3)\bigr)$};
			
			% S^{-1} nodes outside the box
			\node[circ] (SL) at (-3.7,2.25) {$\mathcal S^{-1}$};
			\node[circ] (SR) at ( 3.7,2.25) {$\mathcal S^{-1}$};
			
			% Three attachment points on the top side of the box
			\coordinate (L) at (-1.9,1.4);
			\coordinate (C) at ( 0.0,1.4);
			\coordinate (R) at ( 1.9,1.4);
			
			% Central outgoing line
			\draw[line] (C) -- ++(0,1.4);
			
			% Outer outgoing lines, starting from the top side
			\draw[line]
			(L)
			to[out=90,in=-20]
			(SL.south east);
			
			\draw[line]
			(R)
			to[out=90,in=200]
			(SR.south west);
			
			% Free-hanging legs
			\draw[line]
			(SL.south west)
			to[out=210,in=90]
			(-4.45,-0.25);
			
			\draw[line]
			(SR.south east)
			to[out=-30,in=90]
			(4.45,-0.25);
			
\end{tikzpicture}}} .\ee

This example illustrates the general rule: any splice piece may be represented as a link-complement state in $S^3$, with $\mathcal S^{-1}$-maps glued to the boundary components that are to be interpreted as incoming, and possibly Dehn-filling some boundaries.  Many more examples of splicing links in $S^3$ can be found in \cite{budney2007jsjdecompositionsknotlinkcomplements}.

For ambient manifolds other than $S^3$, the surgery prescription, described earlier, has the advantage that even manifolds which are not embeddable in $S^3$ can be decomposed into link complements in $S^3$, together with surgery data and $\mathcal S^{-1}$-maps. The disadvantage is that this decomposition depends on the chosen surgery presentation, and therefore no longer reflects the uniqueness of the intrinsic JSJ decomposition of the manifold. On the other hand, if it is computationally feasible, one can apply the JSJ decomposition directly to the
full complement manifold and write the TQFT network in terms of the canonical
splice pieces themselves, which by application of the $\mathcal S$-map can be viewed as link complements in various closed manifolds. In this latter case, equation (\ref{splice_operator}) can still be used after replacing $S^3$ by the appropriate closed manifold.

\subsection{Seifert-fibered manifolds}

Seifert-fibered manifolds form one of the basic classes of manifolds appearing in the JSJ decomposition. Informally, a Seifert-fibered manifold is a three-manifold which can be decomposed into disjoint circles, called \emph{fibers}, in such a way that locally it looks like a circle fibration over a two-dimensional surface or orbifold.

More precisely, a compact orientable manifold $\mathcal X$ is Seifert-fibered if it admits a projection
$$S^1\hookrightarrow \mathcal X\xlongrightarrow{\pi} \Sigma,$$
where $\Sigma$ is a compact two-dimensional orbifold. The orbifold $\Sigma$ is called the base of the fibration. In this paper we will restrict to the case where the base orbifold is orientable. Seifert-fibered pieces with non-orientable base can appear in orientable three-manifolds, but then the circle fibration must reverse the orientation of the $S^1$ fiber when one goes around an one-sided loop in the base. Such pieces do not occur for link complements in $S^3$, and we leave their TQFT treatment for future work.

The simplest examples are ordinary circle bundles over surfaces. If $\Sigma$ is a compact surface, then
$
\Sigma\times S^1
$
is Seifert-fibered, with projection onto $\Sigma$. More generally, one may have nontrivial $S^1$-bundles over $\Sigma$. These are the Seifert fibrations with only regular fibers. A genuinely Seifert-fibered manifold may also contain finitely many \emph{exceptional} fibers. In the base orbifold, exceptional fibers appear as singular points.

Near an exceptional fiber, the fibration is not locally the product $D^2\times S^1$, but it is modeled on a quotient of $D^2\times S^1$ by a finite cyclic action. To construct this model fibration over the disk, one may start from the trivial fibration $D^2\times S^1$, cut along a meridional disk, and reglue after a fractional twist $2\pi p/q$. The resulting central fiber is exceptional, since a small disk transverse to this fiber intersects any nearby regular fiber $q$ times. This integer is called the multiplicity of the exceptional fiber.

For an orientable base, we use the same conventions as \cite{Hatcher2001NotesOB}, but we use a slightly different notation
\be\Sigma_{g,n}\left[(\alpha_1,\beta_1),\dots,(\alpha_k,\beta_k)\right] \ee for a Seifert fibration whose base has genus $g$, $n$ boundary components, and $k$ marked fibers. Each boundary circle of the base gives a torus boundary component of the total space $\mathcal X$.  The coprime pairs $(\alpha_i,\beta_i)$ encode the local Seifert data of these marked points. The fibers with $|\alpha_i|>1$ are exceptional with multiplicity $|\alpha_i|$ and project to singular points of the orbifold base. 

A useful construction of this manifold is the following. Start with $\Sigma_{g,n+k}\times S^1$. Then glue a solid torus to each of the last $k$ boundary components with slope $\beta_i/\alpha_i$. The direction $\alpha_i=0$, with infinite slope, is excluded as it kills the fibration. When $|\alpha_i|>1$, this filling produces an exceptional fiber. The case $(\alpha,\beta)=(1,0)$ corresponds to a trivial marked fiber. More generally, a marked fiber with $\alpha=1$ is regular rather than exceptional, but it encodes the global twisting of the fibration. Allowing up to one such marked regular fiber is a convenient way to absorb the Euler number of the fibration into the same notation, so that it need not be tracked separately.

From the point of view of link complements, Seifert-fibered manifolds are the basic non-hyperbolic JSJ pieces. For example, $T^2\times I$ (which can be re-embedded in $S^3$ as the complement of the Hopf link), is a trivial circle fibration over an annulus. Torus knot and torus link complements in $S^3$ are also Seifert-fibered with two exceptional fibers. Lens spaces are examples of closed Seifert-fibered manifolds.

A link-complement manifold whose JSJ pieces are all Seifert-fibered is called a \emph{graph manifold}. Equivalently, graph manifolds are obtained by gluing Seifert-fibered pieces along torus boundaries. The gluing maps need not preserve the fiber slopes on the two sides, and the interesting case is precisely when the fiber direction of one piece is glued to a slope which is not the fiber direction of the adjacent piece. Thus a graph manifold has only a piecewise Seifert-fibered structure: each JSJ component has its own fiber direction, and these directions may be mismatched across the gluing tori. In the TQFT description, this means that graph manifolds are not represented by a single Seifert operator, but by a network of Seifert operators connected by $\SL(2,\mathbb Z)$ transformations encoding the gluing maps.

Seifert-fibered manifolds admit a 2D description by dimensional reduction along the $S^1$ fibers. The resulting 2D TQFT has point defect operators associated with the marked fibers. These defects encode the Dehn-filling data of the marked fibers and can be labelled by the corresponding elements of $\SL(2,\mathbb Z)$.

Graph manifolds admit a related 2D description, but the reduction must be performed piecewise. Each Seifert-fibered piece is reduced along its own fiber direction, while the gluing maps between adjacent pieces become codimension-one defects in the 2D theory. Thus graph manifolds are described by a 2D TQFT with loop defects labelled by elements of $\SL(2,\mathbb Z)$. Contractible defect loops represent marked fibers within a single Seifert-fibered piece, whereas non-contractible defect loops encode gluings along boundaries with incompatible fiber directions. This 2D TQFT description of graph manifolds will be developed in Section \ref{2Dgraph}.

\subsection{Hyperbolic manifolds}

The second basic class of JSJ pieces consists of hyperbolic manifolds. A compact orientable manifold with torus boundary is called hyperbolic if its interior admits a complete, finite-volume Riemannian metric of constant negative curvature.
A fundamental feature of hyperbolic manifolds is rigidity. The hyperbolic metric is uniquely determined by the topology of the manifold. Consequently, geometric quantities constructed from this metric are topological invariants. In particular, if $\mathcal L\subset S^3$ is a hyperbolic link, the volume
$
\operatorname{Vol}(S^3\setminus \nu(\mathcal L))
$
is a topological invariant of the link complement. More generally, for an arbitrary link complement, the sum of the volumes of the hyperbolic JSJ pieces is proportional to the Gromov norm of the complement.

Hyperbolic geometry is related to quantum topology through the volume conjecture and its refinements. In its original form, the volume conjecture predicts that the large-color asymptotics of the colored Jones polynomial of a hyperbolic knot grows exponentially with the hyperbolic volume of the knot complement \cite{Kashaev1997,MurakamiMurakami2001}. From the physics viewpoint, this exponential behaviour is related to the semiclassical limit of Chern-Simons theory with non-compact gauge group $\SL(2,\mathbb C)$ \cite{dimofte2010quantumfieldtheoryvolume,Gukov2005,Witten2011}. 

For a general knot, the corresponding geometric quantity is the sum of the hyperbolic volumes of its JSJ pieces. Seifert-fibered pieces contribute zero volume, while the hyperbolic pieces contribute their finite hyperbolic volumes. This provides another indication that the JSJ decomposition is a natural decomposition from the TQFT perspective.

Hyperbolic pieces are qualitatively different from Seifert-fibered pieces. A Seifert-fibered piece has a preferred circle-fibration structure and therefore admits a reduced 2D description in terms of its base orbifold, possibly with defect insertions. A hyperbolic piece, however, is regarded as a genuinely 3D piece in the JSJ network.

This distinction is also visible at the level of quantum invariants. States and operators built only from Seifert-fibered pieces are controlled by the modular data of the underlying modular tensor category. Hyperbolic links, in contrast, generally probe more detailed categorical information. For example, the Whitehead link is hyperbolic, and its associated $W$-matrix is known to contain information beyond modular data \cite{beyond}. Similarly, systematic searches show that invariants of small hyperbolic knots and links can distinguish modular categories with identical modular data \cite{delaney2018systematicsearchknotlink}. Thus hyperbolic pieces are precisely where one expects link-complement states to depend on categorical data beyond the $S$- and $T$-matrices.

\section{2D TQFT description of graph manifolds\label{2Dgraph}}

In this section we explain how the Seifert-fibered pieces of a graph manifold admit an effective 2D TQFT description. The basic idea is to reduce each Seifert-fibered piece along its $S^1$ fibers. For a single Seifert-fibered manifold, this produces a 2D TQFT on the base orbifold. The smooth surface topology is encoded by the Frobenius algebra of the torus Hilbert space, which is simply the Verlinde algebra of the MTC, while marked fibers appear as point defects labelled by elements of $\SL(2,\mathbb Z)$. For graph manifolds, the same reduction must be performed piecewise. Each JSJ component has its own fiber direction, and the gluing maps between adjacent Seifert-fibered pieces need not identify these fiber directions. After dimensional reduction, these mismatches become loop defects in the 2D theory, again labelled by elements of $\SL(2,\mathbb Z)$. The contractible loops correspond to marked fibers of a Seifert-fibered piece, while non-contractible loops are genuine gluing data. 

This distinction will be important for the entanglement structure of the associated link-complement states. The Frobenius-algebra structure appearing in the dimensional reduction of a Seifert-fibered piece is precisely the algebraic mechanism that produces GHZ-like states \cite{coecke2010compositionalstructuremultipartitequantum}. Meanwhile, non-contractible loop defects in graph manifolds can produce more general states.

\subsection{Seifert-fibered manifolds and Frobenius algebra}

We now describe the effective 2D TQFT obtained by reducing an orientable Seifert-fibered manifold along its $S^1$ fibers. We restrict to Seifert-fibered manifolds with orientable base. The extension to orientable three-manifolds with non-orientable base is possible, using the unoriented 2D TQFT formalism of \cite{Turaev_2006}, but we leave this case for future work.

For the Seifert-fibered manifold $\mathcal X$ with
\be S^1\hookrightarrow \mathcal X\xlongrightarrow{\pi} \Sigma_{g,n}[(\alpha_1,\beta_1),\ldots,(\alpha_k,\beta_k)],\ee
we define the 2D TQFT functor $\Z_{2D}$ by
\be \Z_{2D}\left(\Sigma_{g,n}[(\alpha_1,\beta_1),\ldots,(\alpha_k,\beta_k)]\right):=\Z(\mathcal X)\ee
and the circle vector space is the reduction of the torus space
\be \Z_{2D}\left(S^1\right)=\Z(T^2).\ee

 The ordinary surface bordisms $\Sigma_{g,n}$ give the usual commutative Frobenius-algebra structure, while the marked fibers become point defects labelled by the corresponding Seifert data, equivalently by elements of $\SL(2,\mathbb Z)$. Since these are point defects in the reduced 2D theory, they may be moved freely within each connected component of the bordism.

 When $|\alpha_i|>1$, the image of the fiber under $\pi$ is a conical orbifold point of multiplicity $|\alpha_i|$. The case $(\alpha,\beta)=(1,0)$ corresponds to a trivial defect. Data with $|\alpha|=1$ and $\beta\neq 0$ do not describe an exceptional fiber, rather, they encode a nontrivial twisting of the global fibration. We do not allow the label $(0,1)$, which corresponds to a Dehn filling with infinite slope that kills the fibration and generally produces non-prime manifolds.

The generating bordisms are therefore disks, pairs of pants, and disks carrying point defects. The Frobenius algebra is isomorphic to the Verlinde algebra of the modular tensor category. There is, however, no canonical choice of basis for this algebra: conjugating all generators by an element of $\SL(2,\mathbb Z)$ gives an equivalent description. We choose the basis in which the algebra takes the following form
\begin{subequations}\begin{align}
	\Z_{2D}\left(\vcenter{\hbox{\begin{tikzpicture}
				\filldraw[left color=lightgray, right color=white] (-0.25,0) to [out=-90,in=180] (0,-0.33) to [in=-90,out=0] (0.25,0) to  (0.75,0) to [in=90,out=-90] (0.25,-1) to [out=-90,in=-90] (-0.25,-1) to [in=-90,out=90] (-0.75,0);
				\filldraw[left color=white,right color=lightgray] (-0.5,0) ellipse (0.25 and 0.1);
				\filldraw[left color=white,right color=lightgray] (0.5,0) ellipse (0.25 and 0.1);
				\draw[dotted] (0.25,-1) arc (0:180:0.25 and 0.1);
	\end{tikzpicture}}}\right) &= \sum N_{ab}^c |c\rangle\langle ab|=\sum_a {1\over S_{\id a}}|\tilde a\rangle\langle\tilde a\tilde a|\\
	\Z_{2D}\left(\vcenter{\hbox{\begin{tikzpicture}
				\filldraw[left color=lightgray, right color=white] (-0.25,-1) to [out=90,in=180] (0,-0.66) to [in=90,out=0] (0.25,-1) to [out=-90,in=-90] (0.75,-1) to [in=-90,out=90] (0.25,0) to (-0.25,0) to [in=90,out=-90] (-0.75,-1) to [out=-90,in=-90] (-0.25,-1);
				\filldraw[left color=white,right color=lightgray] (0,0) ellipse (0.25 and 0.1);
				\draw[dotted] (-0.25,-1) arc (0:180:0.25 and 0.1);
				\draw[dotted] (0.75,-1) arc (0:180:0.25 and 0.1);
	\end{tikzpicture}}}\right)&= \sum_a {1\over S_{\id a}}|\tilde a\tilde a\rangle\langle\tilde a|\\
	\Z_{2D}\left(\vcenter{\hbox{\begin{tikzpicture}
				\filldraw[right color=white,left color=lightgray] (-0.25,0) to [out=90,in=180] (0,0.33) to [in=90,out=0] (0.25,0) to [in=-90,out=-90] (-0.25,0);
				\draw[dotted] (0.25,0) arc (0:180:0.25 and 0.1);
	\end{tikzpicture}}}\right)&=|\id\rangle\\
	\Z_{2D}\left(\vcenter{\hbox{\begin{tikzpicture}
				\filldraw[right color=white,left color=lightgray] (-0.25,0) to [out=-90,in=180] (0,-0.33) to [in=-90,out=0] (0.25,0);
				\filldraw[left color=white,right color=lightgray] (0,0) ellipse (0.25 and 0.1);
	\end{tikzpicture}}}\right)&=\langle \id|.
\end{align}\label{Frob_generators}\end{subequations}
The first line is the Verlinde coproduct. The remaining generators are the product, the counit, and the unit. They satisfy the standard relations of a commutative Frobenius algebra \cite{Kock_2003}.

Marked fibers give additional point-defect generators
\be
	\Z_{2D}\left(\vcenter{\hbox{\begin{tikzpicture}
				\filldraw[right color=white,left color=lightgray] (-0.25,0) to [out=90,in=180] (0,0.33) to [in=90,out=0] (0.25,0) to [in=-90,out=-90] (-0.25,0);
				\draw[dotted] (0.25,0) arc (0:180:0.25 and 0.1);
				\fill (0,0.25) circle (1.5pt);
				\node[above] at (0,0.25) {$({\alpha},{\beta})$};
	\end{tikzpicture}}}\right)=\sum_a {( S U_{(\alpha,\beta)})_{a\id}}| \tilde a\rangle .\label{capcup} \ee
Above, the $\SL(2,\mathbb Z)$ matrix is chosen (non-uniquely) so that $U_{(\alpha,\beta)}\left(\begin{smallmatrix}
	1\\
	0
\end{smallmatrix}\right)=\left(\begin{smallmatrix}
\alpha\\
\beta
\end{smallmatrix}\right)$. In particular, $(1,0)$ is the trivial defect, giving rise to the trivial fibration $D^2\times S^1$, while $(1,\beta)$ with $\beta\neq 0$ describes a regular non-trivial fibration over the disk. For $|\alpha|>1$, this generator is a model Seifert fibration that locally describes an exceptional fiber.

The new relations among the generators are those imposed by isotopy invariance. In particular, point defects may be moved freely within each connected component of the bordism. 
There is, however, a further subtlety that we will not treat systematically here. Certain Seifert-fibered manifolds admit more than one Seifert fibration (see Theorem 2.3 of \cite{Hatcher2001NotesOB}). These include the ``small'' Seifert manifolds, such as lens spaces, $S^3$, $S^2\times S^1$, and fibrations over $D^2$ with at most two exceptional fibers.
There is also a related redundancy in the Seifert data, due to our choice to allow up to one marked regular fiber, in order to absorb the Euler number of the fibration. This gives rise to the possibility that, after gluing, two marked points have $|\alpha_i|=1$. Such points should be combined into a single marked point, since such points encode the global twisting of the fibration. Consequently, distinct presentations in terms of the above generators may describe diffeomorphic manifolds. A complete formulation of the associated 2D theory should therefore include additional relations identifying these equivalent presentations. We leave a systematic treatment of these relations for future work.

We can now write the general Seifert-fibered operator. Consider an orientable Seifert-fibered manifold with orientable base of genus $g$, with $k$ marked points of types $(\alpha_i,\beta_i)$, and with $r+s$ torus boundaries. Treat $r$ of these boundaries as incoming and $s$ as outgoing. Up to local unitaries acting on the boundary Hilbert spaces, this evaluates to
\be \sum_a {(SU_{(\alpha_1,\beta_1)})_{ a\id}\cdots (SU_{(\alpha_k,\beta_k)})_{a\id}\over S_{a\id}^{2g+(r+s)+k-2}}|\tilde a\rangle^{\otimes r}\langle \tilde a|^{\otimes s}.\label{seif}\ee

Strictly speaking, a Seifert-fibered manifold with boundary is specified not only by its base, but also by parametrizations of its boundary tori. In the 2D description these boundary parametrizations are represented by $\SL(2,\mathbb Z)$ loop defects parallel to the boundary components, acting as local unitaries. For a single Seifert-fibered piece, we suppress these local boundary operators. They become essential, however, when gluing the Seifert-fibered piece to another manifold.

\iffalse{
Let us finally consider some examples, including the boundary maps explicitly. First, the $(p,q)$-cabling operator of (\ref{torus_pattern}) is a Seifert-fibered bordism with a single singular point, that embeds in $S^3$
\be \P_{(p,q)}=\sum_l {\langle \tilde l|U^\dagger|\id\rangle\over S_{\id l}^n} U|\tilde l\rangle\langle \tilde l|^{\otimes n},\ee
where $n=\gcd(p,q)$ and $U$ is chosen such that $(1,0)U=(p/n,q/n)$. The untwisted re-embedding of this cabling operator is the Seifert link $\mathcal S_{(p,q)}$ (see figure \ref{seifertlink}).

A closely related example is the complement of the torus link
\be \Z(S^3\setminus\nu(\mathcal T_{(p,q)}))=\sum_l {\langle \tilde l|U^\dagger|\id\rangle\langle \tilde\id|U|\tilde l\rangle\over S_{\id l}^n} \langle \tilde l|^{\otimes n}.\ee
This is a Seifert-fibered manifold with two exceptional fibers $(\alpha_1,\beta_1)$, $(\alpha_2,\beta_2)$, satisfying $|\alpha_1\beta_2-\alpha_2\beta_1|=1$.

Finally, the connected sum operator (associated with the Hopf keychain $\mathcal H^{n+1}$) is
\be \P_{\#,n}=\sum_a {1\over S_{\id a}^{n-1}}|a\rangle^{\otimes n}\langle a|.\ee
Above the $n$ ``keys" of the keychain are treated as in-going boundaries, while the keyholder is the outgoing boundary.
}\fi

\subsection{Graph manifolds and the extended algebra}

A graph manifold is a compact manifold whose JSJ pieces are all Seifert-fibered. In the 2D description developed above, Seifert-fibered pieces are described by a Frobenius algebra with point defects. To describe a graph manifold, however, we must also keep track of the gluing maps between adjacent Seifert-fibered pieces. These gluing maps need not preserve the fiber directions on the two sides, and this is what prevents a general graph manifold from being globally Seifert-fibered.

After dimensional reduction, such gluing maps appear as loop defects. Thus the 2D description of graph manifolds is obtained from the Frobenius algebra \eqref{Frob_generators} by allowing codimension-one defects labelled by elements of $\SL(2,\mathbb Z)$. The basic defect generators are the modular transformations
\be 
\Z_{2D}\left(~~~\vcenter{\hbox{\begin{tikzpicture}
	\filldraw[left color=lightgray, right color=white] (-0.25,0) -- (0.25,0) -- (0.25,-1) to [in=-90,out=-90] (-0.25,-1) -- (-0.25,0);
	\filldraw[left color=white,right color=lightgray] (0,0) ellipse (0.25 and 0.1);, 
	\draw[dotted] (0.25,-1) arc (0:180:0.25 and 0.1);
	\draw[dotted,thick] (0.25,-0.5) arc (0:180:0.25 and 0.1);
	\draw[thick,postaction={decorate},
	decoration={markings, mark=at position 0.6 with {\arrow{<}}}] (0.25,-0.5) arc (0:-180:0.25 and 0.1);
	\node[right] at (0.25,-0.5) {$S$};
\end{tikzpicture}}}\right)=\sum_{a,b} S_{ab}| a\rangle\langle  b|,~~~\Z\left(~~~\vcenter{\hbox{\begin{tikzpicture}
\filldraw[left color=lightgray, right color=white] (-0.25,0) -- (0.25,0) -- (0.25,-1) to [in=-90,out=-90] (-0.25,-1) -- (-0.25,0);
\filldraw[left color=white,right color=lightgray] (0,0) ellipse (0.25 and 0.1);, 
\draw[dotted] (0.25,-1) arc (0:180:0.25 and 0.1);
\draw[dotted,thick] (0.25,-0.5) arc (0:180:0.25 and 0.1);
\draw[thick,postaction={decorate},
decoration={markings, mark=at position 0.6 with {\arrow{<}}}] (0.25,-0.5) arc (0:-180:0.25 and 0.1);
\node[right] at (0.25,-0.5) {$T$};
\end{tikzpicture}}}\right)=\sum_{a,b} T_{ab}| a\rangle\langle  b|.\ee
More generally, a loop defect labelled by $U\in\SL(2,\mathbb Z)$ acts by the corresponding modular transformation.

The defining relations are the Frobenius algebra relations, together with the rules following from isotopy and fusion of lines. In particular, parallel defect loops fuse by multiplying their labels
\be \Z\left(~~~\vcenter{\hbox{\begin{tikzpicture}
			\filldraw[left color=lightgray, right color=white] (-0.25,0) -- (0.25,0) -- (0.25,-1) to [in=-90,out=-90] (-0.25,-1) -- (-0.25,0);
			\filldraw[left color=white,right color=lightgray] (0,0) ellipse (0.25 and 0.1);, 
			\draw[dotted] (0.25,-1) arc (0:180:0.25 and 0.1);
			\draw[thick,postaction={decorate},
			decoration={markings, mark=at position 0.6 with {\arrow{<}}}] (0.25,-0.35) arc (0:-180:0.25 and 0.1);
			\draw[dotted,thick] (0.25,-0.35) arc (0:180:0.25 and 0.1);
			\node[right] at (0.25,-0.3) {$U_1$};
			\draw[thick,postaction={decorate},
			decoration={markings, mark=at position 0.6 with {\arrow{<}}}] (0.25,-0.7) arc (0:-180:0.25 and 0.1);
			\draw[dotted,thick] (0.25,-0.7) arc (0:180:0.25 and 0.1);
			\node[right] at (0.25,-0.9) {$U_2$};
\end{tikzpicture}}}\right)= \Z_{2D}\left(~~~\vcenter{\hbox{\begin{tikzpicture}
\filldraw[left color=lightgray, right color=white] (-0.25,0) -- (0.25,0) -- (0.25,-1) to [in=-90,out=-90] (-0.25,-1) -- (-0.25,0);
\filldraw[left color=white,right color=lightgray] (0,0) ellipse (0.25 and 0.1);, 
\draw[dotted] (0.25,-1) arc (0:180:0.25 and 0.1);
\draw[dotted,thick] (0.25,-0.5) arc (0:180:0.25 and 0.1);
\draw[thick,postaction={decorate},
decoration={markings, mark=at position 0.6 with {\arrow{<}}}] (0.25,-0.5) arc (0:-180:0.25 and 0.1);
\node[right] at (0.25,-0.5) {$U_2U_1$};
\end{tikzpicture}}}\right),~~~U_1,U_2\in\SL(2,\mathbb Z) .\ee
The remaining relations follow from isotopy of defect lines. For example
\be \Z_{2D}\left(\vcenter{\hbox{\begin{tikzpicture}
	\filldraw[left color=lightgray, right color=white] (-0.25,0) to [out=-90,in=180] (0,-0.33) to [in=-90,out=0] (0.25,0) to  (0.75,0) to [out=-90,in=0] (0, -0.83) to [out=180,in=-90]
	(-0.75,0);
	\filldraw[left color=white,right color=lightgray] (-0.5,0) ellipse (0.25 and 0.1);
	\filldraw[left color=white,right color=lightgray] (0.5,0) ellipse (0.25 and 0.1);
	\draw[thick,postaction={decorate},
	decoration={markings, mark=at position 0.6 with {\arrow{<}}}] (-0.2,-0.25) arc (0:-180:0.25 and 0.1);
	\draw[dotted,thick] (-0.2,-0.25) arc (0:180:0.25 and 0.1);
	\node[left] at (-0.8,-0.2) {$U$};
	\end{tikzpicture}}}~~\right)=\Z_{2D}\left(~~\vcenter{\hbox{\begin{tikzpicture}
	\filldraw[left color=lightgray, right color=white] (-0.25,0) to [out=-90,in=180] (0,-0.33) to [in=-90,out=0] (0.25,0) to  (0.75,0) to [out=-90,in=0] (0, -0.83) to [out=180,in=-90]
	(-0.75,0);
	\filldraw[left color=white,right color=lightgray] (-0.5,0) ellipse (0.25 and 0.1);
	\filldraw[left color=white,right color=lightgray] (0.5,0) ellipse (0.25 and 0.1);
	\draw[thick,postaction={decorate},
	decoration={markings, mark=at position 0.6 with {\arrow{>}}}] (0.70,-0.25) arc (0:-180:0.25 and 0.1);
	\draw[dotted,thick] (0.70,-0.25) arc (0:180:0.25 and 0.1);
	\node[right] at (0.72,-0.2) {$U$};
\end{tikzpicture}}}\right)=\Z_{2D}\left(~~\vcenter{\hbox{\begin{tikzpicture}
	\filldraw[left color=lightgray, right color=white] (-0.25,0) to [out=-90,in=180] (0,-0.33) to [in=-90,out=0] (0.25,0) to  (0.75,0) to [out=-90,in=0] (0, -0.83) to [out=180,in=-90]
	(-0.75,0);
	\filldraw[left color=white,right color=lightgray] (-0.5,0) ellipse (0.25 and 0.1);
	\filldraw[left color=white,right color=lightgray] (0.5,0) ellipse (0.25 and 0.1);
	\draw[thick,postaction={decorate},
	decoration={markings, mark=at position 0.6 with {\arrow{<}}}] (0.70,-0.25) arc (0:-180:0.25 and 0.1);
	\draw[dotted,thick] (0.70,-0.25) arc (0:180:0.25 and 0.1);
	\node[right] at (0.72,-0.2) {$U^T$};
\end{tikzpicture}}}\right).\ee
Note that due to the Frobenius pairing, line reversal does not produce the inverse operator, but the linear dual. This is an involution of the modular group. In practice, $U^T$ can be constructed by expressing $U$ in terms of $S,T$ and then reversing their order.

The point defects that appeared in the previous subsection do not need to be introduced independently, as they are the special case of contractible loops
\be \Z_{2D}\left(\vcenter{\hbox{\begin{tikzpicture}
			\filldraw[right color=white,left color=lightgray] (-0.25,0) to [out=90,in=180] (0,0.33) to [in=90,out=0] (0.25,0) to [in=-90,out=-90] (-0.25,0);
			\draw[dotted] (0.25,0) arc (0:180:0.25 and 0.1);
			\fill (0,0.25) circle (1.5pt);
			\node[above] at (0,0.25) {$({\alpha},{\beta})$};
\end{tikzpicture}}}\right)=\Z_{2D}\left(\vcenter{\hbox{\begin{tikzpicture}
\filldraw[right color=white,left color=lightgray] (-0.25,0) to [out=90,in=180] (0,0.33) to [in=90,out=0] (0.25,0) to [in=-90,out=-90] (-0.25,0);
\draw[dotted] (0.25,0) arc (0:180:0.25 and 0.1);
\draw[thick,postaction={decorate},
decoration={markings, mark=at position 0.6 with {\arrow{<}}}] (0.25,0.1) arc (0:-180:0.25 and 0.1);
\draw[dotted,thick] (0.25,0.1) arc (0:180:0.25 and 0.1);
\node[above] at (0,0.25) {$U_{(\alpha,\beta)}$};
\end{tikzpicture}}}\right).\ee
 
\begin{figure}
	\centering\includegraphics[width=1\textwidth]{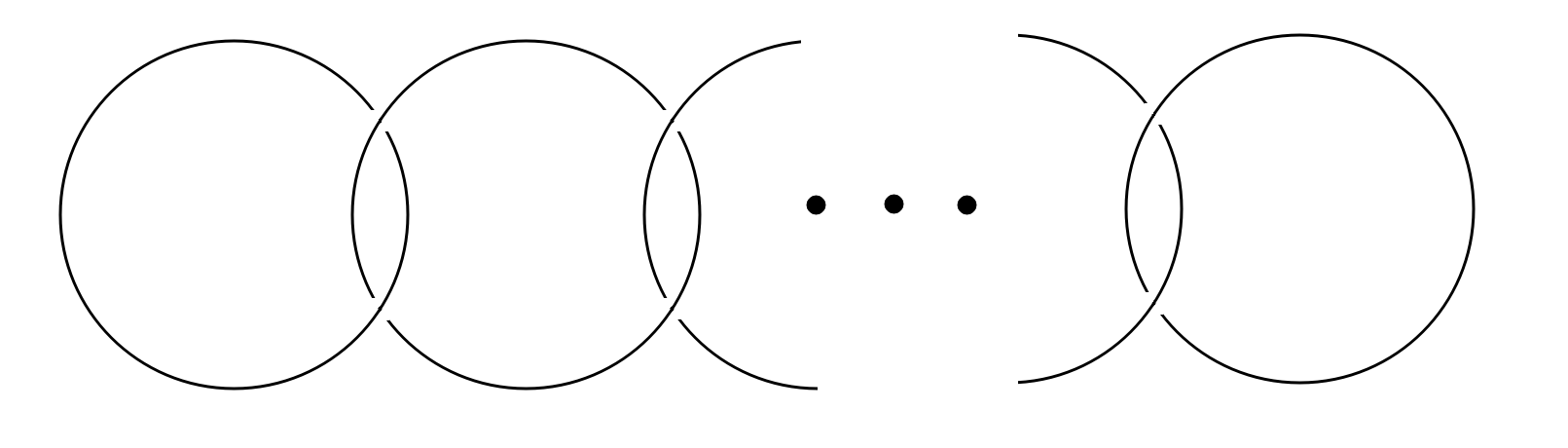}
	\caption{The Hopf chain.\label{hopfchainfig}}
\end{figure}

 Let us now illustrate the formalism with the open Hopf chain (see figure \ref{hopfchainfig}). Denote by $\mathcal C^n$ the $n$-component Hopf chain. Note that $\mathcal C^2$ is the Hopf link, while $\mathcal C^3$ is the same as the three-component Hopf keychain $\mathcal H^3$, both of which have Seifert-fibered complements in $S^3$. The smallest non-trivial example is therefore the $4$-component chain. The JSJ decomposition of $S^3\setminus\nu(\mathcal C^n)$ is built from Hopf keychain $\mathcal H^3$ complements, connected sum operators (which can be re-embedded in $S^3$ as $\mathcal H^3$ complements), together possibly with one Hopf-link $\mathcal H^2$ complement depending on the parity of $n$.

The elementary building blocks are as follows. The Hopf keychain $\mathcal H^3$ complement has state
\be \Z(S^3\setminus\nu(\mathcal H^3))=\Z_{2D}\left(~\vcenter{\hbox{\begin{tikzpicture}
			\filldraw[left color=lightgray, right color=white]
			(-0.75,0)
			-- (-0.25,0)
			to[out=-90,in=180] (0,-0.33)
			to[out=0,in=-90] (0.25,0)
			-- (0.75,0)
			to[out=-90,in=180] (1,-0.33)
			to[out=0,in=-90] (1.25,0)
			-- (1.75,0)
			to[out=-90,in=0] (0.5,-0.95)
			to[out=180,in=-90] (-0.75,0);
			
			\filldraw[left color=white,right color=lightgray] (-0.5,0) ellipse (0.25 and 0.1);
			\filldraw[left color=white,right color=lightgray] (0.5,0) ellipse (0.25 and 0.1);
			\draw[red,thick,postaction={decorate},
			decoration={markings, mark=at position 0.8 with {\arrow{>}}}] (0.5,0) ellipse (0.25 and 0.1);
			\filldraw[left color=white,right color=lightgray] (1.5,0) ellipse (0.25 and 0.1);
			
\end{tikzpicture}}}~\right)=\sum_a {1\over S_{\id a}}\langle \tilde a a \tilde a|,\ee
while the complement of the Hopf link is given by (\ref{Smap})
\be \Z(S^3\setminus\nu(\mathcal H^2))=\Z_{2D}\left(~\vcenter{\hbox{\begin{tikzpicture}
			\filldraw[left color=lightgray, right color=white] (-0.25,0) to [out=-90,in=180] (0,-0.33) to [in=-90,out=0] (0.25,0) to  (0.75,0) to [out=-90,in=0] (0, -0.83) to [out=180,in=-90]
			(-0.75,0);
			\filldraw[left color=white,right color=lightgray] (-0.5,0) ellipse (0.25 and 0.1);
			\filldraw[left color=white,right color=lightgray] (0.5,0) ellipse (0.25 and 0.1);
			\draw[thick,red,postaction={decorate},
			decoration={markings, mark=at position 0.8 with {\arrow{>}}}] (-0.25,0) arc (0:360:0.25 and 0.1);
\end{tikzpicture}}}~\right)=\sum_a \langle a\tilde a|.\ee
The connected-sum operator is the splice (\ref{csum})
\be \P_{\#}=\Z_{2D}\left(~\vcenter{\hbox{\begin{tikzpicture}
			\filldraw[left color=lightgray, right color=white] (-0.25,-1) to [out=90,in=180] (0,-0.66) to [in=90,out=0] (0.25,-1) to [out=-90,in=-90] (0.75,-1) to [in=-90,out=90] (0.25,0) to (-0.25,0) to [in=90,out=-90] (-0.75,-1) to [out=-90,in=-90] (-0.25,-1);
			\filldraw[left color=white,right color=lightgray] (0,0) ellipse (0.25 and 0.1);
			\draw[red,thick,,postaction={decorate},
			decoration={markings, mark=at position 0.8 with {\arrow{>}}}] (0,0) ellipse (0.25 and 0.1);
			\draw[dotted,thick,blue] (-0.25,-1) arc (0:180:0.25 and 0.15);
			\draw[thick,blue,postaction={decorate},
			decoration={markings, mark=at position 0.6 with {\arrow{<}}}] (-0.25,-1) arc (0:-180:0.25 and 0.15);
			\draw[dotted,thick,blue] (0.75,-1) arc (0:180:0.25 and 0.15);
			\draw[thick,blue,postaction={decorate},
			decoration={markings, mark=at position 0.6 with {\arrow{<}}}] (0.75,-1) arc (0:-180:0.25 and 0.15);
\end{tikzpicture}}}~\right)=\sum_a {1\over S_{\id a}}|aa\rangle\langle a|.\ee
It can be obtained from the $\mathcal H^3$ complement by treating two of the boundary components as incoming and one as outgoing.

 In the diagrams above we use a red loop to denote the modular transformation $S^{-1}$ and a blue loop for $S$. We push boundary-parallel loops to the corresponding boundary. Note that since $S$ and $S^{-1}$ are symmetric, we do not need to keep track of the line's orientation, but we will keep the arrows on the boundaries for better readability.

By composing these elementary pieces one obtains the Hopf-chain complement state. For example, the $9$-component Hopf chain is obtained by gluing four Hopf-keychain pieces with three connected-sum maps, as shown in the diagram
\be \Z(S^3\setminus \nu(\mathcal C^9))=\Z_{2D}\left(\vcenter{\hbox{\begin{tikzpicture}[thick]
	
	% Orientation style
	\tikzset{
		orient/.style={
			red,
			thick,
			postaction={decorate},
			decoration={
				markings,
				mark=at position 0.8 with {\arrow{>}}
			}
		}
	}
	
	% First three-output surface
	\begin{scope}[shift={(0,0)}]
		\filldraw[left color=lightgray, right color=white]
		(-0.75,0)
		-- (-0.25,0)
		to[out=-90,in=180] (0,-0.33)
		to[out=0,in=-90] (0.25,0)
		-- (0.75,0)
		to[out=-90,in=180] (1,-0.33)
		to[out=0,in=-90] (1.25,0)
		-- (1.75,0)
		to[out=-90,in=0] (0.5,-0.95)
		to[out=180,in=-90] (-0.75,0);
	\end{scope}
	
	% Second three-output surface
	\begin{scope}[shift={(3,0)}]
		\filldraw[left color=lightgray, right color=white]
		(-0.75,0)
		-- (-0.25,0)
		to[out=-90,in=180] (0,-0.33)
		to[out=0,in=-90] (0.25,0)
		-- (0.75,0)
		to[out=-90,in=180] (1,-0.33)
		to[out=0,in=-90] (1.25,0)
		-- (1.75,0)
		to[out=-90,in=0] (0.5,-0.95)
		to[out=180,in=-90] (-0.75,0);
	\end{scope}
	
	% Third three-output surface
	\begin{scope}[shift={(6,0)}]
		\filldraw[left color=lightgray, right color=white]
		(-0.75,0)
		-- (-0.25,0)
		to[out=-90,in=180] (0,-0.33)
		to[out=0,in=-90] (0.25,0)
		-- (0.75,0)
		to[out=-90,in=180] (1,-0.33)
		to[out=0,in=-90] (1.25,0)
		-- (1.75,0)
		to[out=-90,in=0] (0.5,-0.95)
		to[out=180,in=-90] (-0.75,0);
	\end{scope}
	
	% Fourth three-output surface
	\begin{scope}[shift={(9,0)}]
		\filldraw[left color=lightgray, right color=white]
		(-0.75,0)
		-- (-0.25,0)
		to[out=-90,in=180] (0,-0.33)
		to[out=0,in=-90] (0.25,0)
		-- (0.75,0)
		to[out=-90,in=180] (1,-0.33)
		to[out=0,in=-90] (1.25,0)
		-- (1.75,0)
		to[out=-90,in=0] (0.5,-0.95)
		to[out=180,in=-90] (-0.75,0);
	\end{scope}
	
	% Pair of pants between surfaces 1 and 2
	\begin{scope}[shift={(2,1)}]
		\filldraw[left color=lightgray, right color=white] (-0.25,-1) to [out=90,in=180] (0,-0.66) to [in=90,out=0] (0.25,-1) to [out=-90,in=-90] (0.75,-1) to [in=-90,out=90] (0.25,0) to (-0.25,0) to [in=90,out=-90] (-0.75,-1) to [out=-90,in=-90] (-0.25,-1);
		\filldraw[left color=white,right color=lightgray] (0,0) ellipse (0.25 and 0.1);
		\draw[red,thick,,postaction={decorate},
		decoration={markings, mark=at position 0.8 with {\arrow{>}}}] (0,0) ellipse (0.25 and 0.1);
		\draw[dotted,thick,blue] (-0.25,-1) arc (0:180:0.25 and 0.15);
		\draw[thick,blue,postaction={decorate},
		decoration={markings, mark=at position 0.6 with {\arrow{<}}}] (-0.25,-1) arc (0:-180:0.25 and 0.15);
		\draw[dotted,thick,blue] (0.75,-1) arc (0:180:0.25 and 0.15);
		\draw[thick,blue,postaction={decorate},
		decoration={markings, mark=at position 0.6 with {\arrow{<}}}] (0.75,-1) arc (0:-180:0.25 and 0.15);
	\end{scope}
	
	% Pair of pants between surfaces 2 and 3
	\begin{scope}[shift={(5,1)}]
		\filldraw[left color=lightgray, right color=white] (-0.25,-1) to [out=90,in=180] (0,-0.66) to [in=90,out=0] (0.25,-1) to [out=-90,in=-90] (0.75,-1) to [in=-90,out=90] (0.25,0) to (-0.25,0) to [in=90,out=-90] (-0.75,-1) to [out=-90,in=-90] (-0.25,-1);
		\filldraw[left color=white,right color=lightgray] (0,0) ellipse (0.25 and 0.1);
		\draw[red,thick,,postaction={decorate},
		decoration={markings, mark=at position 0.8 with {\arrow{>}}}] (0,0) ellipse (0.25 and 0.1);
		\draw[dotted,thick,blue] (-0.25,-1) arc (0:180:0.25 and 0.15);
		\draw[thick,blue,postaction={decorate},
		decoration={markings, mark=at position 0.6 with {\arrow{<}}}] (-0.25,-1) arc (0:-180:0.25 and 0.15);
		\draw[dotted,thick,blue] (0.75,-1) arc (0:180:0.25 and 0.15);
		\draw[thick,blue,postaction={decorate},
		decoration={markings, mark=at position 0.6 with {\arrow{<}}}] (0.75,-1) arc (0:-180:0.25 and 0.15);
	\end{scope}
	
	% Pair of pants between surfaces 3 and 4
	\begin{scope}[shift={(8,1)}]
		\filldraw[left color=lightgray, right color=white] (-0.25,-1) to [out=90,in=180] (0,-0.66) to [in=90,out=0] (0.25,-1) to [out=-90,in=-90] (0.75,-1) to [in=-90,out=90] (0.25,0) to (-0.25,0) to [in=90,out=-90] (-0.75,-1) to [out=-90,in=-90] (-0.25,-1);
		\filldraw[left color=white,right color=lightgray] (0,0) ellipse (0.25 and 0.1);
		\draw[red,thick,,postaction={decorate},
		decoration={markings, mark=at position 0.8 with {\arrow{>}}}] (0,0) ellipse (0.25 and 0.1);
		\draw[dotted,thick,blue] (-0.25,-1) arc (0:180:0.25 and 0.15);
		\draw[thick,blue,postaction={decorate},
		decoration={markings, mark=at position 0.6 with {\arrow{<}}}] (-0.25,-1) arc (0:-180:0.25 and 0.15);
		\draw[dotted,thick,blue] (0.75,-1) arc (0:180:0.25 and 0.15);
		\draw[thick,blue,postaction={decorate},
		decoration={markings, mark=at position 0.6 with {\arrow{<}}}] (0.75,-1) arc (0:-180:0.25 and 0.15);
	\end{scope}
	
	% Free boundary circles of the lower surfaces
	\filldraw[left color=white,right color=lightgray]
	(-0.5,0) ellipse (0.25 and 0.1);
	
	\filldraw[left color=white,right color=lightgray]
	(0.5,0) ellipse (0.25 and 0.1);
	\draw[orient] (0.5,0) ellipse (0.25 and 0.1);
	
	\filldraw[left color=white,right color=lightgray]
	(3.5,0) ellipse (0.25 and 0.1);
	\draw[orient] (3.5,0) ellipse (0.25 and 0.1);
	
	\filldraw[left color=white,right color=lightgray]
	(6.5,0) ellipse (0.25 and 0.1);
	\draw[orient] (6.5,0) ellipse (0.25 and 0.1);
	
	\filldraw[left color=white,right color=lightgray]
	(9.5,0) ellipse (0.25 and 0.1);
	\draw[orient] (9.5,0) ellipse (0.25 and 0.1);
	
	\filldraw[left color=white,right color=lightgray]
	(10.5,0) ellipse (0.25 and 0.1);
	
	% Free top circles of the three pairs of pants
	\filldraw[left color=white,right color=lightgray]
	(2,1) ellipse (0.25 and 0.1);
	\draw[orient] (2,1) ellipse (0.25 and 0.1);
	
	\filldraw[left color=white,right color=lightgray]
	(5,1) ellipse (0.25 and 0.1);
	\draw[orient] (5,1) ellipse (0.25 and 0.1);
	
	\filldraw[left color=white,right color=lightgray]
	(8,1) ellipse (0.25 and 0.1);
	\draw[orient] (8,1) ellipse (0.25 and 0.1);
	
\end{tikzpicture}}}\right).\ee
For a Hopf chain of even length, one of the $\mathcal H^3$ must be replaced by a Hopf link $\mathcal H^2$. 

In general, the Hopf keychain complement can be written as follows
\be \Z(S^3\setminus\nu(\mathcal C^n))=\Z_{2D}\left(\vcenter{\hbox{\begin{tikzpicture}
	\filldraw[left color=lightgray, right color=white]
	(-1.75,0)
	-- (-1.25,0)
	to[out=-90,in=180] (-1,-0.33)
	to[out=0,in=-90] (-0.75,0)
	-- (-0.25,0)
	to[out=-90,in=180] (0,-0.33)
	to[out=0,in=-90] (0.25,0)
	-- (0.75,0)
	to[out=-90,in=180] (1,-0.33)
	to[out=0,in=-90] (1.25,0)
	-- (1.75,0)
	to[out=-90,in=180] (2,-0.33)
	to[out=0,in=-90] (2.25,0)
	-- (2.75,0)
	to[out=-90,in=0] (0.5,-1.25)
	to[out=180,in=-90] (-1.75,0);
	
	% erase the top line where the fourth circle would have been
	\draw[white,line width=10pt,line cap=round] (1.22,0) -- (1.78,0);
	
	\filldraw[left color=white,right color=lightgray] (-1.5,0) ellipse (0.25 and 0.1);
	\filldraw[left color=white,right color=lightgray] (-0.5,0) ellipse (0.25 and 0.1);
	\draw[red,thick,postaction={decorate},
	decoration={markings, mark=at position 0.8 with {\arrow{>}}}] (-0.5,0) ellipse (0.25 and 0.1);
	\filldraw[left color=white,right color=lightgray] (0.5,0) ellipse (0.25 and 0.1);
	\draw[red,thick,postaction={decorate},
	decoration={markings, mark=at position 0.8 with {\arrow{>}}}] (0.5,0) ellipse (0.25 and 0.1);
	\filldraw[left color=white,right color=lightgray] (2.5,0) ellipse (0.25 and 0.1);
	
	% dots replacing the fourth circle
	\fill (1.34,0) circle (0.03);
	\fill (1.50,0) circle (0.03);
	\fill (1.66,0) circle (0.03);
	
	% vertical circles
	
	\draw[thick,dotted,blue] (0.1,-0.78) arc (0:360:0.1 and 0.45);
	\draw[thick,blue,
	decoration={markings, mark=at position 0.1 with {\arrow{<}}}] (0.1,-0.78) arc (0:90:0.1 and 0.45);
	\draw[thick,blue] (0.1,-0.78) arc (0:-90:0.1 and 0.45);
	
	\draw[thick,dotted,blue] (1.15,-0.78) arc (0:360:0.1 and 0.45);
	\draw[thick,blue,
	decoration={markings, mark=at position 0.1 with {\arrow{>}}}] (1.15,-0.78) arc (0:90:0.1 and 0.45);
	\draw[thick,blue] (1.15,-0.78) arc (0:-90:0.1 and 0.45);
	
\end{tikzpicture}}}\right)=\sum_{a_i} Q_{a_1a_2}\cdots Q_{a_{n-2}a_{n-1}}\langle a_1 a_2\cdots a_{n-1}\tilde a_{n-1}|,\label{Hopfchainstate}\ee
where
\be Q_{ab}:={S_{ab}\over S_{\id b}}.\ee

It is important not to confuse the open Hopf chain with the closed cyclic Hopf chain. One cannot construct a cyclic Hopf chain by applying the connected-sum operator to the two endpoints of the open chain, since the connected-sum operator takes as input two distinct link-complement manifolds, and does not perform a self-gluing inside a single complement. In fact, such a self-gluing would produce a genus-one base surface and would not describe a 3D manifold embeddable in $S^3$. Moreover, for $n\geq 4$, the closed cyclic Hopf chain complement in $S^3$ is hyperbolic, thus not a graph manifold. The exceptional cases $n=2,3$ are the torus links $\mathcal T_{(2,2)}$, $\mathcal T_{(3,3)}$, whose complements are Seifert-fibered.

We conclude with some comments. First, this 2D language can be generalized to include satellite operators with one incoming and one outgoing boundary. In that case, a satellite operator may be represented as a line defect, and the group of invertible defects $\SL(2,\mathbb Z)$ is enlarged to the monoid of homology cylinders \cite{Levine_2001}. However, once one includes genuine splice pieces with multiple incoming and outgoing boundaries, this 2D picture is no longer useful.

Second, although the discussion in this section decomposes Seifert-fibered manifolds into elementary building blocks, this should not be confused with the JSJ decomposition. The purpose of this finer decomposition is to study the algebraic structure behind Seifert-fibered manifolds. This is not canonical as a decomposition of the three-manifold. From the JSJ point of view, a Seifert-fibered component should be treated as a single piece. On the other hand, the decomposition of a graph manifold into Seifert-fibered JSJ pieces is canonical.

Finally, we comment on 3D abelian TQFTs. Abelian theories are insensitive to much of the refined information carried by hyperbolic pieces. In particular, states constructed by complements of hyperbolic links in an abelian TQFT are determined only by linking data and can be simulated by a suitable graph-manifold construction. Thus, from the perspective of abelian TQFT, hyperbolic pieces are indistinguishable from graph-manifold pieces. In this sense, abelian TQFTs are closer to 2D theories, than their non-abelian counterparts.

\section{Entanglement entropy\label{sec:EE}}

We now study the entanglement entropy of link-complement states. The goal of
this section is to illustrate the formalism developed above and to relate the entanglement patterns to the JSJ decomposition of the complement
manifold. For Seifert-fibered pieces, the effective 2D TQFT description
immediately implies that the states have a GHZ-like form. The reduced density
matrix is diagonal in the idempotent basis of the Frobenius algebra, and the
entropy is independent of the chosen bipartition. This gives a direct
explanation of the entropy formula for Seifert-fibered link complements \cite{Balasubramanian:2025kaf}, as
well as the suppression of non-abelian sectors in the limit of many boundaries \cite{Large-party}.

We then turn to graph manifolds.
The Hopf chain is a simple graph-manifold example. Although it is built
entirely from Seifert-fibered pieces, the non-contractible defect loops produced
by the gluing maps spoil the simple GHZ structure. Instead the entanglement is controlled by a transfer matrix $B$, determined by the modular $S$-matrix. In the limit of many boundaries, the entropy is determined by the Frobenius-Perron eigenvector of $B$, and we show that all sectors of the theory contribute to the entanglement. 

Finally, we study Whitehead doubles of the Hopf chain, whose complements contain hyperbolic JSJ pieces. The Whitehead pattern suppresses abelian sectors and leads to entropy behaviour qualitatively different from the Seifert-fibered and graph-manifold cases.

\subsection{Seifert-fibered manifolds}

We begin with Seifert-fibered manifolds. We treat them in generality, while keeping in mind the main examples discussed earlier: complements of torus-links, Seifert-links, and Hopf keychains. The entanglement entropy of Seifert-fibered link-complement states was studied in \cite{Balasubramanian:2025kaf}, where it was observed that these states have a GHZ-like structure. This is natural from the viewpoint developed here. In quantum information theory, GHZ-type entanglement is intimately related to Frobenius algebras \cite{coecke2010compositionalstructuremultipartitequantum}, and in Section~\ref{2Dgraph} we identified precisely the Frobenius-algebra structure arising naturally from dimensional reduction of Seifert-fibered manifolds.

From the effective 2D TQFT description, the reduced density matrix for any bipartition is diagonal in the idempotent basis of the Frobenius algebra. Consequently, the entanglement entropy is independent of the chosen bipartition. For a Seifert-fibered manifold (\ref{seif}) with base of genus $g$, $n$ boundaries, and $k$ defect points, this entropy is
\be E=-\sum_a p_a\log p_a,~~~{p_a}={\abs{(SU_{(\alpha_1,\beta_1)})_{a\id}\cdots (SU_{(\alpha_k,\beta_k)})_{a\id}\over S_{\id a}^{2g+n+k-2}}^2\over {\sum_b \abs{(SU_{(\alpha_1,\beta_1)})_{b\id}\cdots (SU_{(\alpha_k,\beta_k)})_{b\id}\over S_{\id b}^{2g+n+k-2}}^2}}.\label{EEseif}\ee

It is useful to compare this expression with the large-party limit studied in \cite{Large-party}. For torus-link complements in $S^3$, it was shown there that the large-$n$ limit suppresses the contribution of non-abelian sectors. Equation (\ref{EEseif}) shows that the same mechanism applies to general Seifert-fibered manifolds. Namely, the denominator $S_{\id a}$ suppresses simple objects with quantum dimension greater than $1$, therefore
\be p_a\xlongrightarrow{n\to\infty} \begin{cases}
	0 & a\text{ non-abelian}\\
	{\abs{(SU_{(\alpha_1,\beta_1)})_{a\id}\cdots (SU_{(\alpha_k,\beta_k)})_{a\id}}^2\over {\sum_{b\in\text{abelian}} \abs{(SU_{(\alpha_1,\beta_1)})_{b\id}\cdots (SU_{(\alpha_k,\beta_k)})_{b\id}}}^2} & a\text{ abelian}.
\end{cases}\ee
The same conclusion hold in the limit $g\to\infty$.
 This behavior can be understood as a consequnce of the effective 2D TQFT description. In a semisimple Frobenius algebra, connected surfaces with large negative Euler number suppress idempotents with larger Frobenius trace and localize on the idempotents with minimal Frobenius trace. In the present theory, the Frobenius trace is controlled by $S_{\id a}$, and the minimal-trace idempotents are precisely the abelian anyons. The point defects do not change this suppresion mechanism, but they do modify the distribution among the surviving abelian sectors.

More generally, consider a link complement whose outermost JSJ piece is a single Seifert-fibered manifold, then from (\ref{seif})
\be \vcenter{\hbox{\scalebox{0.7}{\begin{tikzpicture}[
				tensor/.style={
					draw,
					rectangle,
					minimum width=5.5cm,
					minimum height=2.0cm,
					align=center
				},
				line/.style={thick}
				]
				
				% Main box
				\node[tensor] (Z) at (0,0) {$\mathcal Z(\text{Seifert-fibered})$};
				
				% Coordinates for input/output legs
				\foreach \x in {-1.8,0,1.8} {
					% bottom legs
					\draw[line] (\x,-1) -- (\x,-2);
					\node at (\x,-2.35) {$\vdots$};
					
					% top legs
					\draw[line] (\x,1) -- (\x,2);
				}
				
				% Horizontal dots indicating more legs
				\node at (0.9,-1.5) {$\cdots$};
				\node at (0.9, 1.5) {$\cdots$};
				\node at (0.9, 1.8) {$n$};
				
\end{tikzpicture}}}}=\sum_a {\lambda_a\over S_{\id a}^{2g+n}}\langle \tilde a|^{\otimes n}.\ee
The remaining pieces of the manifold attach to the incoming boundary components of this Seifert-fibered piece. After evaluating those inner pieces, their effect is absorbed into coefficients $\lambda_a$, along with the contributions from the marked fibers of the Seifert-fibered piece. Except in special cases where all but one of $\lambda_a$ vanish, this is again a GHZ-like state. Its entanglement entropy is
\be E=-\sum_a p_a\log p_a,~~~p_a={\abs{{\lambda_a}\over S_{\id a}^{2g+n}}^2\over \sum_b \abs{{\lambda_b}\over S_{\id b}^{2g+n}}^2}.\ee
Therefore, in the large-$n$ or large-$g$ limit,
\be p_a\xlongrightarrow{n\to\infty}\begin{cases}
	0 & a\text{ non-abelian}\\
	{\abs{\lambda_a}^2\over \sum_{b\in\text{abelian}} |\lambda_b|^2} & a\text{ abelian}.
\end{cases}\ee

In summary, both the GHZ-like form of Seifert-fibered link-complement states and the suppression of non-abelian sectors in the large-boundary (or large-genus of base surface) limits follow directly from the effective 2D TQFT description. The GHZ-like structure comes from the Frobenius algebra obtained by dimensional reduction, due to the absence of extended defects. The large-$n$ and large-$g$ limits suppress idempotents with non-minimal Frobenius trace, leaving only the abelian sectors. These GHZ-like features persist more generally whenever the outermost layer of the JSJ decomposition is a single Seifert-fibered piece.

\subsection{Hopf chains (Graph manifolds)\label{EEchain}}

We now consider the Hopf chain as a simple example of a graph manifold. This example illustrates that, although graph manifolds are built entirely from Seifert-fibered pieces, they need not inherit the simple GHZ-like structure of a single Seifert-fibered manifold. In particular, the suppression of non-abelian sectors in the large-boundary limit is no longer automatic. The reason is that the gluing maps between Seifert-fibered pieces produce non-contractible defect loops in the effective 2D description, and these loops couple the Frobenius idempotents of adjacent pieces.

Using the 2D framework developed in Section \ref{2Dgraph}, we derive analytic expressions for the entanglement entropy in a general TQFT. In this subsection we focus on the endpoint entropy, obtained by tracing out all components except one endpoint. More general contiguous bipartitions are treated in Appendix \ref{appnm}, where we also compare the analytic formulas with numerical results in the \(\mathfrak{su}(2)_k\) modular tensor categories. Related numerical studies of Hopf-chain entanglement in $\mathfrak{su}(2)_k$ appear in \cite{ramirezvaldez2024remarksentanglemententropyhopf}.

The Hopf chain $\mathcal C^n$ complement in $S^3$ is given by (\ref{Hopfchainstate})
\be \Z(S^3\setminus\nu(\mathcal C^n))=\Z_{2D}\left(~\vcenter{\hbox{\begin{tikzpicture}
			\filldraw[left color=lightgray, right color=white]
			(-1.75,0)
			-- (-1.25,0)
			to[out=-90,in=180] (-1,-0.33)
			to[out=0,in=-90] (-0.75,0)
			-- (-0.25,0)
			to[out=-90,in=180] (0,-0.33)
			to[out=0,in=-90] (0.25,0)
			-- (0.75,0)
			to[out=-90,in=180] (1,-0.33)
			to[out=0,in=-90] (1.25,0)
			-- (1.75,0)
			to[out=-90,in=180] (2,-0.33)
			to[out=0,in=-90] (2.25,0)
			-- (2.75,0)
			to[out=-90,in=0] (0.5,-1.25)
			to[out=180,in=-90] (-1.75,0);
			
			% erase the top line where the fourth circle would have been
			\draw[white,line width=10pt,line cap=round] (1.22,0) -- (1.78,0);
			
			\filldraw[left color=white,right color=lightgray] (-1.5,0) ellipse (0.25 and 0.1);
			\filldraw[left color=white,right color=lightgray] (-0.5,0) ellipse (0.25 and 0.1);
			\draw[red,thick,postaction={decorate},
			decoration={markings, mark=at position 0.8 with {\arrow{>}}}] (-0.5,0) ellipse (0.25 and 0.1);
			\filldraw[left color=white,right color=lightgray] (0.5,0) ellipse (0.25 and 0.1);
			\draw[red,thick,postaction={decorate},
			decoration={markings, mark=at position 0.8 with {\arrow{>}}}] (0.5,0) ellipse (0.25 and 0.1);
			\filldraw[left color=white,right color=lightgray] (2.5,0) ellipse (0.25 and 0.1);
			
			% dots replacing the fourth circle
			\fill (1.34,0) circle (0.03);
			\fill (1.50,0) circle (0.03);
			\fill (1.66,0) circle (0.03);
			
			% vertical circles
			
			\draw[thick,dotted,blue] (0.1,-0.78) arc (0:360:0.1 and 0.45);
			\draw[thick,blue,
			decoration={markings, mark=at position 0.1 with {\arrow{<}}}] (0.1,-0.78) arc (0:90:0.1 and 0.45);
			\draw[thick,blue] (0.1,-0.78) arc (0:-90:0.1 and 0.45);
			
			\draw[thick,dotted,blue] (1.15,-0.78) arc (0:360:0.1 and 0.45);
			\draw[thick,blue,
			decoration={markings, mark=at position 0.1 with {\arrow{>}}}] (1.15,-0.78) arc (0:90:0.1 and 0.45);
			\draw[thick,blue] (1.15,-0.78) arc (0:-90:0.1 and 0.45);
			
\end{tikzpicture}}}~\right)=\sum_{a_i} Q_{a_1a_2}\cdots Q_{a_{n-2}a_{n-1}}\langle a_1 a_2\cdots a_{n-1}\tilde a_{n-1}|.\ee
To compute the endpoint entanglement entropy, we begin with the following operator
\be (\Z(S^3\setminus\nu(\mathcal C^n)))^\dagger \Z(S^3\setminus\nu(\mathcal C^n))=\Z_{2D}\left(~\vcenter{\hbox{\begin{tikzpicture}[baseline=(current bounding box.center)]
	
	% vertical separation
	\def\sep{1.4}
	
	% the original diagram
	\def\hopfchaindiagram{
		\filldraw[left color=lightgray, right color=white]
		(-1.75,0)
		-- (-1.25,0)
		to[out=-90,in=180] (-1,-0.33)
		to[out=0,in=-90] (-0.75,0)
		-- (-0.25,0)
		to[out=-90,in=180] (0,-0.33)
		to[out=0,in=-90] (0.25,0)
		-- (0.75,0)
		to[out=-90,in=180] (1,-0.33)
		to[out=0,in=-90] (1.25,0)
		-- (1.75,0)
		to[out=-90,in=180] (2,-0.33)
		to[out=0,in=-90] (2.25,0)
		-- (2.75,0)
		to[out=-90,in=0] (0.5,-1.25)
		to[out=180,in=-90] (-1.75,0);
		
		% erase the top line where the fourth circle would have been
		\draw[white,line width=10pt,line cap=round] (1.22,0) -- (1.78,0);
		
		\filldraw[left color=white,right color=lightgray] (-1.5,0) ellipse (0.25 and 0.1);
		\filldraw[left color=white,right color=lightgray] (-0.5,0) ellipse (0.25 and 0.1);
		\draw[blue,thick,postaction={decorate},
		decoration={markings, mark=at position 0.8 with {\arrow{>}}}] (-0.5,0) ellipse (0.25 and 0.1);
		\filldraw[left color=white,right color=lightgray] (0.5,0) ellipse (0.25 and 0.1);
		\draw[blue,thick,postaction={decorate},
		decoration={markings, mark=at position 0.8 with {\arrow{>}}}] (0.5,0) ellipse (0.25 and 0.1);
		\filldraw[left color=white,right color=lightgray] (2.5,0) ellipse (0.25 and 0.1);
		
		% dots replacing the fourth circle
		\fill (1.34,0) circle (0.03);
		\fill (1.50,0) circle (0.03);
		\fill (1.66,0) circle (0.03);
		
		% vertical circles
		\draw[thick,dotted,red] (0.1,-0.78) arc (0:360:0.1 and 0.45);
		\draw[thick,red,
		decoration={markings, mark=at position 0.1 with {\arrow{>}}}] (0.1,-0.78) arc (0:90:0.1 and 0.45);
		\draw[thick,red] (0.1,-0.78) arc (0:-90:0.1 and 0.45);
		
		\draw[thick,dotted,red] (1.15,-0.78) arc (0:360:0.1 and 0.45);
		\draw[thick,red,
		decoration={markings, mark=at position 0.1 with {\arrow{<}}}] (1.15,-0.78) arc (0:90:0.1 and 0.45);
		\draw[thick,red] (1.15,-0.78) arc (0:-90:0.1 and 0.45);
	}
	
	% top copy
	\begin{scope}[yshift=\sep cm]
		\filldraw[left color=lightgray, right color=white]
		(-1.75,0)
		-- (-1.25,0)
		to[out=-90,in=180] (-1,-0.33)
		to[out=0,in=-90] (-0.75,0)
		-- (-0.25,0)
		to[out=-90,in=180] (0,-0.33)
		to[out=0,in=-90] (0.25,0)
		-- (0.75,0)
		to[out=-90,in=180] (1,-0.33)
		to[out=0,in=-90] (1.25,0)
		-- (1.75,0)
		to[out=-90,in=180] (2,-0.33)
		to[out=0,in=-90] (2.25,0)
		-- (2.75,0)
		to[out=-90,in=0] (0.5,-1.25)
		to[out=180,in=-90] (-1.75,0);
		
		% erase the top line where the fourth circle would have been
		\draw[white,line width=10pt,line cap=round] (1.22,0) -- (1.78,0);
		
		\filldraw[left color=white,right color=lightgray] (-1.5,0) ellipse (0.25 and 0.1);
		\filldraw[left color=white,right color=lightgray] (-0.5,0) ellipse (0.25 and 0.1);
		\draw[red,thick,postaction={decorate},
		decoration={markings, mark=at position 0.8 with {\arrow{>}}}] (-0.5,0) ellipse (0.25 and 0.1);
		\filldraw[left color=white,right color=lightgray] (0.5,0) ellipse (0.25 and 0.1);
		\draw[red,thick,postaction={decorate},
		decoration={markings, mark=at position 0.8 with {\arrow{>}}}] (0.5,0) ellipse (0.25 and 0.1);
		\filldraw[left color=white,right color=lightgray] (2.5,0) ellipse (0.25 and 0.1);
		
		% dots replacing the fourth circle
		\fill (1.34,0) circle (0.03);
		\fill (1.50,0) circle (0.03);
		\fill (1.66,0) circle (0.03);
		
		% vertical circles
		\draw[thick,dotted,blue] (0.1,-0.78) arc (0:360:0.1 and 0.45);
		\draw[thick,blue,
		decoration={markings, mark=at position 0.1 with {\arrow{<}}}] (0.1,-0.78) arc (0:90:0.1 and 0.45);
		\draw[thick,blue] (0.1,-0.78) arc (0:-90:0.1 and 0.45);
		
		\draw[thick,dotted,blue] (1.15,-0.78) arc (0:360:0.1 and 0.45);
		\draw[thick,blue,
		decoration={markings, mark=at position 0.1 with {\arrow{>}}}] (1.15,-0.78) arc (0:90:0.1 and 0.45);
		\draw[thick,blue] (1.15,-0.78) arc (0:-90:0.1 and 0.45);
	\end{scope}
	
	% bottom copy (mirrored)
	\begin{scope}[yshift=-\sep cm, yscale=-1]
		\hopfchaindiagram
	\end{scope}
	
\end{tikzpicture}}}\right).\ee
We trace out all but the last component of the chain to obtain the following reduced density matrix
\be \rho_{(1|n-1)}\propto \Z_{2D}\left(~\vcenter{\hbox{
		\begin{tikzpicture}
			
			% Adjustable vertical positions
			\def\middleheight{1.15}
			\def\topheight{2.85}
			
			% =========================================================
			% Bottom piece: genus-one cup
			% =========================================================
			\begin{scope}[shift={(0,0)}]
				\filldraw[right color=white,left color=lightgray]
				(-0.4,0)
				to[out=-90,in=180] (0,-0.5)
				to[in=-90,out=0] (0.4,0);
				
				\filldraw[left color=white,right color=lightgray]
				(0,0) ellipse (0.4 and 0.1);
				
				\fill[white]
				(0.15,-0.2)
				arc (0:-180:0.15 and 0.08)
				to[out=50,in=170] (-0.115,-0.23)
				arc (160:20:0.12 and 0.08)
				to[out=10,in=130] (0.15,-0.2)
				-- cycle;
				
				\draw
				(0.15,-0.2)
				arc (0:-180:0.15 and 0.08);
				
				\draw[thin]
				(0.115,-0.23)
				arc (20:160:0.12 and 0.08);
			\end{scope}

			% =========================================================
			% Middle piece: genus-one cylinder
			% =========================================================
			\begin{scope}[shift={(0,\middleheight)}]
				\filldraw[left color=lightgray, right color=white]
				(-0.4,0)
				-- (0.4,0)
				-- (0.4,-0.5)
				to[in=-90,out=-90] (-0.4,-0.5)
				-- (-0.4,0);
				
				\filldraw[left color=white,right color=lightgray]
				(0,0) ellipse (0.4 and 0.1);
				
				\draw[dotted]
				(0.4,-0.5)
				arc (0:180:0.4 and 0.1);
				
				\fill[white]
				(0.15,-0.2)
				arc (0:-180:0.15 and 0.08)
				to[out=50,in=170] (-0.115,-0.23)
				arc (160:20:0.12 and 0.08)
				to[out=10,in=130] (0.15,-0.2)
				-- cycle;
				
				\draw
				(0.15,-0.2)
				arc (0:-180:0.15 and 0.08);
				
				\draw[thin]
				(0.115,-0.23)
				arc (20:160:0.12 and 0.08);
				
				\draw[
				red,
				thick,
				decoration={
					markings,
					mark=at position 0.55 with {\arrow{<}}
				}
				]
				(-0.10,-0.25)
				arc (0:-180:0.15 and 0.05);
				
				\draw[red,thick,dotted]
				(-0.10,-0.25)
				arc (0:180:0.15 and 0.05);
				
				\draw[
				blue,
				thick,
				decoration={
					markings,
					mark=at position 0.6 with {\arrow{>}}
				}
				]
				(0.40,-0.25)
				arc (0:-180:0.15 and 0.05);
				
				\draw[blue,thick,dotted]
				(0.40,-0.25)
				arc (0:180:0.15 and 0.05);
			\end{scope}
			
			% Curly brackets around the middle piece
			\node[scale=2.5] at (-0.72,\middleheight-0.25) {$\{$};
			\node[scale=2.5] at ( 1.4,\middleheight-0.23) {$\}^{n-3}$};

			% =========================================================
			% Top piece: pair of pants
			% =========================================================
			\begin{scope}[shift={(0,\topheight)}]
				\filldraw[left color=lightgray, right color=white]
				(-0.25,0)
				to[out=-90,in=180] (0,-0.33)
				to[in=-90,out=0] (0.25,0)
				-- (0.75,0)
				to[in=90,out=-90] (0.35,-1)
				to[out=-90,in=-90] (-0.35,-1)
				to[in=-90,out=90] (-0.75,0);
				
				\filldraw[left color=white,right color=lightgray]
				(-0.5,0) ellipse (0.25 and 0.1);
				
				\filldraw[left color=white,right color=lightgray]
				(0.5,0) ellipse (0.25 and 0.1);
				
				\draw[dotted]
				(0.35,-1)
				arc (0:180:0.35 and 0.1);
			\end{scope}
			
			\begin{scope}[shift={(1,\topheight)},rotate=180]
			\filldraw[left color=lightgray, right color=white] (-0.25,0) to [out=-90,in=180] (0,-0.33) to [in=-90,out=0] (0.25,0) to  (0.75,0) to [out=-90,in=0] (0, -0.83) to [out=180,in=-90]
			(-0.75,0);
			\filldraw[left color=white,right color=lightgray] (-0.5,0) ellipse (0.25 and 0.1);
			\filldraw[left color=white,right color=lightgray] (0.5,0) ellipse (0.25 and 0.1);
			\end{scope}
		\end{tikzpicture}
}}~\right).\ee
Above we denoted that the middle piece repeats $n-3$ times.
We define the middle piece as the transfer matrix
\be B:=\Z_{2D}\left(~\vcenter{\hbox{\begin{tikzpicture}
			\filldraw[left color=lightgray, right color=white] (-0.4,0) -- (0.4,0) -- (0.4,-0.5) to [in=-90,out=-90] (-0.4,-0.5) -- (-0.4,0);
			\filldraw[left color=white,right color=lightgray] (0,0) ellipse (0.4 and 0.1);, 
			\draw[dotted] (0.4,-0.5) arc (0:180:0.4 and 0.1);
			\fill[white]
			(0.15,-0.2)
			arc (0:-180:0.15 and 0.08)
			to[out=50,in=170] (-0.115,-0.23)
			arc (160:20:0.12 and 0.08)
			to[out=10,in=130] (0.15,-0.2)
			-- cycle;
			\draw (0.15,-0.2) arc (0:-180:0.15 and 0.08);
			\draw[thin] (0.115,-0.23) arc (20:160:0.12 and 0.08);
			
			\draw[red,thick,
			decoration={markings, mark=at position 0.6 with {\arrow{<}}}] (-0.10,-0.25) arc (0:-180:0.15 and 0.05);
			\draw[red,thick,dotted] (-0.10,-0.25) arc (0:180:0.15 and 0.05);
			\draw[blue,thick,
			decoration={markings, mark=at position 0.6 with {\arrow{>}}}] (0.40,-0.25) arc (0:-180:0.15 and 0.05);
			\draw[blue,thick,dotted] (0.40,-0.25) arc (0:180:0.15 and 0.05);
	\end{tikzpicture}}}~\right)=\sum_{a,b} {|S_{ab}|^2\over S_{\id a}S_{\id b}}|\tilde a\rangle\langle\tilde b|.\label{Bmatrix}\ee
Putting the pieces together and normalizing, the reduced density matrix for the endpoint is diagonal
\be \rho_{(1|n-1)}={\sum_b |\tilde b\rangle\langle \tilde b| \sum_a {(B^{n-3})_{ab}\over S_{\id a} S_{\id b}}\over \sum_{a,b}{(B^{n-3})_{ab}\over S_{\id a} S_{\id b}}}.\ee
Thus the entanglement entropy is given by the probabilities
\be p_a={\sum_a {(B^{n-3})_{ab}\over S_{\id a} S_{\id b}}\over \sum_{c,b} {(B^{n-3})_{cb}\over S_{\id c} S_{\id b}}}.\ee

We now consider the large-$n$ limit. The matrix $B$ is real, symmetric, with non-negative entries. In a unitary modular tensor category, it is primitive, so the Frobenius-Perron theorem implies that it has a unique eigenvector $\xi$ with strictly positive entries and maximal positive eigenvalue $\lambda$
\be B\xi=\lambda\xi.\label{PFvector}\ee
For large $n$, we can approximate $B^n\approx \lambda^n \xi^T\xi$. Inserting this expression into the endpoint probabilities
\be p_a\approx{\xi_a/S_{\id a}\over \sum_b \xi_b/S_{\id b}}.\ee
Therefore the infinite-chain endpoint entropy is determined by the Frobenius-Perron eigenvector of $B$. This behaviour is qualitatively different from the Seifert-fibered case. Since $\xi_a>0$, the distribution has full support on all simple objects. Thus non-abelian sectors are not suppressed to zero in this limit.

In a purely abelian theory all entries of the $S$-matrix have modulus $1$, so that $B_{ab}=1$. The FP eigenvector is the unique eigenvector with the non-zero eigenvalue
\be \xi_a=1.\ee
In the large-$n$ limit we therefore have
\be p_a={1\over N},\ee
which achieves the maximal entanglement.

For a general contiguous bipartition $(m|n-m)$, with both $m$ and $n$ large, we can again obtain an analytic expression in terms of the FP eigenvector $\xi$. We derive this formula in Appendix~\ref{appnm} and compare it with numerical results in $\mathfrak{su}(2)_k$.

\subsection{Including hyperbolic pieces: Whitehead doubles}

Finally, we consider Whitehead-doubled links in $S^3$ as a simple class of examples whose complements contain hyperbolic pieces in their JSJ decomposition. We will see that the inclusion of these pieces leads to behaviour that is qualitatively different from both the Seifert-fibered and the graph-manifold cases discussed above.

The Whitehead doubling operator is (\ref{Pw})
\be \P_{\mathcal W}=\sum_{a,c} |c\rangle\langle a| {S^c_{aa}\over S_{\id a}}.\ee
As discussed in section \ref{subsec:W}, viewing this as a linear map $\P_{\mathcal W}:\V_*\to \V_*$, its image has dimension at most $1+\mathcal N$, where $\mathcal N$ is the number of non-abelian anyons. Thus, Whitehead doubling suppresses the contribution of the abelian anyons (other than the vacuum). This is, in some sense, a complementary behavior to the large-$n$ limit in Seifert-fibered manifolds, where the non-abelian sectors are suppressed instead. Ultimately, this reflects the fact that the Whitehead link has vanishing linking number, so abelian anyons cannot distinguish it from the unlink.

An immediate consequence is an upper bound on the entanglement entropy. For a bipartition $(m|n-m)$ of a Whitehead double of an $n$-component link, the state on each side effectively takes values in a Hilbert space of dimension at most $1+\mathcal N$ per component. Therefore
\be E\leq \min(m,n-m) \log(1+\mathcal N),~~~\mathcal N=\#\text{ of non-abelian anyons}.\ee

\begin{figure}
	$$\includegraphics[width=0.8\textwidth]{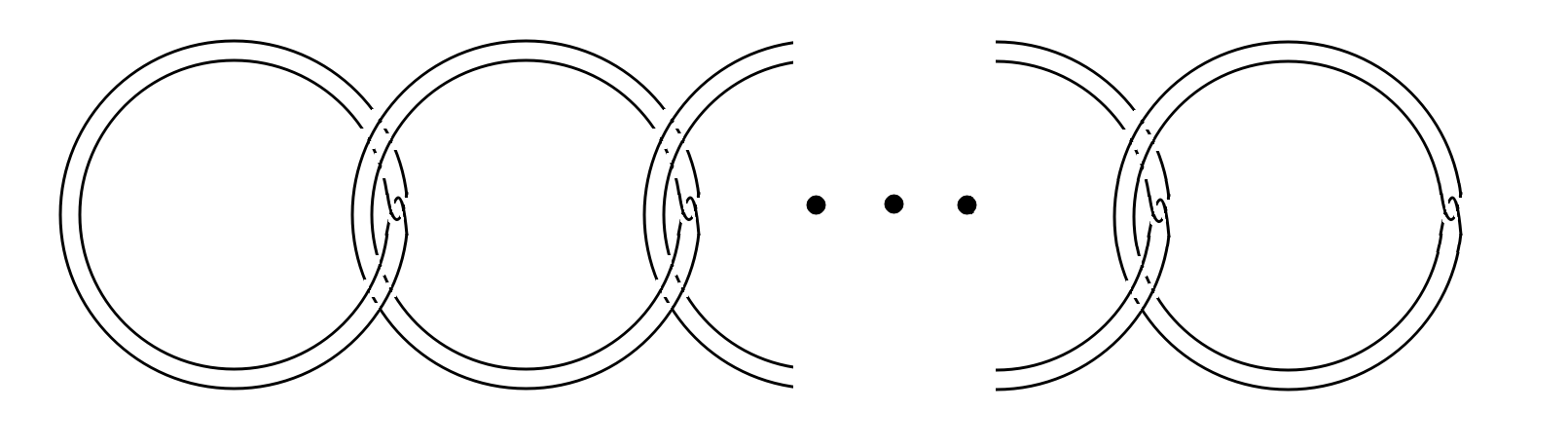}$$
	\caption{The Whitehead double of the Hopf chain.\label{hopfchaindouble}}
\end{figure}

We now turn to the Whitehead double of the Hopf chain (see figure \ref{hopfchaindouble}). There are several variations one could consider, such as alternating chains or twisted Whitehead doubles, but for simplicity we focus on the iterated untwisted Whitehead double. If $l$ denotes the number of iterations, then the corresponding complement state is obtained by acting with the Whitehead operator $l$ times on each boundary component
\be \Z(S^3\setminus\nu(\mathcal C^n))(\P_{\mathcal W}^l\otimes\cdots\otimes \P_{\mathcal W}^{l}).\ee

Tracing out all but the first component of the chain, we obtain a reduced density matrix of the form
\be \rho_l={\sum_{a,b}L^{(l)}_{a,b}|a\rangle\langle b|\over \tr(L^{(l)})},\ee
where we defined
\be L^{(l)}:=(\P_{\mathcal W}^l)^\dagger A_n \mathcal \P_{\mathcal W}^l ,\ee
\be A_k:=Q((\mathcal P_{\mathcal W}^l \mathcal P_{\mathcal W}^{l\dagger})\circledcirc A_{k-1})Q^\dagger, ~~~A_2=S\P_{\mathcal W}^l\P_{\mathcal W}^{l\dagger}\overline S .\ee
Above $\circledcirc$ denotes the Hadamard product of two matrices and $A_k$ were defined recursively.

\begin{figure}
	\includegraphics[width=1\textwidth]{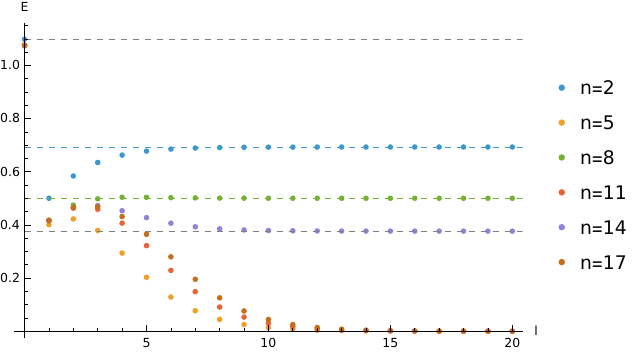}
	\includegraphics[width=1\textwidth]{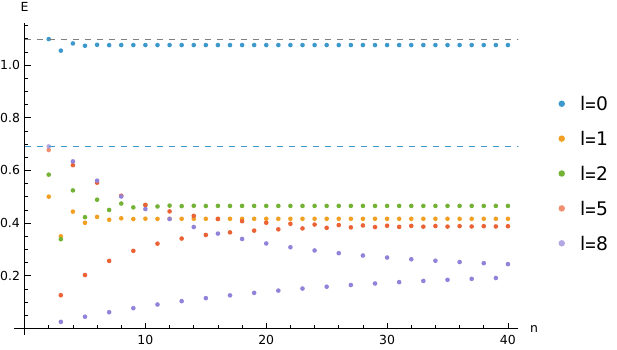}
	\caption{The endpoint entanglement entropy of the iterated Whitehead-double of the Hopf chain in the Ising category. On the top plot, the entanglement entropy is plotted against the iteration index $l$ for various chain lengths $n$. On the bottom plot, the entanglement entropy is plotted against the chain length $n$ for various values of the iteration index $l$. The grey dashed line denotes the maximal entropy $\log(3)$. The colored dashed lines denote the $l\to\infty$ asymptotic value (\ref{EEasym}). The blue dashed line also denotes the maximal entropy of the Whitehead double $\log(2)$.  The regular Hopf chain corresponds to $l=0$.  \label{figWchain}}
\end{figure}

 The minimal theory that illustrates the behavior of Whitehead doubling is the Ising category, which contains one non-abelian anyon $\sigma$ and one abelian $\psi$, in addition to the vacuum. Details of this category are found in appendix \ref{appWH}.
In this example, the Whitehead pattern is
\be \P_{\mathcal W}=|\id\rangle\langle\id|+|\id\rangle\langle \psi|+\sqrt 2 e^{-i\pi/4}|\sigma\rangle\langle\sigma|.\ee
Its image is two-dimensional, as expected from the suppression of the non-vacuum abelian sector $\psi$. At first sight, one might expect that iterating the Whitehead pattern would eventually project entirely onto the non-abelian sector, since
\be {\P_{\mathcal W}^l\over \tr(\P_{\mathcal W}^l)}\xrightarrow{l\to\infty} |\sigma\rangle\langle\sigma|.\ee
This does happen in some cases, but not in general. The point is that the limit must be taken at the level of the reduced density matrix. The interaction with the other linear maps in the chain can suppress the leading behavior of $\P_{\mathcal W}^l$, so that subleading contributions, including components along the vacuum direction, may dominate in the final reduced state.

This subtlety is visible already in the iterated Whitehead double of the Hopf chain. As shown in figure \ref{figWchain}, the behaviour depends on the parity of the number of components. For odd $n$, iterating the Whitehead pattern drives the endpoint entanglement entropy to zero, indicating that the state asymptotically collapses onto a one-dimensional subspace. In contrast, for even $n$, the entanglement entropy converges to a finite positive value\footnote{In fact, this value can be calculated analytically \be E\xrightarrow{l\to\infty} -{1\over 1+n/2}\log{1\over 1+n/2}-{n/2\over 1+n/2}\log{n/2\over 1+n/2}\label{EEasym}.\ee}.
 Thus both of these asymptotic behaviors occur within this set of examples.

More concretely, figure \ref{figWchain} compares the endpoint entanglement entropy of the ordinary Hopf chain, the Whitehead double of the Hopf chain, and higher iterated Whitehead doubles, as a function of the chain length $n$ and the iteration number $l$. A single application of the Whitehead pattern already lowers the entropy below the bound $\log 2$, reflecting the fact that the image of $\P_{\mathcal W}$ in the Ising theory is $2$-dimensional. Further iterations then separate the even and odd chains. For odd $n$, the entropy is driven to zero, whereas for even $n$ it approaches a nonzero asymptotic value.

\section{Summary}

One of the main purposes of this paper was to describe the satellite and splice constructions in the language of 3D TQFT, and to explain how they act on link-complement states. The satellite formula \eqref{pattern_operator} gives a prescription for constructing the TQFT satellite operator, while the more general splice construction is given by formula \eqref{splice_operator}. These formulas provide a practical way to build complex link-complement states from simpler pieces, provided that the individual link invariants can be evaluated by diagrammatic techniques in the modular tensor category, or by other methods.

This point of view also gives a systematic way to organize the structure of link-complement states. The JSJ decomposition cuts a three-manifold along tori into Seifert-fibered and hyperbolic pieces. Under the TQFT, these pieces become (multi)linear operators, and the original link-complement state is decomposed into a network of TQFT operators. In this sense, Seifert-fibered and hyperbolic pieces should be regarded not only as building blocks of the manifold, but also as elementary TQFT operators from which any link-complement state can be assembled.

We applied this perspective to the entanglement structure of link-complement states. For Seifert-fibered manifolds, we showed that the GHZ-like form of the state follows directly from the structure of the effective 2D theory obtained by dimensional reduction along the circle fibers. This reduced theory is governed by a Frobenius algebra, with additional point defects. Since these defects are not extended, they preserve the underlying GHZ-like structure. This gives a conceptual explanation of the GHZ-like behaviour observed in \cite{Balasubramanian:2025kaf}, and relates it to the known connection between Frobenius algebras and GHZ-type entanglement \cite{coecke2010compositionalstructuremultipartitequantum}. It also explains why, in the limit of many boundaries, non-abelian sectors are suppressed for Seifert-fibered link complements.

We then considered graph manifolds, which are built entirely from Seifert-fibered pieces. Their effective 2D description includes non-contractible loop defects, which can spoil the GHZ-like form of the state. The Hopf chain provides a concrete example of this phenomenon. We showed analytically that in the large-chain limit, both abelian and non-abelian sectors contribute to the entanglement entropy.

Finally, we studied the effect of inserting hyperbolic pieces by Whitehead doubling. The Whitehead pattern has vanishing linking number, and therefore abelian anyons cannot distinguish it from the unlink. At the level of the satellite operator, this appears as a projection of all abelian sectors to the vacuum sector. As a result, Whitehead-doubled links have a reduced upper bound on their entanglement entropy.

The broader lesson is the JSJ decomposition of a link complement manifold gives useful information about the structure of the corresponding quantum state. Seifert-fibered manifolds lead to effectively 2D states with GHZ-like behaviour. Combining Seifert-fibered pieces leads to more complex behavior, but they are still effectively 2D theories with extended defects. On the other hand, hyperbolic pieces not only introduce genuine 3D topology, but also probe deeper categorical data of the TQFT. This suggests that the JSJ decomposition provides a natural organizing principle for the study of link-complement states.

\appendix

\renewcommand{\theequation}{\thesection.\arabic{equation}}

\section{Punctured $S$-matrix in terms of $R$ and $F$ symbols\label{app:SRF}}

The punctured $S$-matrix is determined by the diagram	
	\be \begin{tikzpicture}[baseline=0, thick,scale=.5, shift={(0,-4.8)}]
		
		%	\draw[help lines] (0,0) grid (5,10);
		
		\begin{scope}[decoration={markings, mark=at position 0.35 with {\arrow{<}}}]
			%\draw (2,5) circle (1.25cm);
			%\draw (4,5) circle (1.25cm);
			\draw[postaction={decorate}] (2, 5+1.25) arc (90:360:1.25cm);
			\draw (2, 5+1.25) arc (90:45:1.25cm);
			\draw (2+1.25, 5) arc (0:30:1.25cm);
			\draw (2-1.25,5) node[left] {\small $a$};
		\end{scope}
		
		\begin{scope}[decoration={markings, mark=at position 0.005 with {\arrow{>}}}]
			\draw[postaction={decorate}] (4-1.25, 5) arc (180:-90:1.25cm);
			\draw (4, 5-1.25) arc (270:225:1.25cm);
			\draw (4-1.25, 5) arc (180:210:1.25cm);
			\draw (4-1.25,5) node[left] {\small $b$};
		\end{scope}
		
		\begin{scope}[decoration={markings, mark=at position 0.5 with {\arrow{<}}}]
			\draw[black, postaction=decorate] (2,5-1.25) to [out=-90, in=-90] (4, 5-1.25); 
			\draw (3,3-0.2) node {\small $c$};
		\end{scope}
	\end{tikzpicture}=\sum_d {\sqrt{d_d\over d_ad_b}} R_d^{\bar a b}R_{d}^{b\bar a}  \begin{tikzpicture}[baseline=0, thick,scale=.5, shift={(0,-4.8)}]
		
		%\draw[help lines] (0,0) grid (5,10);
		
		\begin{scope}[decoration={markings, mark=at position 0.35 with {\arrow{<}}}]
			%\draw (2,5) circle (1.25cm);
			%\draw (4,5) circle (1.25cm);
			\draw[postaction={decorate}] (2, 5+1.25) arc (90:330-5:1.25cm);
			\draw (2, 5+1.25) arc (90:30+5:1.25cm);
			\draw (2-1.25,5) node[left] {\small $a$};
		\end{scope}
		
		\begin{scope}[decoration={markings, mark=at position 0.4 with {\arrow{<}}}]
			\draw[postaction={decorate}] (4, 5-1.25) arc (-90:150-5:1.25cm);
			\draw (4, 5-1.25) arc (-90:-150+5:1.25cm);
			\draw (6-0.75,5) node[left] {\small $b$};
		\end{scope}
		
		\begin{scope}[decoration={markings, mark=at position 0.5 with {\arrow{<}}}]
			\draw[black, postaction=decorate] (2,5-1.25) to [out=-90, in=-90] (4, 5-1.25); 
			\draw (3,5) node[right] {\small $d$};
		\end{scope}
		
		\begin{scope}[decoration={markings, mark=at position 0.5 with {\arrow{>}}}]
			\draw[black, postaction=decorate] (3,4+0.25) to (3, 6-0.25); 
			\draw (3,3-0.2) node {\small $c$};
		\end{scope}
	\end{tikzpicture}=\sum_d {\sqrt{d_d d_ad_b}} R_d^{\bar a b}R_{d}^{b\bar a} F_{\bar ba\bar c}^{\bar ab\bar d},\ee
	from which we obtain
	\be S_{ab}^c={1\over \mathcal D\sqrt{d_c}}\begin{tikzpicture}[baseline=0, thick,scale=.5, shift={(0,-4.8)}]
		
		%	\draw[help lines] (0,0) grid (5,10);
		
		\begin{scope}[decoration={markings, mark=at position 0.35 with {\arrow{<}}}]
			%\draw (2,5) circle (1.25cm);
			%\draw (4,5) circle (1.25cm);
			\draw[postaction={decorate}] (2, 5+1.25) arc (90:360:1.25cm);
			\draw (2, 5+1.25) arc (90:45:1.25cm);
			\draw (2+1.25, 5) arc (0:30:1.25cm);
			\draw (2-1.25,5) node[left] {\small $a$};
		\end{scope}
		
		\begin{scope}[decoration={markings, mark=at position 0.005 with {\arrow{>}}}]
			\draw[postaction={decorate}] (4-1.25, 5) arc (180:-90:1.25cm);
			\draw (4, 5-1.25) arc (270:225:1.25cm);
			\draw (4-1.25, 5) arc (180:210:1.25cm);
			\draw (4-1.25,5) node[left] {\small $b$};
		\end{scope}
		
		\begin{scope}[decoration={markings, mark=at position 0.5 with {\arrow{<}}}]
			\draw[black, postaction=decorate] (2,5-1.25) to [out=-90, in=-90] (4, 5-1.25); 
			\draw (3,3-0.2) node {\small $c$};
		\end{scope}
	\end{tikzpicture}={1\over \mathcal D}\sum_d \sqrt{d_d d_ad_b\over d_c} R_d^{\bar a b}R_{d}^{b\bar a} F_{\bar ba\bar c}^{\bar ab\bar d}.\ee
	Using that $R^{ab}_cR^{ba}_c=N_{ab}^c {\theta_c\over\theta_a\theta b}$ we can write
	\be S_{ab}^c={1\over \mathcal D}\sum_d \sqrt{d_d d_ad_b\over d_c} {\theta_d\over\theta_a\theta_b} N_{\bar ab}^d F_{\bar ba\bar c}^{\bar ab\bar d}.\ee
	For $c=\id$, using that $F_{\bar ba\id}^{\bar ab\bar d}=\sqrt{d_d\over d_ad_b}$, we obtain
	\be S_{ab}^{\id}={1\over \mathcal D} {\sum_d N^d_{b\bar a} d_d\theta_d\over \theta_a\theta_b}=S_{ab}.\ee
	In the main text, we only need the entries $a=b$ of the punctured $S$-matrix
	\be S_{aa}^c={d_a\over \mathcal D}\sum_d \sqrt{d_d\over d_c} N^d_{a\bar a}{\theta_d\over \theta_a^2} F_{\bar aac}^{\bar aa\bar d}.\label{Scaa}\ee
	Clearly for $c\neq\id$, this vanishes if $a$ is an invertible anyon.

\section{Entanglement of the Hopf chain\label{appnm}}

In this appendix we study the $(m|n-m)$ bipartition of the Hopf chain, while in subsection \ref{apphopfsu2} we study how the entanglement grows with the length of the chain and bipartition numerically in $su(2)_k$.

 The reduced density matrix for $(m|n-m)$ bipartition of the Hopf chain can be expressed as
\be \rho_{(m|n-m)}={\sum \overline A_m(\vec a) A_m(\vec a')H_{n-m}(a_m,a_m')|a_1\cdots a_m\rangle\langle a_1'\cdots a_m'|\over \sum A_m(\vec a)\overline A_m(\vec a')H_{n-m}(a_m,a_m)},\ee
where we defined
\be \vec a=(a_1,\dots,a_m),\ee
\be A_m(\vec a)=Q_{a_1a_2}\cdots Q_{a_{m-1}a_m},\ee
\be H_{k}(a_m,a_m')=\sum_b {\overline S_{a_m b}S_{a_m'b}}\sum_z {(B^{k-2})_{zb}\over S_{\id z} S_{\id b}}.\ee
For large $k=n-m$, the matrix $B^{k}$ approaches a projector to its FP eigenvector
\be (B^{k})_{ab}\approx \lambda^k \xi_a\xi_b\label{Blarge}\ee
and we have
\be H_k(a_m,a_m')\approx\sum_{b} {\overline S_{a_m b}S_{a_m'b}\over S_{\id b}} \xi_b\sum_z \lambda^{k-2}\xi_z/S_{\id z}.\ee
The density matrix in this limit reads
\be \rho_{(m|n-m)}\approx{\sum \overline A_m(\vec a) A_m(\vec a')\sum_b {\xi_b\over S_{\id b}}\overline S_{a_mb}S_{a_m'b}|a_1\cdots a_{m-1}a_m\rangle\langle a_1'\cdots a_{m-1}'a_{m}'|\over \sum A_m(\vec a)\overline A_m(\vec a)\sum_b {\xi_b\over S_{\id b}}} .\ee
To simplify this expression, define the orthonormal vectors
\be |\chi_b\rangle={1\over \sqrt{g_b}}\sum_{a_1,\dots,a_{m-1}}\overline A_m(a_1,\dots,a_{m-1},b) |a_1,\dots,a_{m-1},b\rangle,\label{chibasis}\ee
where we defined
\be g_b=\sum_a (F^{m-1})_{ab},~~~F_{ab}=|Q_{ab}|^2={S_{\id a}\over S_{\id b}}{B_{ab}}.\label{fbb}\ee
Then we can write
\be  \rho_{(m|n-m)}\approx{ \sum_{b,a,a'} {\xi_b\over S_{\id b}}\overline S_{ab}S_{a'b}\sqrt{g_{a}g_{a'}}|\chi_{a}\rangle\langle \chi_{a'}|\over  \sum_{b,a} {\xi_b\over S_{\id b}}\overline S_{ab}S_{ab}g_a}.\ee
Define
\be G_{ab}=\delta_{ab}{g_a},~~~\Delta_{ab}=\delta_{ab}{\xi_b\over S_{\id b}}.\ee
Then the density matrix can be written in the basis (\ref{chibasis}) as
\be \rho_{(m|n-m)}\approx {G^{1/2}\overline S\Delta S G^{1/2}\over \tr(G\overline S\Delta S)}.\label{EEnmm}\ee

Consider now the bipartition with both $m,n\to\infty$ such that $m/n$ is fixed. We can simplify (\ref{fbb}) using (\ref{Blarge})
\be g_b\approx \lambda^m \sum_a S_{\id a}{\xi_a\xi_b\over S_{\id b}}=\Delta_{bb}\lambda^m \sum_a S_{\id a}\xi_a.\ee
After cancelling out the factors due to normalization we obtain
\be \rho_{(m|n-m)}\approx {\Delta^{1/2}\overline S\Delta S \Delta^{1/2}\over \tr(\Delta\overline S\Delta S)}.\label{EEnminf}\ee

\subsection{Numerics in $su(2)_k$\label{apphopfsu2}}

First we begin with the endpoint entanglement of the chain. We showed in section \ref{EEchain} that the entanglement entropy of the bipartition $(1|n-1)$ of the Hopf chain as $n\to\infty$ converges to
\be E\xrightarrow{n\to\infty}-\sum p_a\log p_a,~~~p_a={\xi_a/S_{\id a}\over \sum_b \xi_b/S_{\id b}},\ee
where $\xi$ is the Frobenius-Perron eigenvector of the matrix $B$ defined in \ref{PFvector}.

\begin{figure}
	\includegraphics{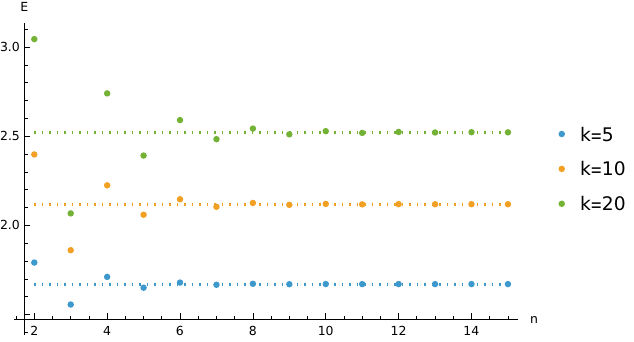}
	\caption{The entanglement entropy for the bipartition $(1|n-1)$ of the Hopf chain for $su(2)_k$. The horizontal axis is the length of the chain $n$. The dotted line corresponds to the asymptotic (large $n$) limit of the entanglement entropy, calculated directly from the FP vector. \label{chain1n}}
\end{figure}

Using the modular matrix of $su(2)_k$, the matrix $B$ can be expressed as
\be B_{ab}={\sin^2\left({(a+1)(b+1)\pi\over k+2}\right)\over \sin\left({(a+1)\pi\over k+2}\right) \sin\left({(b+1)\pi\over k+2}\right)} .\ee
We can find the FP eigenvector numerically, using Mathematica. In figure \ref{chain1n} we plot the endpoint entanglement as a function of the chain length $n$, for various values of the level $k$. We see that they converge quite rapidly to the asymptotic behavior, in an alternating pattern. We believe that this alternating behavior (which was also observed in \cite{ramirezvaldez2024remarksentanglemententropyhopf}) traces back to the JSJ decomposition of the Hopf chain, due to the fact that an odd chain canonically decomposes into $3$-chains plus a Hopf link, while an even chain decomposes into $3$-chains only. To confirm this, one would need to calculate the subleading corrections to the FP eigenvector, which is beyond our scope.

\begin{figure}
	\includegraphics{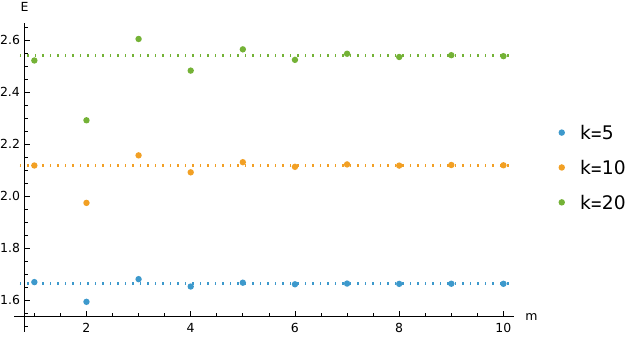}
	\caption{The entanglement entropy for the bipartition $(m|n-m)$ at $n\to\infty$ of the Hopf chain for $su(2)_k$. The horizontal axis is the length of the first subsystem $m$. The dotted line corresponds to large $m$ limit of the entanglement entropy, calculated from (\ref{EEnminf}). \label{chainmn}}
\end{figure}

Let us now turn to the general contiguous bipartitions $(m|n-m)$. For very large $n$ and finite $m$ the entanglement entropy is determined by the eigenvalues of the density matrix (\ref{EEnmm}), while at large $n,m$ and finite ratio $m/n$, it is determined by (\ref{EEnminf}). The entanglement entropy is plotted in figure \ref{chainmn} for an infinite chain ($n\to\infty$) and finite $m$. We see that the convergence to the asymptotic behavior $m\to\infty$ is very rapid.

\section{Whitehead doubling operator in the Fibonacci and Ising categories\label{appWH}}

\subsection{Fibonacci category}
The simplest MTC with non-abelian fusion rules is the Fibonacci category, which has two simple objects $\id,\tau$ and fusion rules
\be \tau\otimes\tau=\id\oplus\tau.\ee
There is a unique solution to the pentagon equations for the $F$-symbols (up to gauge freedom). The non-zero entries are the following
\be F_{\id\id\id}^{\id\id\id}=F_{\tau\tau\tau}^{\id\id\id}=F_{\id\id\tau}^{\tau\tau\id}=F_{\tau\id\tau}^{\tau\id\tau}=F_{\id\tau\tau}^{\id\tau\tau}=1 ,\ee
\be F_{\id\tau\id}^{\tau\id\tau}=F_{\tau\id\id}^{\id\tau\tau}=F_{\id\tau\tau}^{\tau\tau\tau}=F_{\tau\id\tau}^{\tau\tau\tau}=F_{\tau\tau\tau}^{\id\tau\tau}=F_{\tau\tau\tau}^{\tau\id\tau}=1,\ee
\be F_{\tau\tau\id}^{\tau\tau\id}=d_\tau^{-1},~~~F_{\tau\tau\tau}^{\tau\tau\id}=F_{\tau\tau\id}^{\tau\tau\tau}=d_\tau^{-1/2},~~~F_{\tau\tau\tau}^{\tau\tau\tau}=-d_\tau^{-1}.\ee
The quantum dimension is
\be d_\tau=\phi={1+\sqrt 5\over 2}.\ee
There are two solutions for the braiding matrix, corresponding to the left-handed and right-handed Fibonacci anyon. For the right-handed theory
\be R_{\tau}^{\tau\tau}=e^{3\pi i/5},~~~R_{\id}^{\tau\tau}=e^{-4\pi i/5}.\ee
The modular $S$ matrix is
\be S={1\over \sqrt{\phi+2}}\begin{pmatrix}
	1 & \phi\\
	\phi & -1
\end{pmatrix}.\ee
\be \theta_\tau=e^{4\pi i/5}.\ee
From (\ref{Scaa}) we find that the punctured $S$-matrix with puncture labeled by $\tau$, has only a single nonzero entry
\be S_{ab}^\tau=\delta_{a\tau}\delta_{b\tau}e^{3\pi i/10}.\ee

The Whitehead operator is given by
\be \P_{\mathcal W}=|\id\rangle\langle\id|-\phi^{-1}|\id\rangle\langle\tau|+{e^{3\pi i/10}}{\sqrt{2+\phi}\over\phi}|\tau\rangle\langle\tau| .\ee

\subsection{Ising category}

The Ising category has $3$ simple objects $(\id,\sigma,\psi)$ with fusion rules
\be \psi\times\psi=\id,~~~\psi\times\sigma=\sigma,~~~\sigma\times\sigma=\id+\psi.\ee
As a fusion category, this is the same as $su(2)_2$, with non-trivial $F$-symbols 
\be [F_{\sigma\sigma}^{\sigma\sigma}]={1\over \sqrt 2}\begin{pmatrix}
	1 & 1 \\
	1 & -1
\end{pmatrix},~~~F^{\sigma\psi\sigma}_{\sigma\psi\sigma}=F^{\psi\sigma\sigma}_{\psi\sigma\sigma}=-1.\ee
The matrix $[F_{\sigma\sigma}^{\sigma\sigma}]$ above is written in the $\{\id,\psi\}$ basis. The rest of the allowed $F$-symbols can be taken to be equal to $1$.

The hexagon equation with the above fusion data, however, has multiple solutions. The braided fusion category called the right-handed Ising category conventionally has the following choice of braiding
\be R^{\sigma\sigma}_{\id}=e^{-i\pi/8},~~~R_\psi^{\sigma\sigma}=e^{i 3\pi/8},~~~R^{\psi\psi}_{\id}=-1,~~~R^{\sigma\psi}_{\sigma}=R^{\psi\sigma}_\sigma=-i.\ee
The quantum spins are given by
\be \theta_\psi=-1,~~~\theta_\sigma=e^{\pi i/8} .\ee
This theory is also known as $(\overline{E_8})_2$.

From the above data we find the punctured $S$-matrix
\be S^{\id}={1\over 2}\begin{pmatrix}
	1 & \sqrt 2 & 1\\
	\sqrt 2 & 0 & -\sqrt 2\\
	1 & -\sqrt 2 & 1
\end{pmatrix},~~~S_{ab}^{\sigma}=0,~~~S^\psi_{ab}=\delta_{a\sigma}\delta_{b\sigma}e^{-i\pi/4}.\ee
Above, $S^{\id}$ is the modular $S$-matrix written in the basis $(\id,\sigma,\psi)$.

The Whitehead operator can then be written as
\be \P_{\mathcal W}=|\id\rangle\langle\id|+|\id\rangle\langle \psi|+\sqrt 2 e^{-i\pi/4}|\sigma\rangle\langle\sigma|.\ee

\bibliographystyle{unsrt} % This chooses the bibliography style for bibtex.
\bibliography{bibliography} 

\end{document}